\documentclass[preprint]{aastex61}

\usepackage{natbib}
\usepackage{CJK}

\newcommand{\RCS}{$\chi^2/\textrm{dof}$ }
\newcommand{\net}{n_{\rm e}t_{\rm i}}
\newcommand{\kTe}{kT_{\rm e}}
\newcommand{\netunit}{cm$^{-3}$\,s }
\newcommand{\Msun}{M_\odot}

\received{August 7, 2018}
\accepted{December 31, 2018}
\submitjournal{ApJ}

\begin{document}
\begin{CJK}{Bg5}{bsmi}
	
	\title{Spatially Resolved X-ray Spectroscopy of Kepler's Supernova Remnant: Distinct Properties of the Circumstellar Medium and the Ejecta}
	
	\correspondingauthor{Yang Chen}
	\email{ygchen@nju.edu.cn}
	
	\author{Lei Sun (®]½U)}
	\affiliation{Department of Astronomy, Nanjing University, Nanjing 210023, China}
	
	\author{Yang Chen (³¯¶§)}
	\affiliation{Department of Astronomy, Nanjing University, Nanjing 210023, China}
	\affiliation{Key Laboratory of Modern Astronomy and Astrophysics, Nanjing University, Ministry of Education, China}
	
	\begin{abstract}
		Kepler's supernova remnant (SNR) is believed to result from a Type Ia supernova, but be interacting with dense circumstellar material (CSM), which makes its progenitor system a mystery. { Using} the \textit{Chandra} ACIS-S 741\,ks effective exposure data and an advanced adaptive binning algorithm, we analyse the spectra from 
		%4671 
		tessellated regions in Kepler's SNR. For the first time, we map out the detailed spatial distributions of emission measure (EM), electron temperature, ionization parameter, and metal abundances 
		%(O, Mg, Si, S, L-shell emitting and K-shell emitting Fe) 
		{ for both the low- and high-temperature plasma components}. The { low-temperature} plasma diverges into two branches in the temperature v.s.\ ionization parameter { diagram}, which appear to be spatially associated with the warm CSM dusts and the Si- and S-rich ejecta, respectively. We construct the probability distribution functions of abundance ratios 
		{ of O and Mg to L-shell emitting Fe  ([O]/[Fe]$_{\rm  L}$ and [Mg]/[Fe]$_{\rm  L}$)}, and fit them with double Gaussians. Thereby, we distinguish the CSM from the ejecta: the CSM is characterized by [O]/[Fe]$_{\rm  L}$ $= 0.77^{+0.30}_{-0.23}$ and [Mg]/[Fe]$_{\rm  L}$  $= 1.11^{+0.46}_{-0.32}$, while the ejecta by [O]/[Fe]$_{\rm  L}$ $= 0.31^{+0.17}_{-0.10}$ and [Mg]/[Fe]$_{\rm  L}$  $= 0.38^{+0.36}_{-0.19}$. We estimate the total {hydrogen} mass of the shocked CSM as $\sim 1.4 \Msun$ and the EM-weighted mean [Mg]/[O]  $= 1.14 \pm 0.49$, which can be reproduced with an asymptotic giant branch donor star with initial mass of $\sim 4 \Msun$. The abundance ratios { from} the shocked ejecta are well compatible with the predicted results from spherical delayed-detonation models for Type Ia supernovae. We { also} find that the two ``ears'' of the remnant are dominated by Si- and S-rich ejecta, thus favoring a pre-explosion jets scenario.
	\end{abstract}
	
	\section{Introduction} \label{intro}
	Type Ia supernovae (SNe Ia) is the runaway thermonuclear explosion of a white draft (WD). The studies on SNe Ia play important roles in many areas of modern astrophysics and cosmology \citep[e.g.,	][]{1998AJ....116.1009R,1999ApJ...517..565P}.
	%, e.g. the chemical evolution of the Universe, the origin of galactic cosmic rays, the accelerating expansion of the Universe and the properties of the dark energy. 
	However, many basic aspects of the SN Ia physics, such as the nature of their progenitor systems and the detailed explosion mechanism, are still under debate. 
	{Plenty of models have been proposed for the progenitor systems of SN Ia. Most of them fall into two broad scenarios: the single degenerate (SD), where WD accretes mass from a non-degenerate companion, and the double degenerate (DD), where two WDs merge \citep[see, e.g.,][for recent reviews]{2012NewAR..56..122W,2014ARA&A..52..107M}.
	The observational clues to the progenitors of SNe Ia have been found not only from the explosions themselves, but also from their remnants.}
	
	Supernova remnants (SNRs) provide unique perspectives to study SNe physics. Particularly, 
	young SNRs that are hundreds or thousands of years old could be used to probe circumstellar material (CSM) up to several pc around the progenitor system, and thus to trace the evolution history back to thousands or even millions of years before explosion \citep[e.g.,][]{2017hsn..book.2233P}. Young SNRs are best viewed in X-ray band, as both shock-heated thermal plasma emission and non-thermal emission from accelerated particles typically peak at 0.5--10 keV \citep[e.g.,][]{2008ARA&A..46...89R,2012A&ARv..20...49V}. Moreover, the X-ray properties of these emission often show significant spatial variations, which reflect the complex physical environment in SNRs. 
	
	Over the last $\sim 20$ years, advances in X-ray imaging spectroscopy techniques provided us a direct way to map the spatial distribution of physical quantities in SNRs. Some commonly used methods to investigate spatial features are spectral analysis of individual interesting regions and equivalent width (EW) map of emission lines. In order to make full use of the information carried by X-ray data, new techniques have been developed, which can automatically construct small tessellated regions for spectral analysis \citep[e.g.,][]{2003MNRAS.342..345C,2006MNRAS.368..497D,2015MNRAS.453.3953L}. Those techniques have been successfully applied on spatially resolved spectroscopic studies of many extended X-ray sources including SNRs \citep[e.g.,][]{2018A&A...615A.150Z}. In this paper, we adopt an adaptive binning algorithm to investigate the spatial distribution of physical properties of Kepler's SNR (G4.5+6.8; or Kepler for short).
	
	Kepler's SNR is one of the mostly studied SNRs. It was firstly discovered by \citet{1943ApJ....97..119B} as the remnant of historical SN 1604. It's optical spectra show evidences for the emission from high-density material with enhanced N abundance, which was interpreted as the CSM of SN progenitor system \citep[e.g.,][]{1943ApJ....97..128M,1982A&A...112..215D,1983MNRAS.204..273L,1991ApJ...366..484B}, thus prefer a core-collapse (CC) SN. Its location ($590d_5$ pc away from the Galactic Plane, where $d_5$ is the distance to the remnant in units of 5\,kpc) indicates a Type Ia origin. Debates on the origin type of Kepler have lasted for a long time until detailed X-ray studies based on \textit{ASCA} \citep{1999PASJ...51..239K} and \textit{Chandra} \citep{2007ApJ...668L.135R} observations revealed a large amount of Fe, which was a conclusive evidence for a thermonuclear explosion (SN Ia). However, whether it follows a SD or a DD scenario is still an open question. Although there is no surviving companion have been found as yet \citep{2014ApJ...782...27K,2018ApJ...862..124R}, the clear CSM-interaction features require a SD scenario. Recently, several studies concerning the CSM-interaction has been carried out, the observations and hydrodynamic simulations suggested that the companion is most likely an evolved, i.e., an asymptotic giant branch (AGB) star, with a slow and massive wind \citep[e.g.,][]{2007ApJ...662..998B,2012A&A...537A.139C,2012ApJ...755....3W,2013ApJ...764...63B,2014MNRAS.442..229T}. On the other hand, a core-degenerate (CD) scenario has also been discussed in consideration of the missing companion and the ``ears'' feature of the remnant \citep{2013MNRAS.435..320T}.
	
	In this work, we apply an adaptive binning algorithm to the deep \textit{Chandra} observations of Kepler's SNR. With this spatially resolved spectroscopic study, we aim at decomposing the thermal and non-thermal emission components, mapping out detailed distribution of physical quantities, and further characterizing the different properties of shock-heated ejecta and CSM. We describe the spectral analysis and its results in Section \ref{sec:analysis}, and then discuss the physical properties in Section \ref{sec:discussion}, finally we have a summary in Section \ref{sec:conclusion}.
	
	\section{Observations and Data Analysis} \label{sec:analysis}
	
	\subsection{Observations}
	 %{Kepler's SNR has been observed with the \textit{Chandra X-Ray Observatory} for four times: in 2000 (PI: S. Holt), 2004 (PI: L. Rudnick), 2006 (PI: S. Reynolds), and 2014 (PI: K. Borkowski). The time difference between the first observation and the last one is 13.86\,yr. \citet{2017ApJ...845..167S} have investigated the proper motions of 14 compact X-ray-bright knots in Kepler using these observation data. They found that several knots have large proper motions ($\gtrsim0.1''$\,yr$^{-1}$). Thus the shifts can be up to $>1''$ during a time range of more than 10 years, which are comparable with the typical scale of the tessellated bins in our work (usually several arc-seconds, see below). Therefore, we only use the deepest observation taken in 2006 for the spectral analysis in this work.} 
	 
	 {Kepler's SNR has been observed with the \textit{Chandra X-Ray Observatory} for four times: in 2000 (PI: S. Holt), 2004 (PI: L. Rudnick), 2006 (PI: S. Reynolds), and 2014 (PI: K. Borkowski). This work exploits the deepest observation}
	  \footnote{{The time difference between the first observation and the last one is 13.86\,yr. \citet{2017ApJ...845..167S} have investigated the proper motions of 14 compact X-ray-bright knots in Kepler using all the four observations. They found that several knots have large proper motions ($\gtrsim0.1''$\,yr$^{-1}$). In addition, the velocity of the main shock can be up to $\sim 2300$\,km\,s$^{-1}$ \citep{2016ApJ...817...36S}, corresponding also to $\sim0.1''$\,yr$^{-1}$.  Thus the shifts of both the ejecta knots and the main shock can be up to $>1''$ during a time range of more than 13 years, which are comparable with the typical scale of the tessellated bins in our work (usually several arc-seconds, see below). Therefore, the other three observations are not used.} },
	   which is made between 2006 April and July as a Large Project (LP). It contains six installments and has a total effective exposure time of 741\,ks \citep{2007ApJ...668L.135R}. 
	 All the observations were made with the ACIS-S CCD camera and covered the SNR with the backside-illuminated S3 chip. Information of all six observations is summarized in Table \ref{tab:obs}, and the merged image {is} shown in Figure \ref{fig:SNR}. {All these data} were processed using CIAO (version 4.9)\footnote{http://cxc.harvard.edu/ciao/} and calibrated using CALDB (version 4.7.3)\footnote{http://cxc.harvard.edu/caldb/}. We use XSPEC (version 12.9.0u)\footnote{https://heasarc.gsfc.nasa.gov/xanadu/xspec/} for spectral analysis.

	\subsection{Adaptive binning algorithm and spectral fitting model}
	In order to investigate the spatial distribution of physical parameters through out the whole SNR region, we adopt an adaptive image binning method called the Weighted Voronoi Tessellations (WVT) binning algorithm \citep{2006MNRAS.368..497D}, which is a generalization of \citet{2003MNRAS.342..345C} Voronoi binning algorithm. 
	We produce the merged counts image of the SNR region for adaptive binning. With the WVT binning algorithm, the whole spectral analysis region is divided into 4671 bins, and each bin contains $\sim$ 6400 counts (S/N $\sim$ 80). We then extract X-ray spectra in 0.3--8.0\,keV energy band and perform spectral analysis in each of the tessellated regions. For each region and for each observation, separate spectra and response files [redistribution matrix files (RMFs) and ancillary response files (ARFs)] are generated. All the spectra are rebinned to at least 25 counts per bin for chi-square test in subsequent spectral fitting. The background spectra are extracted from an annulus region around the SNR (see Figure \ref{fig:SNR}, several point sources are excluded manually).
	
	We {jointly} fit the background-subtracted spectra in each region with an absorbed ``\texttt{vnei} $+$ \texttt{srcut}'' model. The \texttt{vnei} component describes the thermal emission from the collisional ionized plasma in non-equilibrium ionization (NEI) state, which is characterized by a constant temperature and single ionization parameter. We use the \texttt{srcut} component to account for the synchrotron emission from the shock-accelerated non-thermal electrons, which may be especially dominating on the periphery of the remnant. We also add a Gaussian line at 1.2\,keV to account for the missing Fe L line in \texttt{vnei} model \citep[e.g.,][]{2011PASJ...63S.837Y}. All these components are subjected to foreground absorption described with a \texttt{wabs} model, which is a photo-electric absorption model using Wisconsin cross-sections \citep{1983ApJ...270..119M}.
	
	In regard to the parameter treatment, we assume the absorbing column density through out the whole remnant is constant and fix it at $N_{\rm H} = 5.2 \times 10^{21}$\,cm$^{-2}$ according to previous studies \citep[e.g.,][]{2005ApJ...621..793B,2007ApJ...668L.135R}. For the \texttt{vnei} component, we allow the abundances of O, Mg, Si, S and Fe to vary, tie the abundance of Ni to Fe, and leave the abundances of all the other elements fixed to solar. For the \texttt{srcut} component, we set only the normalization (radio flux density at 1\,GHz, $S_{\rm 1GHz}$) as free parameter. The radio spectral index is fixed at $\Gamma = 0.71$, which is the mean value determined by \citet{2002ApJ...580..914D}. The break frequency is fixed at $\nu_{\rm roll-off} = 3.6 \times 10^{17}$\,Hz based on \citet{2005ApJ...621..793B}. In addition, we note that letting the redshift in \texttt{vnei} component vary freely may help to reduce the residual. This could be understood because the X-ray emitting plasma may have very high radial velocity, hence a velocity component along the line of sight, which can be characterized by redshift. However, in consideration of the possible degeneration between the parameters, we fit the redshift separately with all the other parameters fixed to their best-fitting value in the rest frame.
	
	We caution that this two-component model may still be oversimple and does not take into account the possible multi-temperature structure and the ionization history of the plasma. However, most of bins are small enough thus the physical properties can be assumed to be uniform within each of them. Therefore, the spectral model {described above} 
	%we adopted in this work
	 is aimed at decomposing the thermal and non-thermal emission, and further characterizing the distribution of average thermal and ionization states of the {X-ray emitting} plasma.
	
	\subsection{Spectral fitting results}
	Our ``\texttt{vnei} $+$ \texttt{srcut}'' model gives acceptable fitting results for most of the tessellated regions with reduced chi-square \RCS $\sim$ 1.0--1.2 (Figure \ref{fig:RCS}). 
	{The average degree of freedom in the fits is $\sim 200$.}
	For those regions with large \RCS, residuals could be caused by the potential multi-temperature structure of the plasma.
	{Examples of spectra extracted from four tessellated regions with different features are shown in Figure \ref{fig:spec_example}. The spectra of BIN-1471 and BIN-2402 show significant Mg emission line at $\sim 1.35$\,keV, while those of BIN-715 and BIN-3992 are dominated by Fe-L complex. The best-fit parameters are summarized in Table \ref{tab:spec_example}. }
	
	We construct the spatial distribution maps of the parameters based on the spectral fitting results, demonstrated in Figure \ref{fig:th_par_map}. Panel (a) and (b) show the decomposed spatial distribution of thermal and non-thermal emission, which are characterized by {specific} emission measure (EM\footnote{ Defined as ${\rm EM} = 10^{-14}/(4\pi d^2)\int n_{\rm e} n_{\rm H} f dV$, where $d$ is the distance to the SNR, $n_{\rm e}$ and $n_{\rm H}$ are the electron and hydrogen densities, respectively, and $f$ is the volume filling factor. The specific EM denotes the EM per squared angular size.}) and 1 GHz radio {brightness} $S_{\rm 1GHz}$ respectively. Panel (c) shows the map of electron temperature. Panel (d) shows the map of ionization parameter $\net$. Finally, panel (e)--(i) show the distributions of elemental abundances ([O], [Mg], [Fe]$_{\rm L}$\footnote{ [Fe]$_{\rm L}$ denotes the abundance of L-shell emitting Fe.}, [Si], and [S]).
	\subsubsection{Thermal and non-thermal emission}
	The thermal and non-thermal components are well disentangled with our spatially resolved spectroscopic analysis (Figure \ref{fig:th_par_map}--a,b). Thermal emission mainly arise from the northwest shell as annularly distributed bright knots and the central region as a bar-like structure. In these regions we obtain high volume EM, indicating high density of the shock-heated gas. In addition to these compact features, diffuse thermal emission can be seen all over the remnant with a more symmetric distribution. 
	{We get a total thermal EM $\sim 0.14$\,cm$^{-5}$ for the entire remnant, which corresponds to $n_{\rm e}M_{\rm gas} \sim 49$\,$\Msun$\,cm$^{-3}$ assuming a distance of 5.1\,kpc.\footnote{ Assuming an uniform distribution of the electron density, we can rewrite the definition of EM by ${\rm EM} \approx 10^{-14}/(4\pi d^2)~n_{\rm e} M_{\rm gas}/(\mu_{\rm H} m_{\rm p})$, where $M_{\rm gas}$ is the total gas mass and $\mu_{\rm H}$ is the mass correction factor for metal abundance ($\mu_{\rm H} \sim 1.4$ for a plasma with solar abundance). Sometimes, $n_{\rm e} M_{\rm gas}$ is equivalently referred to as ``emission measure'' with the distance given \citep[e.g.,][]{2007ApJ...662..998B}.}} 
	On the other hand, non-thermal emission is especially dominating on the outskirts and appears as 
	{narrow filaments around the periphery of the remnant}. 
	The 1\,GHz {brightness} of these filaments is typically several mJy\,arcmin$^{-2}$. And we get an integrated flux of $\sim$ 23\,Jy, which basically consistent with the radio observations \citep[$\sim$ 19\,Jy, see, e.g.,][]{2002ApJ...580..914D,2014BASI...42...47G}.
	\subsubsection{Electron temperature and ionization parameter}\label{sec:kT_net}
	The thermal and ionization states of the plasma are characterized by electron temperature $\kTe$ and ionization parameter $\net$. With the spectral analysis, both of $\kTe$ and $\net$ show distinct spatial variations (Figure \ref{fig:th_par_map}--c,d). The outermost layer of the remnant has the highest temperature ($\gtrsim$ 1\,keV) and lowest ionization parameter ($\lesssim~10^{11}$\,\netunit), indicating the gas there {is} recently shocked. The temperature of the inner part is typically lower. We obtain an EM-weighted mean temperature $\kTe = 0.52^{+0.16}_{-0.13}$\,keV and an EM-weighted mean ionization parameter log$_{10}(\net) = 11.14 \pm 0.39$ for the entire remnant.\footnote{The error bar denotes the standard deviation (STD) of the physical quantity, which characterizes its dispersion in the entire remnant.} There are also some high-temperature, low-ionization regions inside the remnant. They can either be outer gas heated by forward-shock which is projected into the center or inner ejecta newly heated by reverse-shock. In consideration of the enhanced metal abundances, we suggest that most of these regions consist of reverse-shock heated ejecta (except for the central bar-like structure which can be interpreted as CSM, see Section \ref{subsec:CSM}). We also note that there are several tessellated regions show strikingly high $\net$ (the white bins in Figure \ref{fig:th_par_map}--d, mainly along the SE rim). Most of them are among the non-thermal-dominated areas, thus the thermal component is poorly constrained. However, a high ionization parameter can also be explained by a rather small filling factor of the thermal emitting gas (see Section \ref{subsec:density}).
	\subsubsection{Metal abundances}
	The abundances and distributions of {different} metal species could help to reveal the progenitor system and the explosion mechanism of SNe. Specifically, iron group elements (IGEs) like Fe, Ni and intermediate mass elements (IMEs) such as Si, S, Ca are the major products in SNe Ia nucleosynthesis, while the lighter elements like O and Mg have relatively low yields \citep[e.g.,][]{2010ApJ...712..624M}. Our spectral analysis suggests that the bulk of the remnant have solar or sub-solar O, Mg abundances and super-solar Si, S, Fe abundances (Figure \ref{fig:th_par_map}--e, f, g, h, i), thus consistent with the scenario of SN Ia ejecta interacting with the CSM. 
	
	We obtain the EM-weighted mean [O] $= 0.68^{+0.36}_{-0.23}$ and [Mg] $= 0.67^{+0.47}_{-0.28}$. High O, Mg abundances mainly appear on the edge of the remnant, especially the outermost layer of northern shell (Figure \ref{fig:th_par_map}--e,f). This has been noticed by \citet{2013ApJ...764...63B}, and is identified as shock heated CSM. 
	
	On the other hand, SN ejecta is probed by high Fe, Si and S abundances. The Fe abundance is mainly in the range of 1 $\sim$ 10, with an EM-weighted mean [Fe]$_{\rm  L}$ $= 1.44^{+1.30}_{-0.68}$ (the Fe abundance here is constrained by its L-shell emission $\sim$ 1 keV, while Fe K line $\sim$ 6.4 keV is not included due to low statistics in such high energy band, see the discussion in Section \ref{subsec:mass} {and Section \ref{subsec:hard_band_spec}}). Different from O and Mg, enhanced Fe abundance mainly appears inside the remnant (Figure \ref{fig:th_par_map}--g), indicating the reverse shock heated Fe-rich ejecta. 
	
	Si and S are produced by oxygen burning process during the explosion. The abundances of Si and S show large dispersions in the remnant, with overall EM-weighted mean [Si] $= 4.4^{+4.8}_{-2.3}$ and [S] $= 10.5^{+13.7}_{-6.0}$. The highest Si, S abundances take place on the southern edge and two ``ears'' of the remnant (Figure \ref{fig:th_par_map}--h,i). And in these regions, Si and S extent to a larger radius than Fe, indicating a stratified structure of {the} ejecta. This consistent with the results obtained by \citet{2004A&A...414..545C} based on radial profile and equivalent width map analysis.

	\section{Discussion} \label{sec:discussion}
	
	\subsection{Electron number density and ionization age} \label{subsec:density}
	The volume emission measure of the thermal plasma can be used to estimate the electron number density in Kepler's SNR. We adopt a distance to Kepler of $\sim$ 5.1 {kpc} according to recent studies \citep[e.g.][]{2016ApJ...817...36S,2017ApJ...842..112R}, and assume a spherical structure for the entire remnant and a prism geometry for the volume of each bin. We use similar method described in \citet{2018A&A...615A.150Z} to estimate the volume of each prism. The depth of the prism is determined as $l(r) = 2 \sqrt{R_{\rm SNR}^2 - r^2}$, where $R_{\rm SNR}$ is the radius of the SNR and $r$ is the projected distance to the SNR center. We take the projection center of Kepler at R.A.=$17^{\rm h}30^{\rm m}41^{\rm s}$, Decl.=$-21^\circ29'30''$ and the radius as $1'.8$ (shown in Figure \ref{fig:th_par_map}-a) according to \citet{2017ApJ...845..167S}. The electron density $n_{\rm e}$ and the hydrogen density $n_{\rm H}$ are linked by $n_{\rm e} = 1.2 n_{\rm H}$. Since the thermal plasma in Kepler has been demonstrated to be highly structured and may concentrate in a part of the prism, we leave the filling factor $f$ in the calculation results. The distribution of $f^{1/2} n_{\rm e}$ is {shown} in Figure \ref{fig:density}. 
	
	With $f^{1/2} n_{\rm e}$ and the ionization parameter $\net$ given, we can further determine the ionization age of the plasma $t_{\rm i}$,  the time past since the plasma was shocked. Shown in Figure \ref{fig:density}, $f^{-1/2} t_{\rm i}$ comes up to thousands of years in several tessellated regions such as those along the SE rim. However, $t_{\rm i}$ is expected to be smaller than the dynamic age of Kepler $\sim400$\,yr. Therefore, we can give an estimation on the filling factor as $f \lesssim [400~{\rm yr} /(f^{-1/2} t_{\rm i})]^2$. As mentioned in Section \ref{sec:kT_net}, several regions along the SE rim show strikingly high $\net$. The ionization age in these regions is calculated to be $t_{\rm i} \sim 4000 f^{1/2}$ yr, which indicates a small filling factor $f \sim 0.01$. Such a small filling factor and the enhanced Si and S abundances of the plasma imply a clumpy structure of the Si- and S-rich ejecta. It may further account for the low gas temperature in these regions due to a low shock velocity in high density clumps.
	
	\subsection{Masses of metal species} \label{subsec:mass}
	The masses of different metal species in each tessellated region are calculated using electron number density and abundances obtained before. We adopt the same geometry described in Section \ref{subsec:density}. The masses are calculated in two cases: (1)  $f = 1$ assumed for all the regions; (2)  $f = 1$ adopted only for the regions with $f^{-1/2} t_{\rm i} < 400$ yr, but $f = [400~{\rm yr} /(f^{-1/2} t_{\rm i})]^2$ adopted for the others. Finally, we obtain the total masses in the remnant by sum over all the regions, as presented in Table \ref{tab:mass}. 
	
	\citet{2015ApJ...808...49K} {had} calculated the ejecta masses for different elements, which are similar to our results, except for Fe mass (they reported a shocked Fe mass of $\sim 0.71$ $\Msun$). The surprisingly low Fe mass ({$\sim$ 0.02-0.04\,$\Msun$}) we got {may} be caused by several reasons. First and foremost, in our spectral fitting, Fe abundance is constrained merely by its L-shell emission lines, which can only come from highly ionized Fe (i.e., Fe XVII $\sim$ Fe XXIV). Although bright Fe K$\alpha$ line has also been observed in Kepler, it is not included in our spectral analysis due to low statistics in high energy band. \citet{2015ApJ...808...49K} {used} a four-component model to reproduce the integral thermal emission of Kepler, which is far more delicate than our single temperature model. Especially, they added a pure Fe component to characterize the high temperature {Fe}-rich ejecta, and took into account the Fe K$\alpha$ emission. As a result, they obtained an overall measurement of shocked Fe. Therefore, a significantly lower Fe mass obtained by our {above} analysis actually implies that the majority of Fe is still in low ionization states {and} thus not emit L-shell emission. On the other hand, \citet{2014ApJ...785L..27Y} reported a centroid energy of Fe K$\alpha$ at 6438 eV for Kepler. Such a low Fe K$\alpha$ centroid energy corresponds to a low ionization level of Fe, which is consistent with our inference. 
	
	{In order to take the K-shell emitting Fe into account and constrain the hot components, we carry out a spatially resolved analysis particularly on the hard band (2.2-8.0\,keV) spectra using the WVT algorithm. The results got from the hot ejecta are also summarized in Table \ref{tab:mass}, which indicate a large amount of Fe $\sim 0.5 \Msun$. The detailed treatment of the hard-band spectra is described below in Section \ref{subsec:hard_band_spec} and APPENDIX \ref{sec:apx}.}
	
	%In Figure \ref{fig:yield}, we compare the metal masses of Kepler to predicted yields of several supernova models, including three different models for Ia SNe \citep{2010ApJ...712..624M} and one for core-collapse SNe \citep{2006NuPhA.777..424N}. Even though the Fe mass is underestimated, a high Si,S,Fe to O,Mg mass ratio still indicates an Ia origin of Kepler.
	
	\subsection{Thermal and ionization state}
	{Despite the potentially missed high temperature components, the spectra analysis above provides new perspectives to investigate the spatial variations of the average thermal and ionization states in Kepler.} As shown in Figure \ref{fig:th_par_map}--a, b, both of electron temperature $\kTe$ and ionization parameter $\net$ show large dispersions and significant spatial variations. $\kTe$ ranges from 0.3 keV to $\gtrsim 1.5$ keV, {and} $\net$ varies in $10^{10}-10^{12}$ \netunit. The hottest gas distributes mainly on the north rim of the remnant, with an average electron temperature $\sim 1.2$ keV. Its low ionization age suggests that the gas is newly shock-heated. In principle, shock velocity can be inferred from the average temperature of the gas immediately behind the shock front:
%	\begin{equation}
%	V_{\rm s} = \left(\frac{16}{3}\frac{\overline{kT}}{\mu m_{\rm p}}\right)^{1/2} \approx 1000~{\rm km~s^{-1}}\left(\frac{\overline{kT}}{1.2~{\rm keV}}\right)^{1/2}.
%	\end{equation}
\(
V_{\rm s}=(16/3)(\overline{kT})/(\mu m_{\rm p})\approx 1000(\overline{kT}/1.2\,{\rm keV})^{1/2}~{\rm km~s^{-1}}.
\)
%	Therefore the shock velocity there is estimated to be $\sim 1000$ km s$^{-1}$. 
This is somehow smaller than the shock velocity determined by optical observations. The  width of thermal broadened H$\alpha$ line corresponds to a shock velocity of 1500--2000 km s$^{-1}$ \citep[e.g.,][]{1989ApJ...338L..13F,1991ApJ...366..484B,2008ApJ...689.1089V}. \citet{2016ApJ...817...36S} recently obtained a typical $V_{\rm s} \sim 1690$ km s$^{-1}$ based on the proper motions of Balmer filaments. However, in our spectral analysis only the electron temperature can be determined, which may be different from the ion temperature or the average temperature of plasma. In fact, the evidences of temperature non-equilibration at the collisionless shock front have been found by both simulations and observations \citep[see, e.g.,][]{1988ApJ...329L..29C,2000ApJ...543L..67S,2003ApJ...587L..31V}. It has been indicated that electrons may be substantially heated but not completely up to the same temperature as ions, especially for high Mach number shocks \citep{2007ApJ...654L..69G}. Moreover, the establishment of the temperature equilibration between electrons and ions may require $\net \gtrsim 10^{12}$ \netunit~\citep{2012A&ARv..20...49V}, which is significantly larger than the typical $\net$ in Kepler. Therefore, the disagreement among electron-temperature-inferred and thermal-broadening-inferred shock velocity potentially reflects the temperature non-equilibration between electrons and ions in Kepler.
	
%	The rim-hotter structure of the northern shell may due to several reasons. By comparing with the density distribution, we note that the rise in temperature mainly takes place at the outer layer of the high density shell. In these regions, shock front has broken through the dense material, into a low density environment (thus a rarefaction scenario). This results in an increasing shock speed and a higher post-shock temperature. On the other hand, shock-heated hot gas may experience extra cooling processes.
	
	For a better understanding of the thermal and ionization state, we produce the EM-weighted probability distribution functions \citep[PDFs, see, e.g.,][]{2015MNRAS.453.3953L} of $\kTe$ and $\net$ (Figure \ref{fig:kT_Tau_hist}). We find that both of them follow unimodal distributions. Electron temperature has a bottom-heavy structure peaked at $\sim 0.48$ keV, with a long tail to the hotter end. {The} ionization parameter shows more symmetric probability distribution and peaks at $\sim 1.6\times10^{11}$ \netunit.
	
	We further investigate the relationship between $\kTe$ and $\net$. In a $\kTe$-$\net$ diagram (Figure \ref{fig:kT_Tau}), electron temperature and ionization parameter follow an anti-correlation: $\kTe$ decreases with the increasing $\net$. Similar feature has been discovered in SN 1006 by \citet{2015MNRAS.453.3953L}, while they attributed to the degeneration between this two parameters in spectral fitting \citep{2016MNRAS.462..158L}. Indeed, parameter degeneration may affect the fitting results of a single bin region. For example, if artificially increase $\net$ while leaving other parameters unchanged, one may expect a lower $\kTe$ in the fitting results. This can also be illuminated by a confidence contour in $\kTe$-$\net$ space \citep[e.g., Fig.3 in][]{2000AdSpR..25..549S}. 
	%But it is unclear that whether this degeneration effect will similarly work on massive spectral fittings among different bin regions. 
	{But it is unclear whether this degeneration will similarly affect the parameter distribution of a large number of bin regions (i.e. an anti-correlation distribution in the $\kTe$-$\net$ diagram).}
	In addition to the anti-correlation, we notice that the X-ray emitting plasma diverge into two branches, one is hot and young (of low $\net$), the other one is cool and old (of high $\net$). This unique feature, however, cannot be explained by degeneration effects. We further compare our results to some previous measurements \citep{2007ApJ...668L.135R,2013ApJ...764...63B,2015ApJ...808...49K} and find that they are roughly consistent with the hot branch. Moreover, we note that all these archival measurements are made from spectral fittings of bright knots which are finally identified as shock-heated CSM. On the other hand, we find that the two branches lie in different parts of the remnant with distinct distributions (Figure \ref{fig:kT_Tau}). Particularly, the hot branch has a spatial association with the \textit{Spitzer} 24 $\mu$m image \citep{2007ApJ...662..998B}, which traces the warm CSM dusts. Thus we infer this two-branch structure may reflect the different nature of the plasma: the hot branch corresponds to the newly shock-heated CSM, while the cool branch is basically ejecta materials (particularly the Si- and S-rich ejecta by comparing to the abundance map).

	\subsection{[O]/[Fe]$_{\rm  L}$ and [Mg]/[Fe]$_{\rm  L}$: distinguish the CSM from SN ejecta}\label{subsec:CSM}
	
	Type Ia SNe typically produce a large amount of IGE, while the yields of light elements such as O, Ne, Mg are relatively low. In the ejecta of SNe Ia, 
	{O-to-Fe and Mg-to-Fe abundance ratios}
	%abundance ratios [O]/[Fe] and [Mg]/[Fe]
	 are expected to be only $\lesssim$ 0.2--0.5 \citep[e.g.,][]{2010ApJ...712..624M}. Therefore a near-solar abundance ratio is often used to probe the existence of CSM \citep[e.g.,][]{2007ApJ...668L.135R,2013ApJ...764...63B}. The spatial distributions of [O]/[Fe]$_{\rm  L}$ and [Mg]/[Fe]$_{\rm  L}$ ratios in Kepler are highly consistent (Figure \ref{fig:O_Mg_Fe_map}), 
	 {and provide a straight forward way to distinguish the CSM from the (low-temperature, Fe-L emitting) ejecta}. 
	 The vast majority of the remnant is suggested to be ejecta-dominated,  with relatively low [O]/[Fe]$_{\rm  L}$, [Mg]/[Fe]$_{\rm  L}$ $\lesssim 0.4$. High abundance ratios ($\sim 1$) mainly appear in NW, central bar-like region and outer regions of the remnant, indicating that these regions are dominated by the shock-heated CSM.
	
	{The differences between CSM and ejecta are also indicated by the EM-weighted PDFs of [O]/[Fe]$_{\rm  L}$ and [Mg]/[Fe]$_{\rm  L}$, which show bimodal features} (Figure \ref{fig:O_Mg_hist}). We use a double-Gaussian model to fit the PDF. The Gaussian centroid corresponds to the peak value of abundance ratio, and the area covered by Gaussian curve stands for the total emission measure of the CSM/ejecta component. Results of the double-Gaussian fitting are summarized in Table \ref{tab:double_gaussian}.  The fitting results indicate an O-rich CSM component with [O]/[Fe]$_{\rm  L}$ $= 0.77^{+0.30}_{-0.23}$ and $n_{\rm e}M_{\rm gas} \sim 15$ $\Msun$ cm$^{-3}$, as well as a Mg-rich CSM component with [Mg]/[Fe]$_{\rm  L}$  $= 1.11^{+0.46}_{-0.32}$ and $n_{\rm e}M_{\rm gas} \sim 6$ $\Msun$ cm$^{-3}$.  We note that \citet{2007ApJ...662..998B} derived a CSM emission measure $n_{\rm e}M_{\rm gas} \sim 10$ $\Msun$ cm$^{-3}$ (they assumed a distance of 4 kpc) based on integral O line fluxes from \textit{XMM-Newton} observations. That corresponds to $\sim 16$ $\Msun$ cm$^{-3}$ at 5.1 kpc, which is very close to the value we got for O-rich CSM. 
	
	\subsubsection{Integral properties of the CSM and their implications}
	Based on PDF fitting results (Table \ref{tab:double_gaussian}), we take [O]/[Fe]$_{\rm  L}$ $> 0.54$ {and} [Mg]/[Fe]$_{\rm  L}$ $> 0.79$ as a filter to select the regions dominated by the CSM and investigate the integral properties of them. 
	{Using the method described in Section \ref{subsec:density} and Section \ref{subsec:mass}, we get a mean electron density of 7--11 cm$^{-3}$. The total hydrogen mass of the CSM-dominated plasma could be $\sim 2.2 \Msun$ for $f = 1$ or be $\sim 1.4 \Msun$ for $f = [400~{\rm yr} /(f^{-1/2} t_{\rm i})]^2$. This would correspond to $n_{\rm e}M_{\rm gas}\sim10$--$20\Msun\,{\rm cm}^{-3}$, essentially consistent with the above estimate for the CSM.}
%	The total {hydrogen} mass of the CSM-dominated plasma is calculated to be $\sim 2.2 \Msun$ assuming a filling factor $f = 1$, while it will be reduced to $\sim 1.4 \Msun$ by taking $f = [400~{\rm yr} /(f^{-1/2} t_{\rm i})]^2$. This indicates a mean electron density of 7--11 cm$^{-3}$. 
	The CSM in Kepler is suggested to be confined in a bow shock formed by its fast moving progenitor system \citep[e.g.,][]{1987ApJ...319..885B,2012A&A...537A.139C}. Thus we assume the CSM has been almost totally shocked by the SN blast wave and take $\ga 1.4 \Msun$ as its total mass. The EM-weighted mean abundances of O and Mg are [O] $= 0.71^{+0.46}_{-0.28}$, [Mg] $= 0.72^{+0.45}_{-0.28}$, and the mean {Mg-to-O} abundance ratio is [Mg]/[O] $= 1.14 \pm 0.49$. 
	
	\citet{2012A&A...537A.139C} argued that the CSM in Kepler comes from the stellar wind of the donor star, which is most likely an AGB star. They constrained the initial mass of the donor star as 4--5 $\Msun$ based on the chemical composition of the CSM shell (mainly an enhanced nitrogen abundance) and the AGB model simulations by \citet{2007PASA...24..103K}. Our results here can provide a separate estimate on the properties of the donor star.  On the basis of the updated AGB yields \citep{2010MNRAS.403.1413K}, the total {hydrogen} mass and {Mg-to-O} abundance ratio of CSM-dominated plasma can be well reproduced by the wind of an AGB donor star of $\sim 4 \Msun$ with solar metallicity (Figure \ref{fig:AGB_mass}). Here, we take $\gtrsim 2 \Msun$ as a proper estimation for the total wind mass, which includes the observed $\ga 1.4 \Msun$ CSM and the assumed $\gtrsim 0.6 \Msun$ stellar wind accreted to the WD \citep[based on an observation-indicated mean WD mass of $\sim$ 0.6--0.8 $\Msun$, see, e.g.][]{2007MNRAS.375.1315K}. Meanwhile, the nitrogen abundance in the wind of a $\sim 4 \Msun$ AGB star is $\sim 2.6$, which agrees with the observation results \citep{1983MNRAS.204..273L,1991ApJ...366..484B}.
	
	\subsubsection{Integral properties of the ejecta and their implications}\label{subsubsec:ejecta}
	On the other hand, we take [O]/[Fe]$_{\rm  L}$ $< 0.48$ and [Mg]/[Fe]$_{\rm  L}$ $< 0.74$ (see Table \ref{tab:double_gaussian}) as a filter to select the ({low-temperature, Fe-L emitting}) ejecta dominated regions. We further compare the abundance ratios of the metal species in these regions to the predicted results of different SNe models (Figure \ref{fig:CC_Ia_ratio}). The metal abundance ratios (with respect to Si) of the ejecta are obtained as [O]/[Si] $= 0.11^{+0.09}_{-0.05}$, [Mg]/[Si] $= 0.10^{+0.14}_{-0.05}$, [S]/[Si] $= 2.37^{+0.98}_{-0.69}$ and [Fe]$_{\rm  L}$/[Si] $= 0.38^{+0.32}_{-0.17}$. The SNe models considered here include one for CC SNe \citep{2006NuPhA.777..424N} and several for SNe Ia: 2-D pure-deflagration (W7, C-DEF), center/off-center delayed-detonation (C-DDT, O-DDT) models in \citep{2010ApJ...712..624M}; 3-D delayed-detonation models with multi-spot ignition (the numbers of ignition spots are 1, 10, 100 and 1600), in \citep{2013MNRAS.429.1156S} and 3-D pure-deflagration models with multi-spot ignition, in \citep{2014MNRAS.438.1762F}. {The abundance ratios of high-temperature ejecta (see Section \ref{subsec:hard_band_spec}) are also {shown} in the figure.} The CC origin of Kepler are clearly ruled out by the over-predicted O and Mg abundances. Among the SNe Ia models, pure-deflagration scenarios may not apply to Kepler since they typically produce too much IGE with respect to IME. Figure \ref{fig:CC_Ia_ratio} shows that the metal abundance in the ejecta of Kepler are well compatible with the spherical delayed-detonation models (e.g., C-DDT model and DDT-N1600 model), except for an under-predicted [S]/[Si] which presents in all the models.
	
	\subsection{[Si]/[Fe]$_{\rm  L}$ and [S]/[Fe]$_{\rm  L}$: properties of the ``ears''}
	
	The abundance maps of Kepler (Figure \ref{fig:th_par_map}) reveal a stratified structure of the ejecta: lighter elements distribute at the outermost layer and heavier elements like iron mainly in the inner regions. This can be better seen in the maps of abundance ratios [Si]/[Fe]$_{\rm  L}$ and [S]/[Fe]$_{\rm  L}$ (Figure \ref{fig:Si_S_Fe_map}). Basically, Si and S extent to further distance than Fe. Especially, the two ``ears'' of the remnant show enhanced metal abundances and high Si, S to Fe abundance ratios, which indicate that the thermal-emitting plasma there is dominated by Si- and S-rich ejecta. 
	
	The morphological features of two opposite ``ears'' have been found in several Ia SNRs such like Kepler, G1.9$+$0.3, G299.2$-$2.9, etc. \citep[see the detailed discussions in][]{2015MNRAS.447.2568T}. \citet{2013MNRAS.435..320T} proposed that the ``ears'' can be formed by pre-explosion jets. The formation of jets may take place either a long time before the explosion (in a SD scenario when the WD accretes mass from the donor star, or in a CD scenario when the WD accretes mass from the common envelope) or immediately before the explosion (in a CD scenario when the WD and core merge). In their models the connection line of two ``ears'' is interpreted as a symmetry axis, and the ejecta will expand to further radius along the axis. On the other hand, \citet{2013ApJ...764...63B} suggested a different topology of the remnant. They interpreted the distribution of the central CSM as a disk seen edge-on, thus the connection line of two ``ears'' represents actually an equatorial plane for pre-SN mass loss of the donor star. According to their hydrodynamic simulations, the ejecta along the plane will be blocked by the disk-like CSM. Our results indicate that the ``ears'' mainly consist of Si-, S-rich ejecta, thus favoring the pre-explosion jets scenario. The central bar-like CSM might imply some other form of asymmetric mass-loss process of the donor star, rather than a disk.
	
	\subsection{High temperature components and K-shell emitting Fe in Kepler}\label{subsec:hard_band_spec}
		
		\citet{2015ApJ...808...49K} revealed that the X-ray emitting plasma in Kepler has a multi-temperature structure. Based on \textit{XMM-Newton}, \textit{Chandra} and \textit{Suzaku} observations, they successfully fitted the spatially integrated spectra of Kepler with a model which consists of four thermal components. Two of these components have relatively high electron temperature $\kTe \gtrsim 2.0$\,keV, and dominate the spectra in $\gtrsim 2.2$\,keV energy band which covers the S, Ar, Ca and Fe K$\alpha$ emission lines. In our above analysis, the X-ray photons are concentrated in the soft band ($\lesssim 2.2$\,keV), and thus the potential high temperature components may be missed in the single-temperature fit. Meantime, the Fe K$\alpha$ lines, which are prominent in the spatially integrated spectra, are also missed in the above analysis.
		In order to constrain the high temperature components and the K-shell emitting Fe in Kepler, we further analyze the 2.2--8.0\,keV spectra particularly. The key results of the analysis are summarized below and the technical details are referred to APPENDIX \ref{sec:apx}.
		
		Applying the WVT algorithm to Fe K line complex, we rebin the whole SNR into 139 larger regions. For each region, the spectra are extracted in 2.2--8.0\,keV and fitted with a ``2 \texttt{vnei} $+$ \texttt{powerlaw}'' model. The hydrogen absorption has little effect on the hard band spectra, and thus is ignored. Similar to \citet{2015ApJ...808...49K}, we assume one of the \texttt{vnei} components is high-temperature ejecta emitting lines of Si, S, Ar, Ca and Fe, and the other one is pure Fe ejecta producing most of the Fe K$\alpha$ emission. Based on the spectral fit results, we obtain an EM-weighted average electron temperature $\kTe = 2.3^{+1.6}_{-1.0}$\,keV. The EM-weighted average abundance of S, Ar, Ca and Fe are obtained as (in units of $10^4$ solar abundance): [S] $= 10.8^{+4.2}_{-3.0}$, [Ar] $= 10.1^{+6.4}_{-3.9}$, [Ca] $= 16.8^{+21.9}_{-9.5}$ and [Fe]$_{\rm K} = 4.8^{+13.2}_{-3.5}$\footnote{ [Fe]$_{\rm K}$ denotes the abundance of K-shell emitting Fe.}, respectively. The EM-weighted average ionization parameters are $\net = 1.9^{+1.4}_{-0.8} \times 10^{10}$\,\netunit for the high-temperature ejecta, and $\net = 2.9^{+12.5}_{-2.3} \times 10^9$\,\netunit for the pure Fe ejecta.
		
		With the EM, electron temperature, and metal abundances given, we can further estimate the masses of metal species in the high-temperature components. We summarize the results in Table \ref{tab:mass} together with those of the low-temperature component. It can be found that the hot, K-shell emitting Fe has a total mass of $\sim 0.5 \Msun$, which is an order of magnitude higher than that of the L-shell emitting Fe. Therefore, the majority of Fe in Kepler is in high temperature plasma and in low ionization states, suggesting it is recently heated by reverse shock. This is also consistent with our inference about Fe in Section \ref{subsec:mass}.
		
		We also derive the EM-weighted metal abundance ratios of the hot ejecta as [S]/[Si] $=1.08^{+0.42}_{-0.30}$, [Ar]/[Si] $=1.01^{+0.64}_{-0.39}$, [Ca]/[Si] $=1.68^{+2.19}_{-0.95}$, and [Fe]$_{\rm K}$/[Si] $=0.48^{+1.32}_{-0.35}$. Together with the values obtained for the low-temperature ejecta, the metal abundance ratios are compared with the predicted results of various SNe models. The addition of the abundance ratios of the hot ejecta enriches the data points in Figure \ref{fig:CC_Ia_ratio}, and  leads to the implication that Kepler likely originated from a delayed-detonation explosion (see Section \ref{subsubsec:ejecta}).
		
		We note that the total EM ($n_{\rm e}M_{\rm gas}$) of the high-temperature components is about four magnitudes smaller than that of the low-temperature component, and thus has little effects on the average properties discussed above.

	\section{Summary} \label{sec:conclusion}
	We carry out a detailed spatially resolved X-ray spectroscopic study of Kepler's SNR by applying the WVT binning algorithm to the 741\,ks deep \textit{Chandra} observations of Kepler. The whole remnant region is adaptively divided into 4671 tessellated bins with $\sim 6400$ counts contained in each bin. The spectra of each bin region are extracted in 0.3--8.0 keV and are fitted with a ``\texttt{vnei $+$ srcut}'' model. {We also apply the WVT algorithm specifically to the Fe K$\alpha$ complex  so as to constrain the physical properties of the K-shell emitting Fe.} With the fitting results, we obtain detailed distribution of physical parameters in the remnant for the first time. Main results of this work are summarized below:
	\begin{enumerate}
		\item The thermal and non-thermal (synchrotron) emission are well disentangled. We obtain a  total emission measure of $n_{\rm e}M_{\rm gas} \sim { 49}$ $\Msun$\,cm$^{-3}$ for the thermal component, and a 1\,GHz radio flux of $\sim 23$\,Jy for the synchrotron component. We also produce the spatial distribution maps of electron temperature, ionization parameter and abundances of O, Mg, Si, S and Fe in the remnant.
		\item We further produce the spatial distributions of electron density $n_{\rm e}$ and ionization age $t_{\rm i}$. In some regions along the SE rim, a small filling factor of $f \sim 0.01$ is suggested for the thermal plasma.
		\item The total masses of different metal species are estimated. {We get a total mass of shocked Fe $\sim 0.5\Msun$, which is comprised of $\la 10\%$ L-shell emitting Fe and $\ga 90\%$ K-shell emitting Fe. This implies that the majority of Fe is still in low ionization states.}
		% A low Fe mass ($\sim 0.02$--$0.04 \Msun$) suggested by our results implies that the majority of Fe is still in low ionization states and thus does not emit L-shell emission.
		\item The highest electron temperature appears at the northern rim of the remnant, with an average $\kTe \sim 1.2$\,keV. This indicates a shock velocity $\sim 1000$\,km s$^{-1}$, which is lower than optical observation results, and can be attributed to the temperature non-equilibration between electrons and ions in Kepler. The X-ray emitting plasma diverge into two branches in the $\kTe$-$\net$ diagram. The hot branch is consistent with previous measurements of CSM knots and has a spatial association with the warm CSM dusts and thus corresponds to newly shocked CSM. The cool branch basically consists of the Si- and S-rich ejecta.
		\item We use abundance ratios [O]/[Fe]$_{\rm  L}$ and [Mg]/[Fe]$_{\rm  L}$ to distinguish the CSM from {the low-temperature, Fe-L emitting} ejecta. The PDFs of [O]/[Fe]$_{\rm  L}$ and [Mg]/[Fe]$_{\rm  L}$ show a bimodal structure ,which reflects the different chemical compositions between the CSM and ejecta. With a double-Gaussian fitting, we reveal an O-rich CSM component with [O]/[Fe]$_{\rm  L}$ $= 0.77^{+0.30}_{-0.23}$ and a Mg-rich CSM component with [Mg]/[Fe]$_{\rm  L}$  $= 1.11^{+0.46}_{-0.32}$. The integral properties of CSM suggests an AGB donor star with initial mass of $\sim 4 \Msun$ for the progenitor system of Kepler. On the other hand, the metal abundance ratios of the ejecta are well compatible with the spherical delayed-detonation models for SNe Ia.
		\item The thermal emitting plasma in two ``ears'' of the remnant mainly consists of Si- and S-rich ejecta, which favors a pre-SN jets scenario for the origin of the ``ears''.
	\end{enumerate}
	
	\acknowledgements
	L. S.\ is indebted to Jiang-Tao Li for the valuable advices and thanks Ning-Xiao Zhang for the technical help.
	This work is supported by the 973 Program under grants 2017YFA0402600 and 2015CB857100 and the NSFC under grants 11773014, 11633007, and 11851305.
	
	\facilities{ Chandra(ACIS-S)}
	
	\software{ CIAO \citep{2006SPIE.6270E..60F}, WVT binning \citep{2006MNRAS.368..497D,2003MNRAS.342..345C},  Xspec \citep{1996ASPC..101...17A}, SPEX \citep{1996uxsa.conf..411K}, DS9\footnote{http://ds9.si.edu/site/Home.html}}
	
	\appendix
	
	\section{Analysis of the hard band spectra \label{sec:apx}}
\begin{figure}
		\plotone{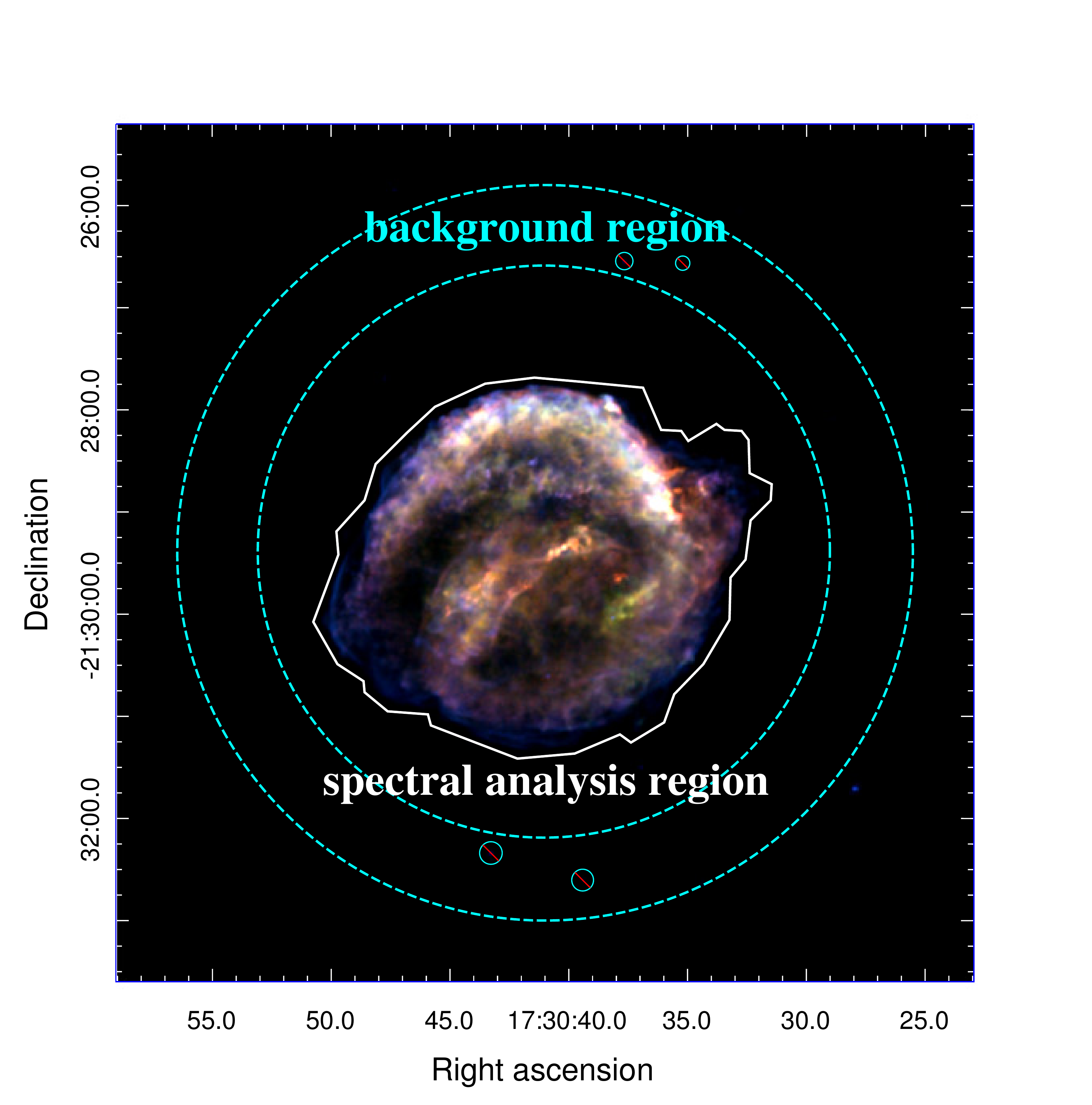}
		\caption{Merged image of the \textit{Chandra} ACIS-S observations with a effective explosure time of 741 ks. \textit{Red}: 0.3--0.72 keV; \textit{green}: 0.72--1.7 keV; \textit{blue}: 1.7--8.0 keV. The white solid contour shows the region for spatially resolved spectroscopy analysis, while the cyan dashed annulus show the region for background subtraction.\label{fig:SNR}}
	\end{figure}

\begin{figure}
	\plotone{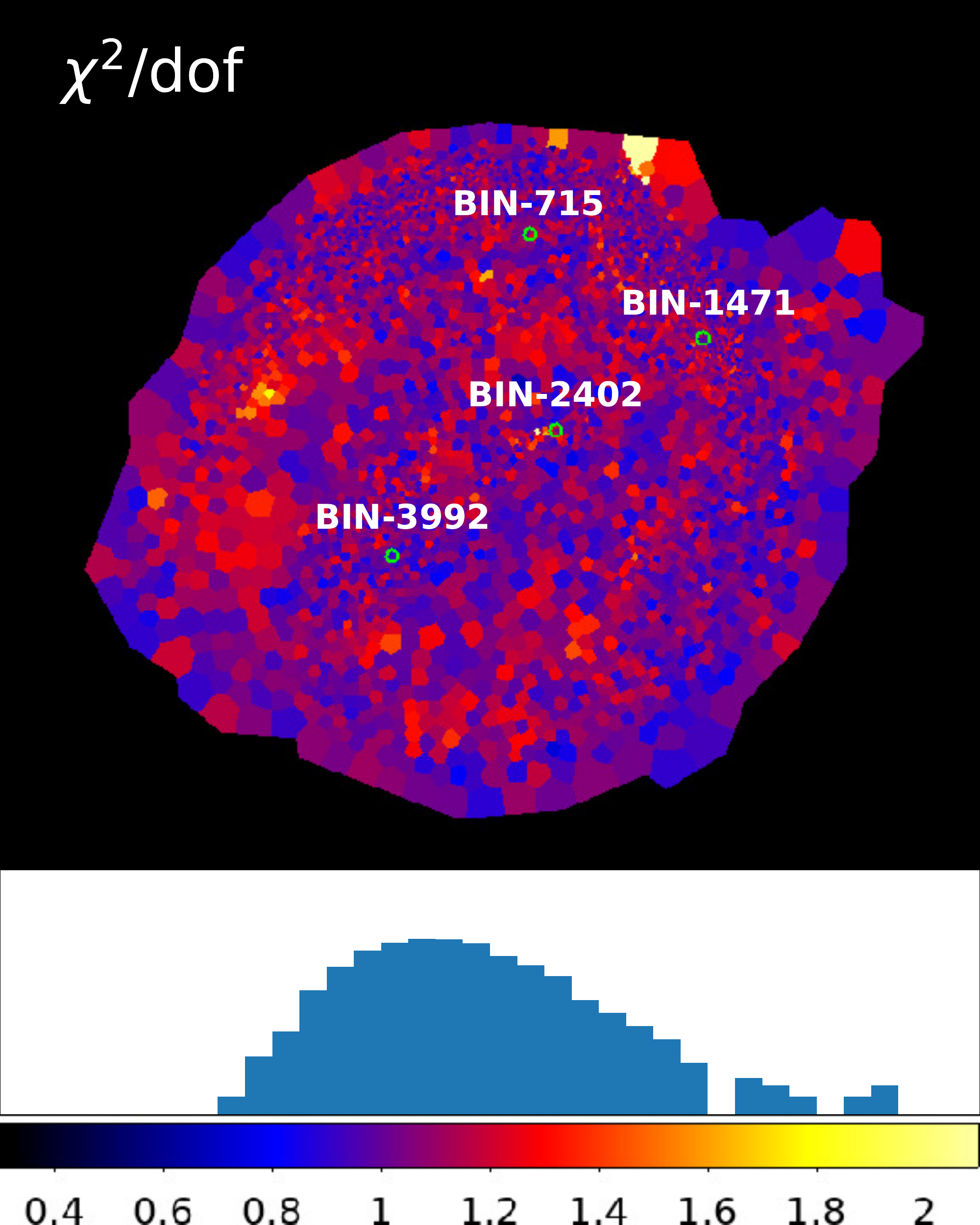}
	\caption{Map of $\chi^2$/dof and its distribution among all the spectral bins. {The four green circles denote the regions used to extract the example spectra in Figure \ref{fig:spec_example}.}\label{fig:RCS}}
\end{figure}

\begin{figure*}
	\gridline{
		\fig{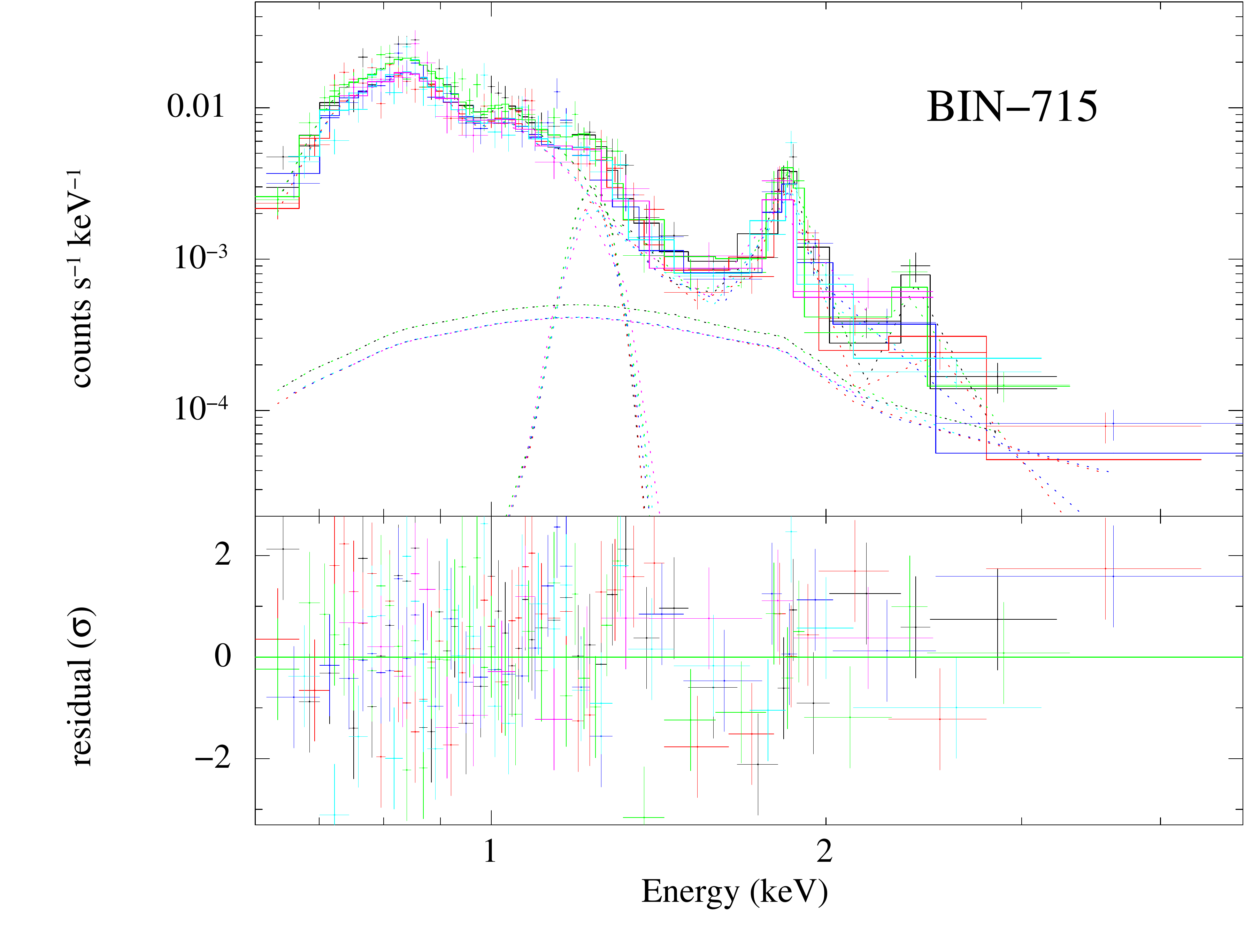}{0.5\textwidth}{}
		\fig{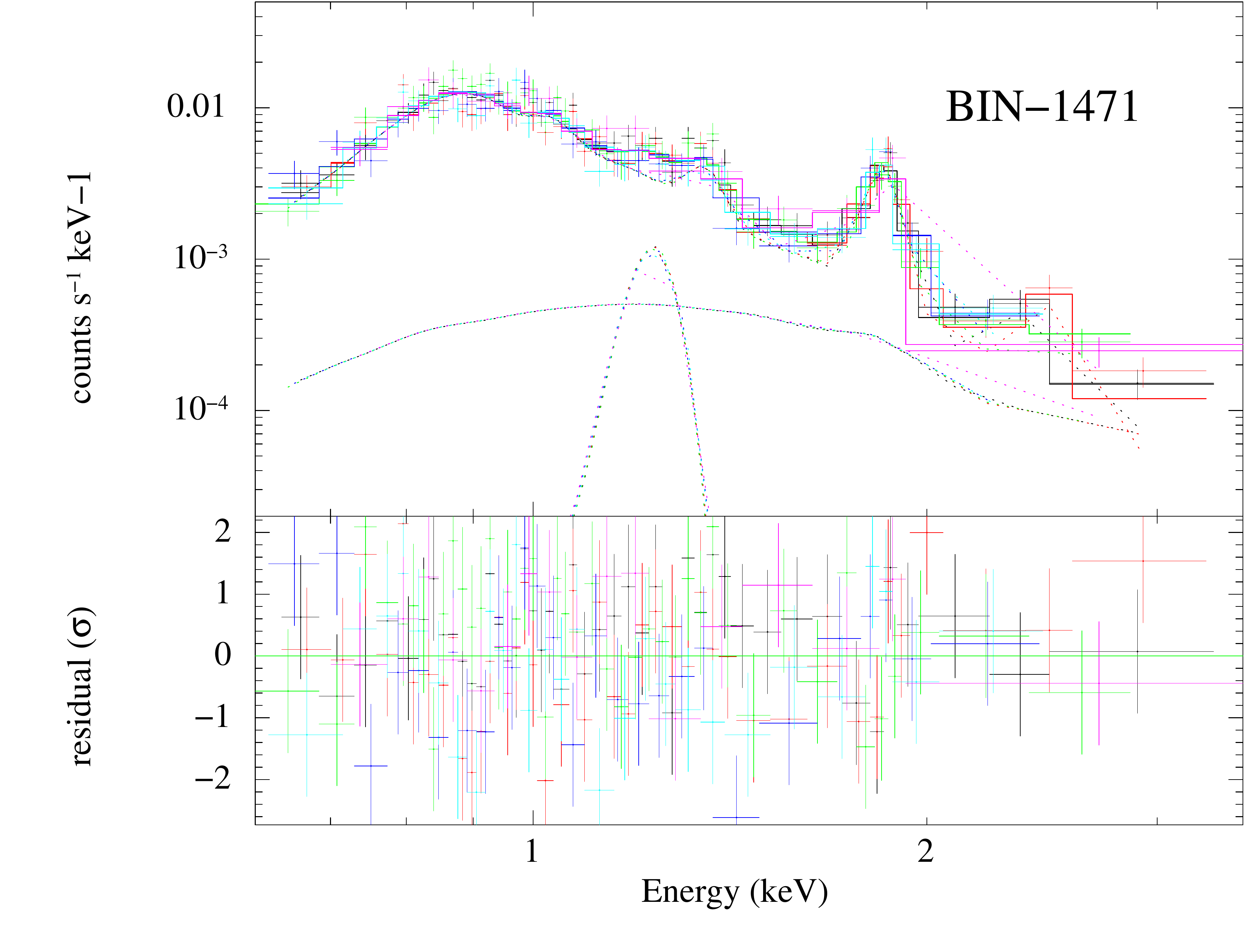}{0.5\textwidth}{}}
	\gridline{
		\fig{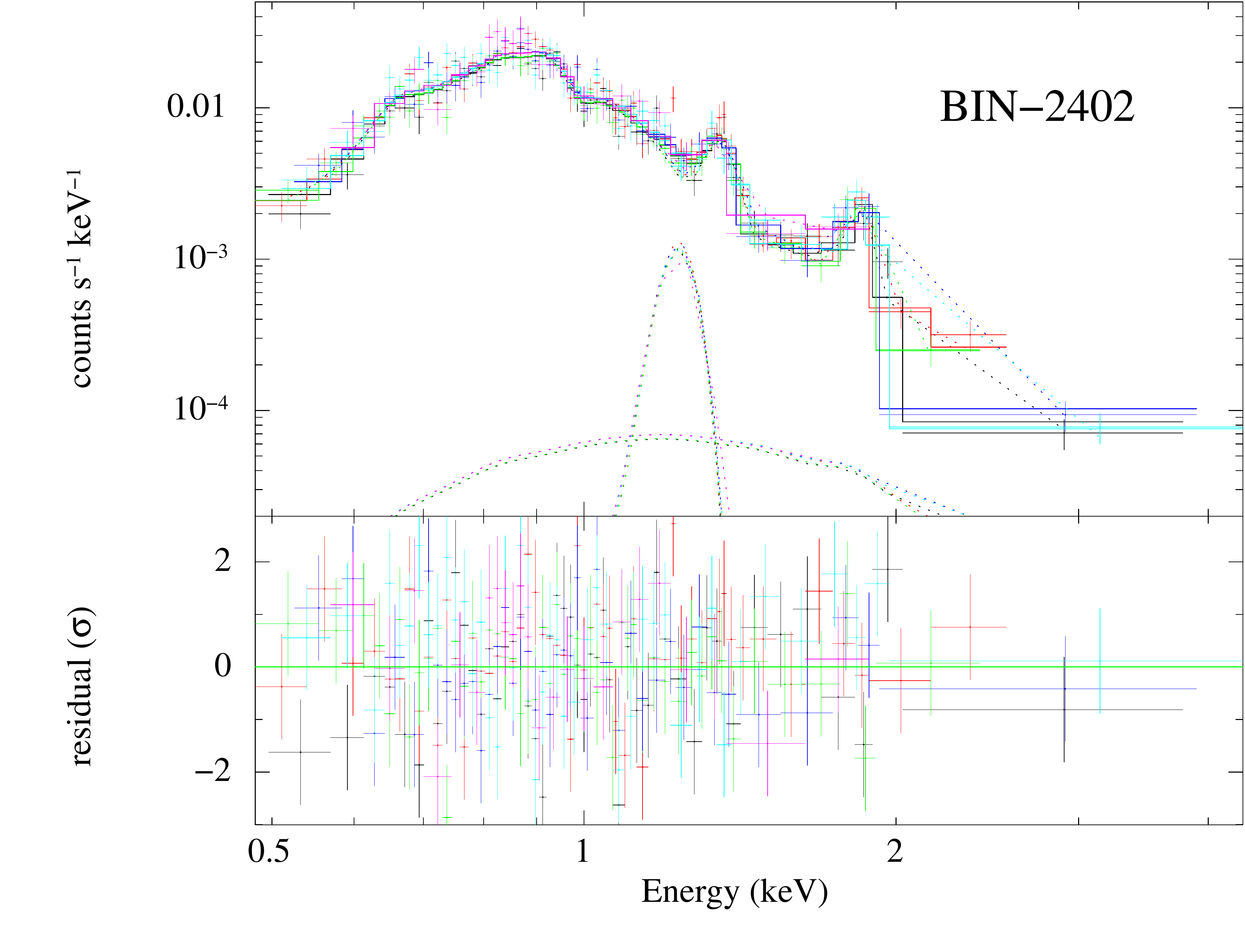}{0.5\textwidth}{}
		\fig{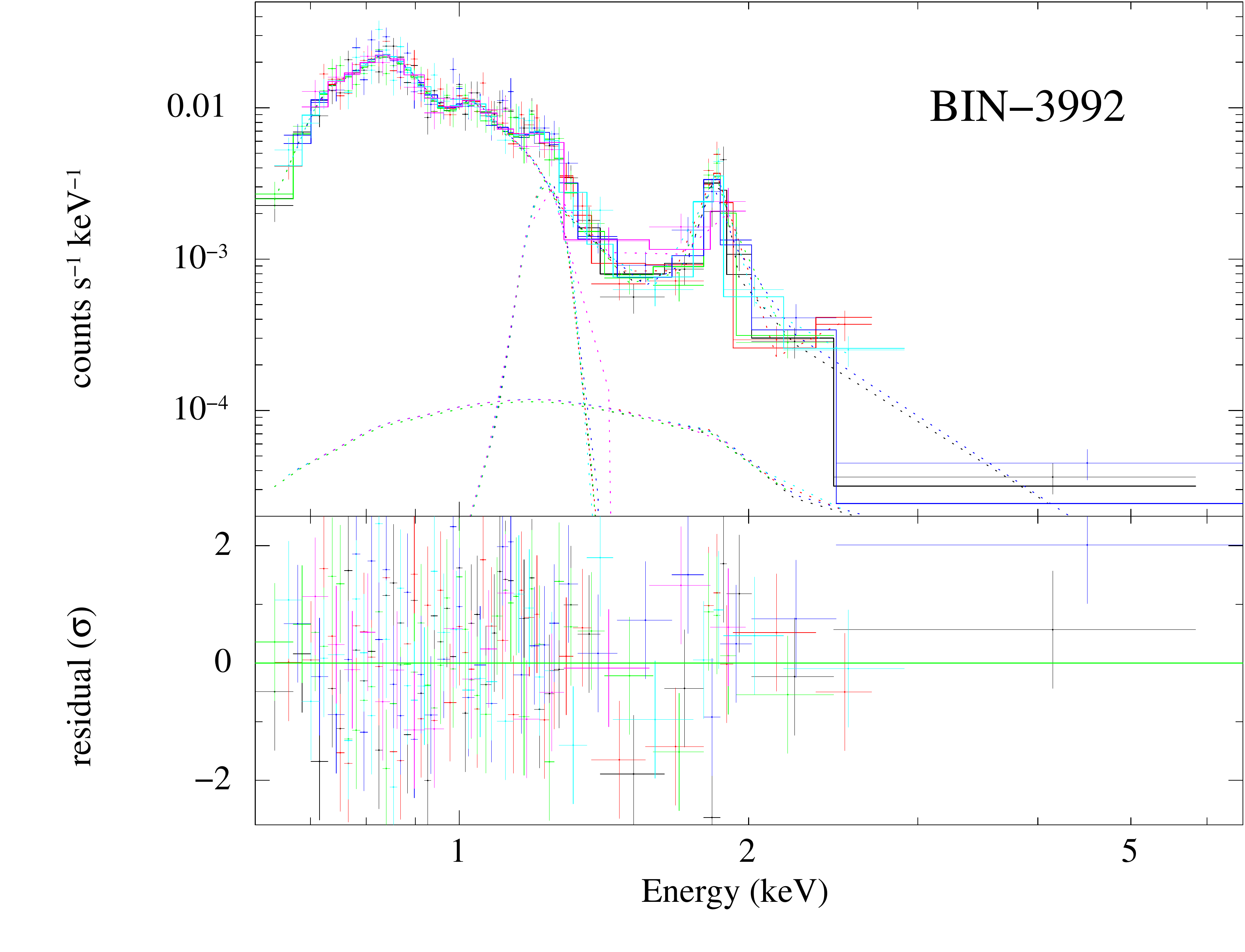}{0.5\textwidth}{}}
	\caption{{Examples of fitted spectra from four tessellated bin regions. The data points in different colors are the background-subtracted spectra from different observation installments, and the thin solid lines with corresponding colors are their best-fit models. Different model components are plotted as dotted lines. The lower panels show the \RCS of the fitting. The spectra extracted from BIN-715 and BIN-3992 show prominent Fe L lines (at 0.7-1.0\,keV) and Si features (at $\sim$1.85\,keV), indicating that they are dominated by ejecta materials. In contrast, the spectra extracted from BIN-1471 and BIN-2402 show bright Mg emission lines (at $\sim$1.35\,keV), suggesting that they are dominated by CSM.}\label{fig:spec_example}}
\end{figure*}

\begin{figure*}
	\gridline{
		\fig{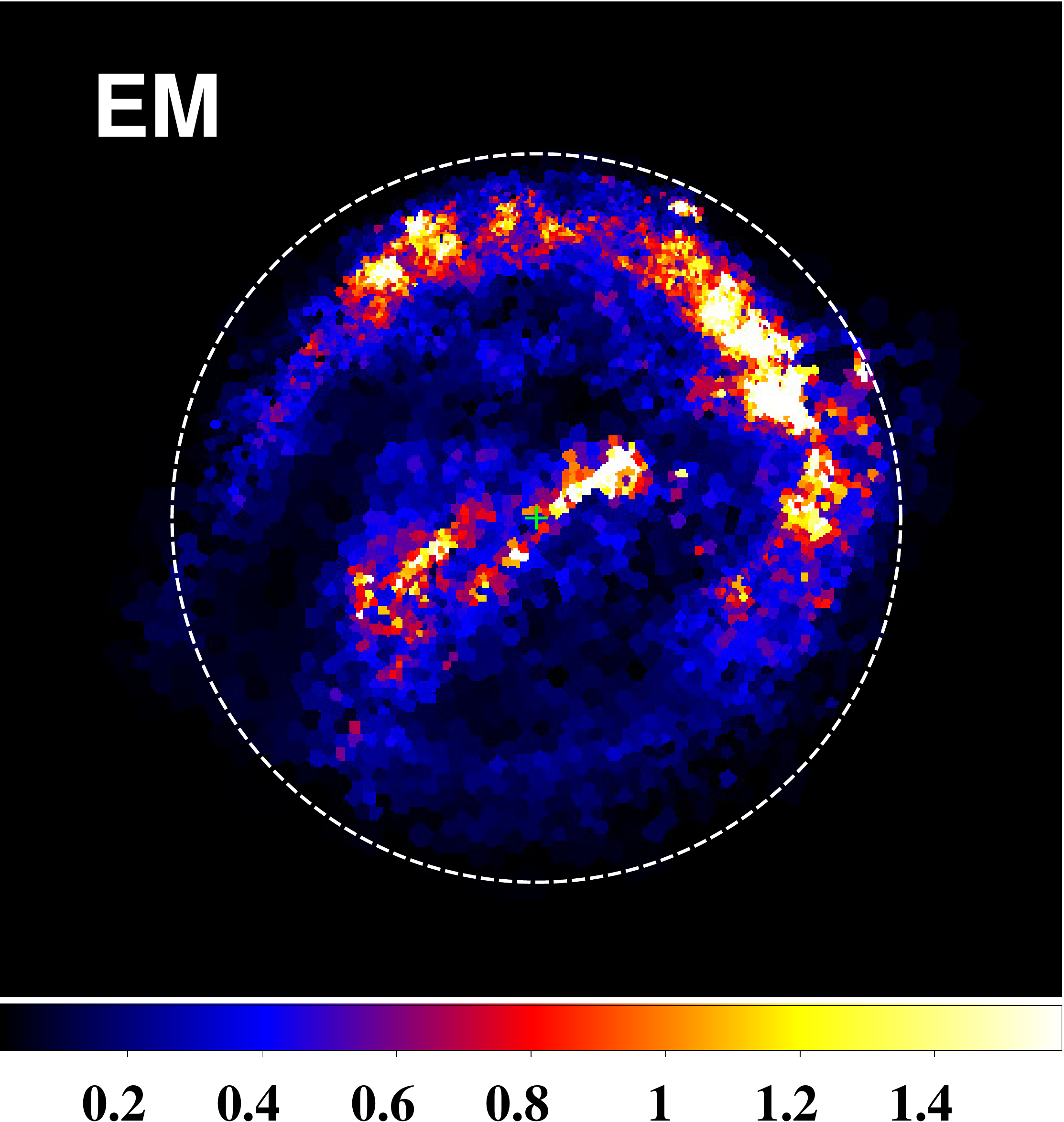}{0.33\textwidth}{(a)}
		\fig{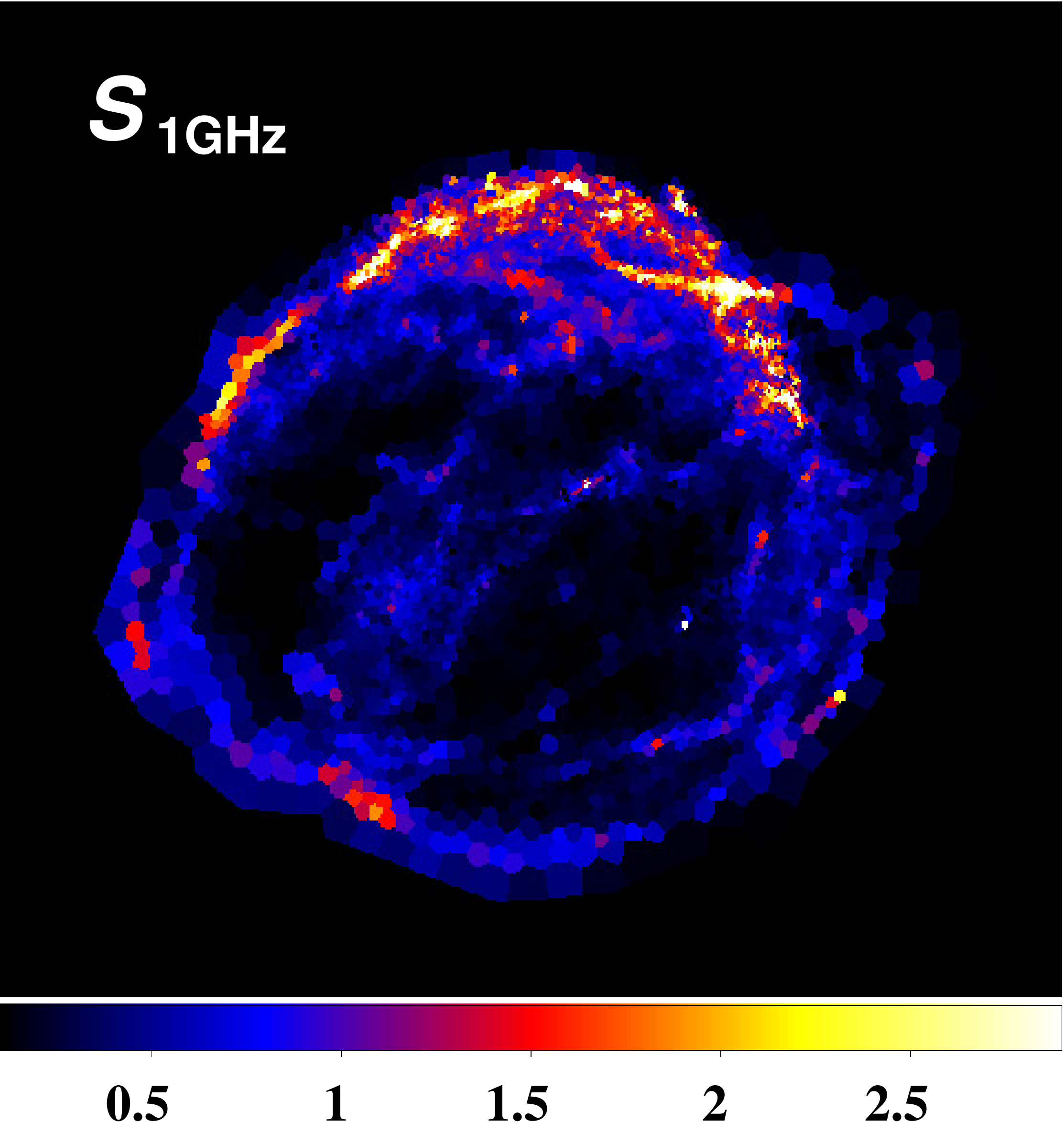}{0.33\textwidth}{(b)}
		\fig{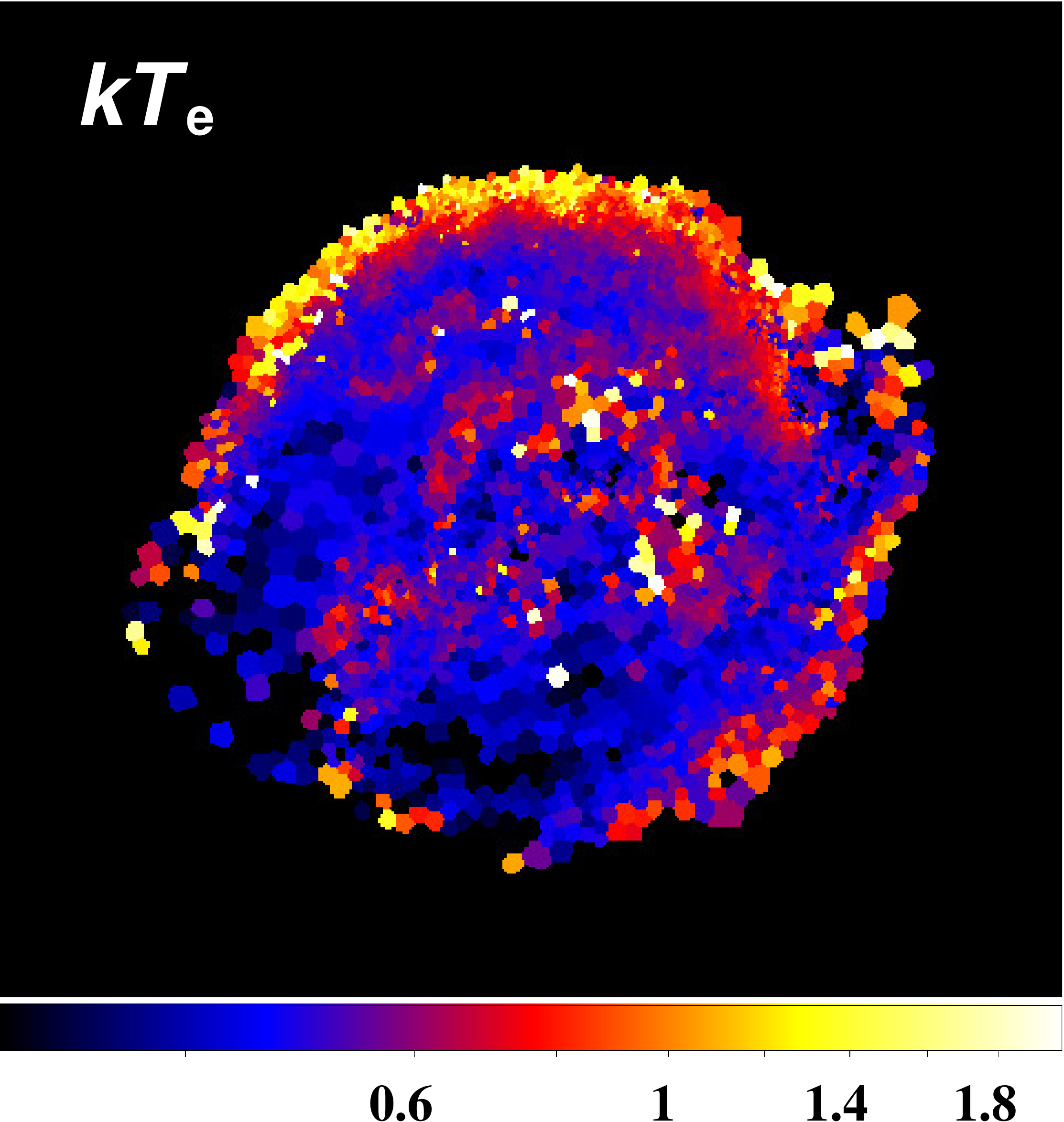}{0.33\textwidth}{(c)}}
	\gridline{
		\fig{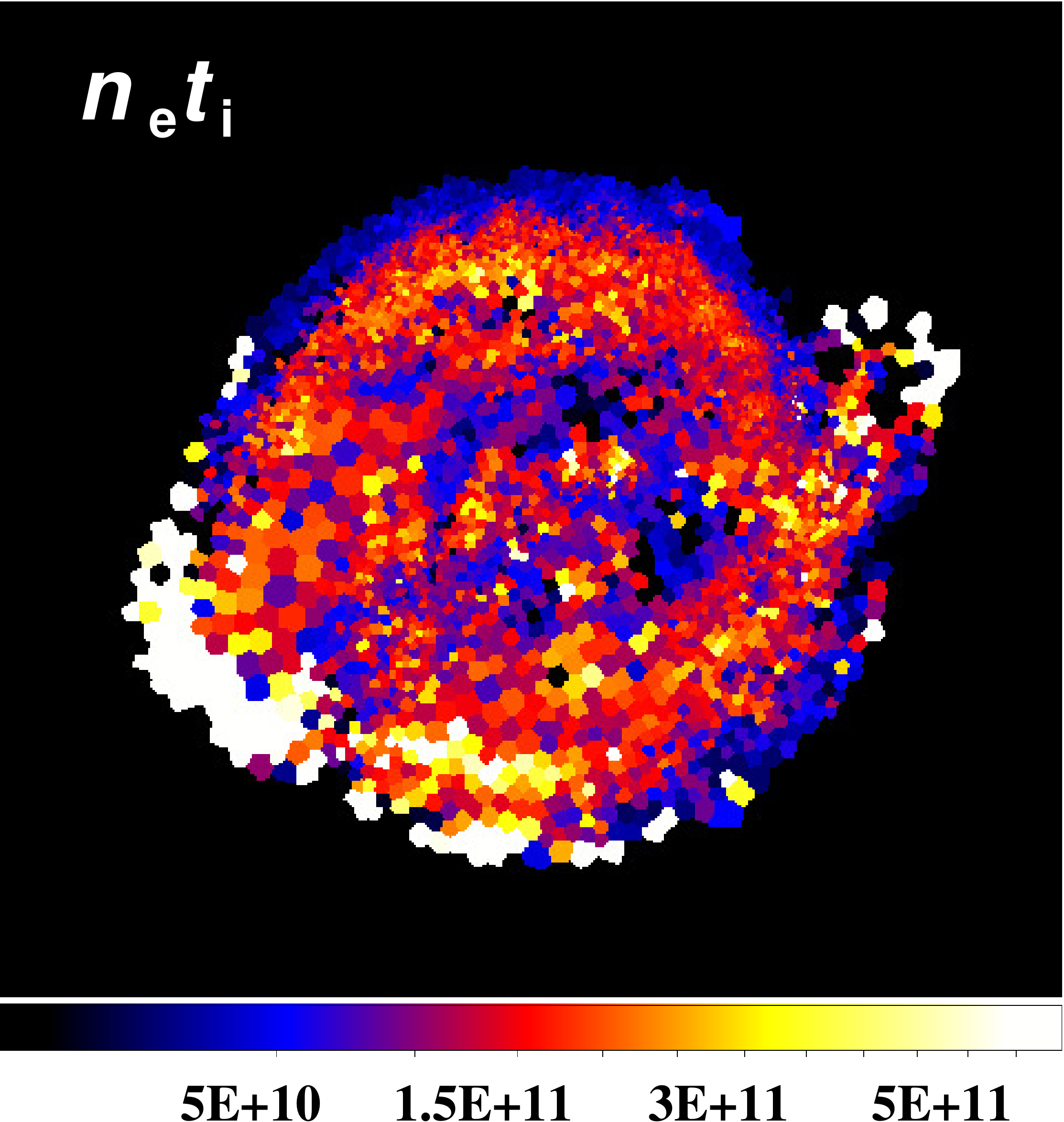}{0.33\textwidth}{(d)}
		\fig{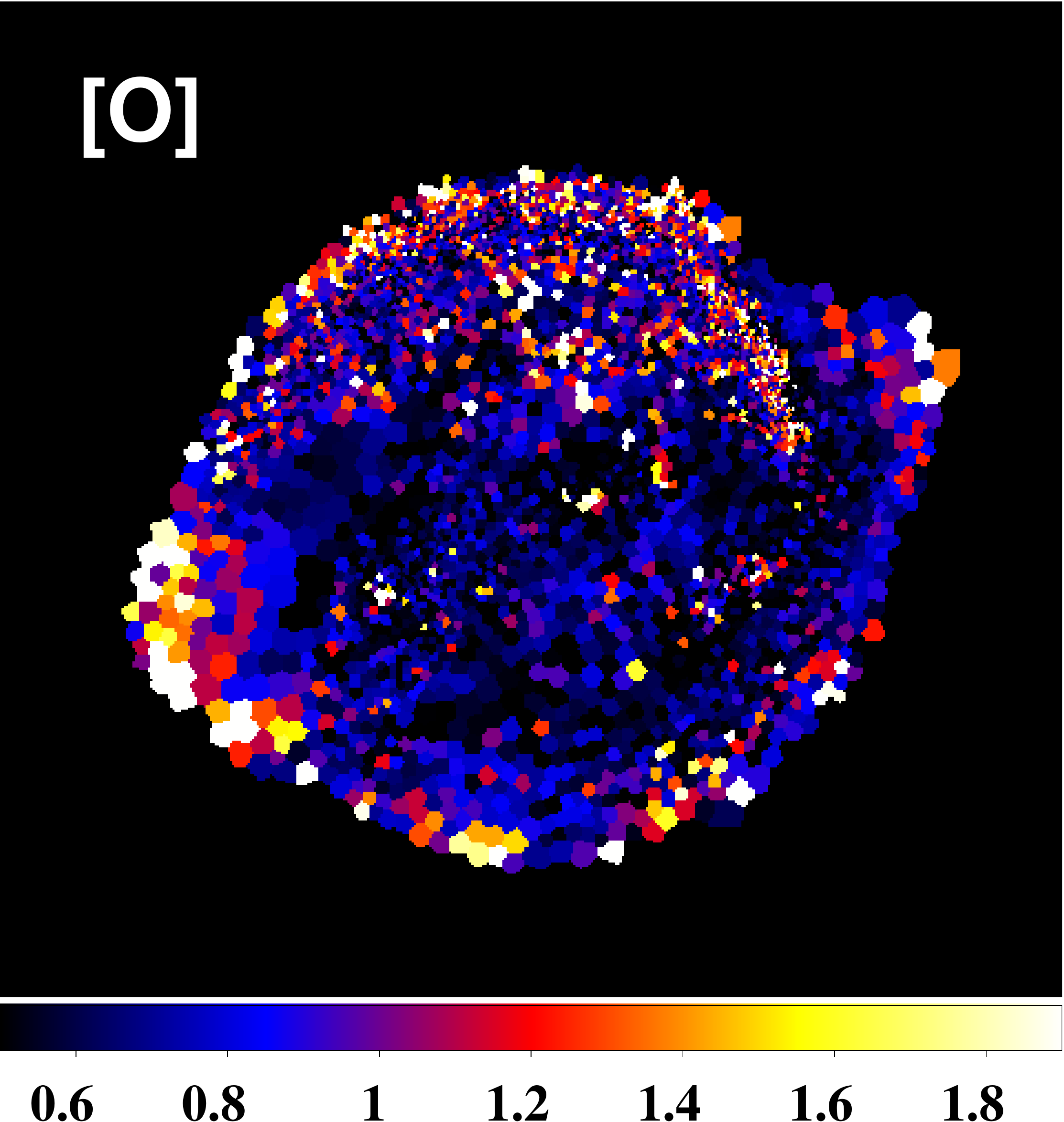}{0.33\textwidth}{(e)}
		\fig{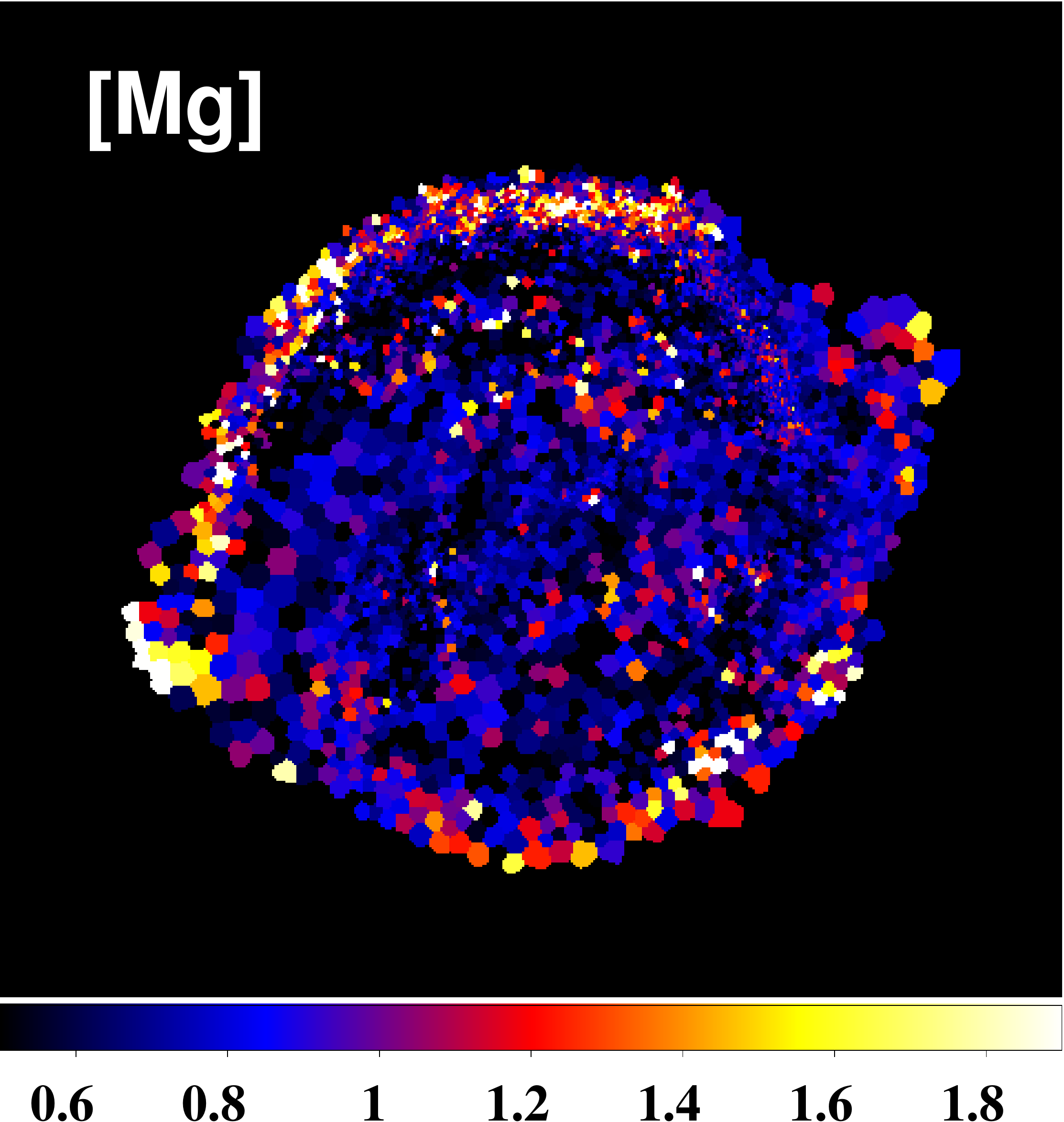}{0.33\textwidth}{(f)}}
	\gridline{
		\fig{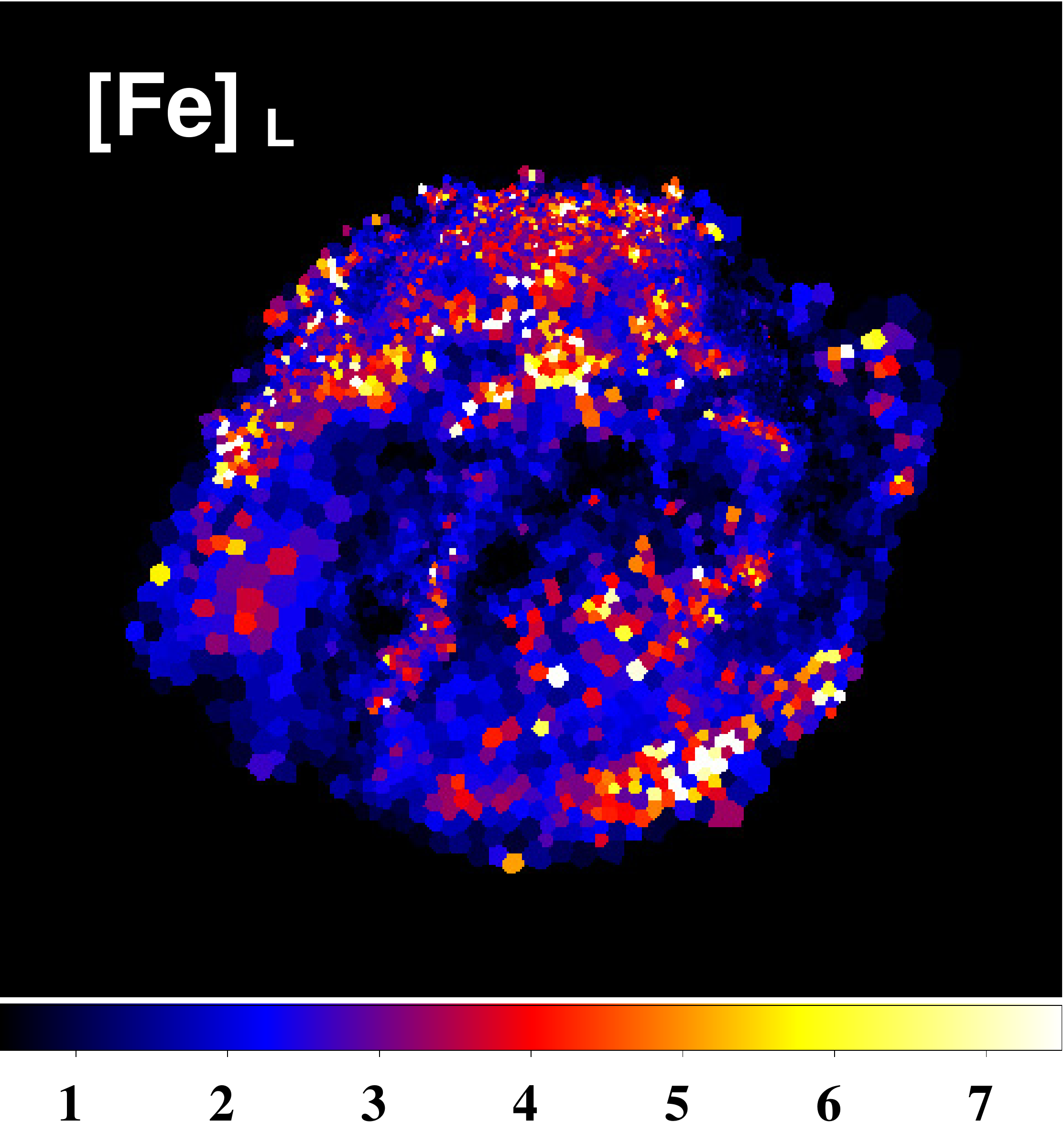}{0.33\textwidth}{(g)}
		\fig{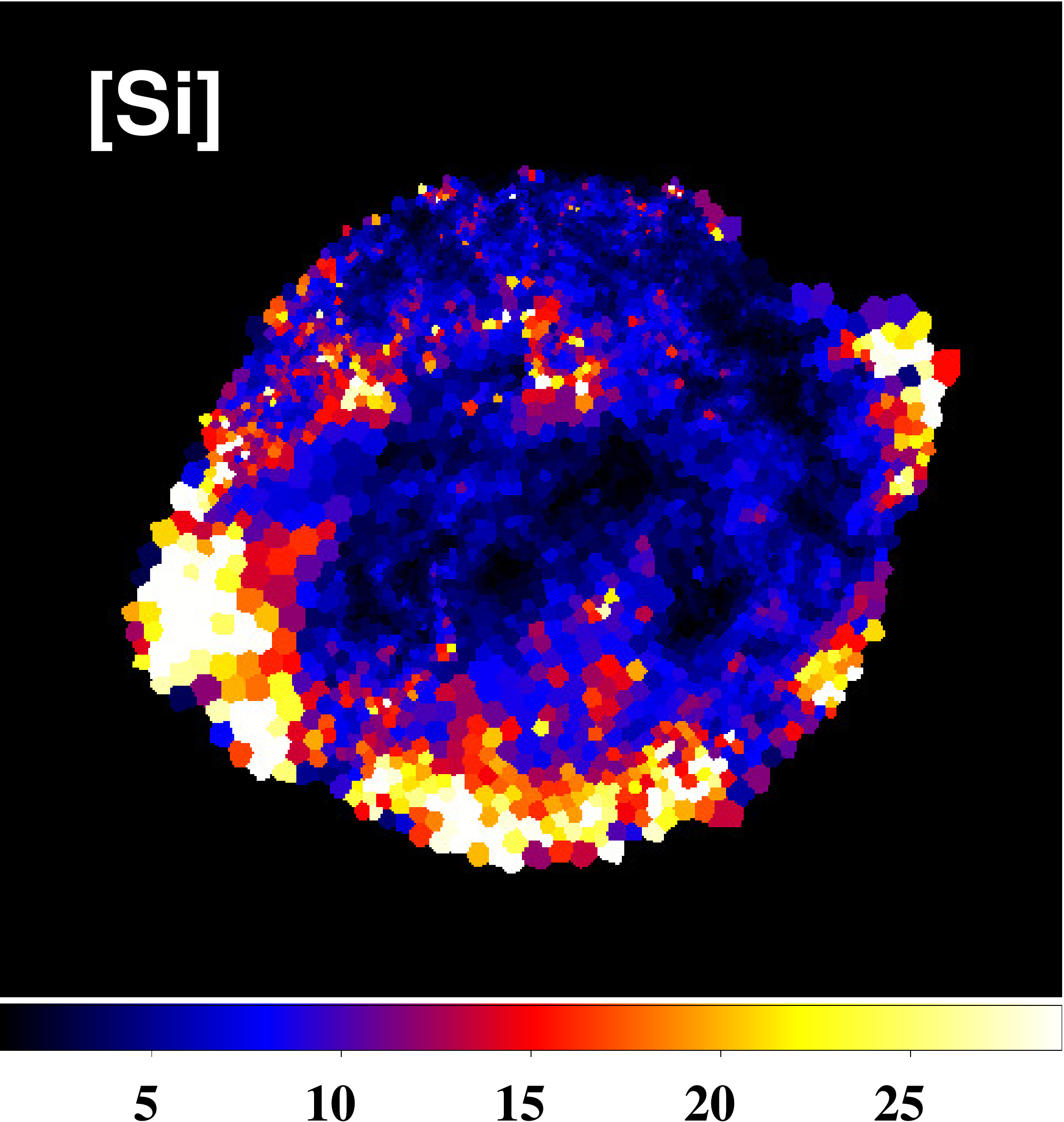}{0.33\textwidth}{(h)}
		\fig{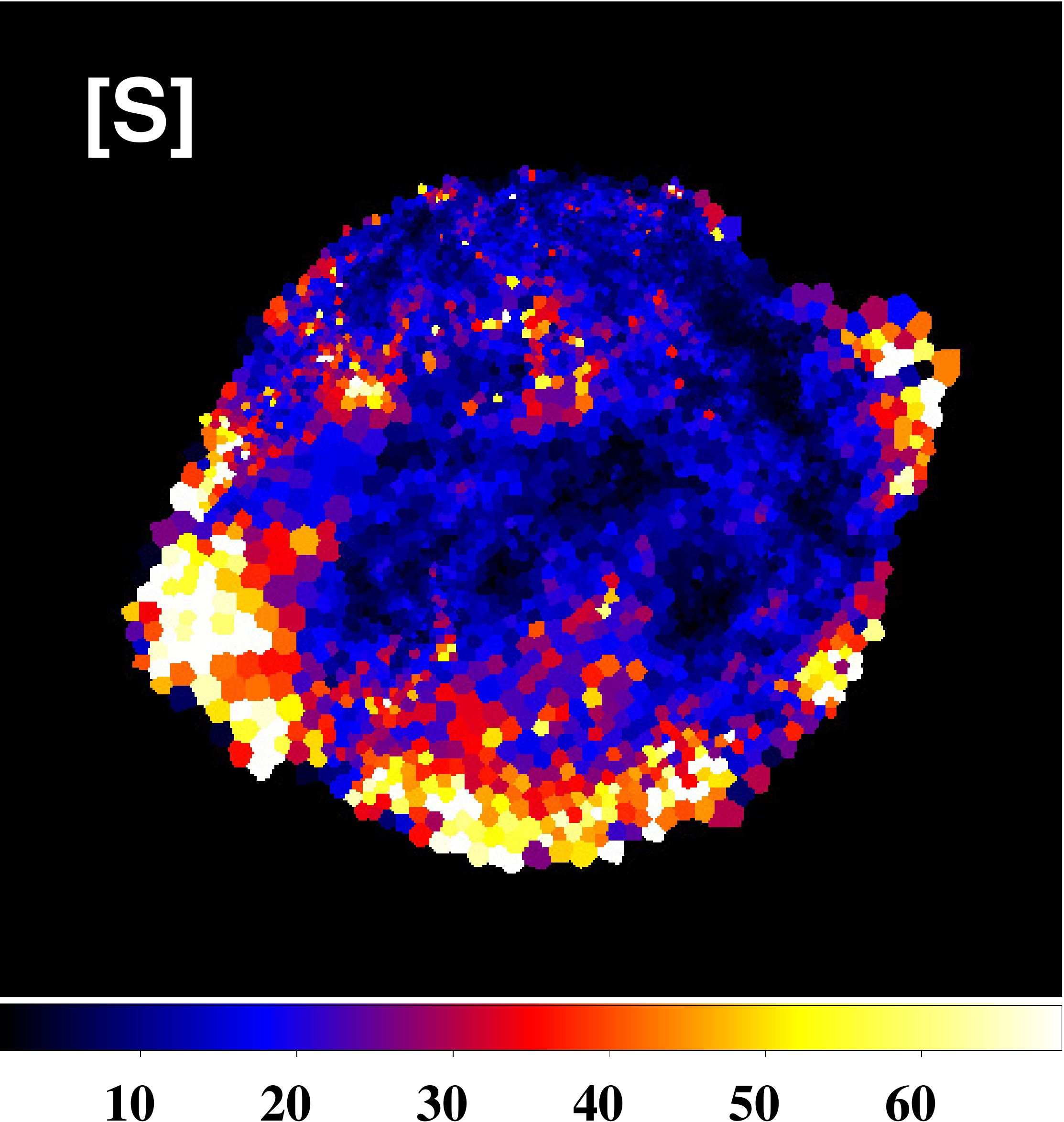}{0.33\textwidth}{(i)}}
	\caption{{Maps of individual spectral fitting parameters: (a) specific EM ($10^{-5}$\,cm$^{-5}$\,arcsec$^{-2}$), (b) Brightness of radio emission at 1\,GHz (mJy\,arcsec$^{-2}$), (c) electron temperature (keV), (d) ionization parameter (\netunit), (e)--(i) metal abundances.} \label{fig:th_par_map}}
\end{figure*}

\begin{figure}
	\begin{center}
		\plottwo{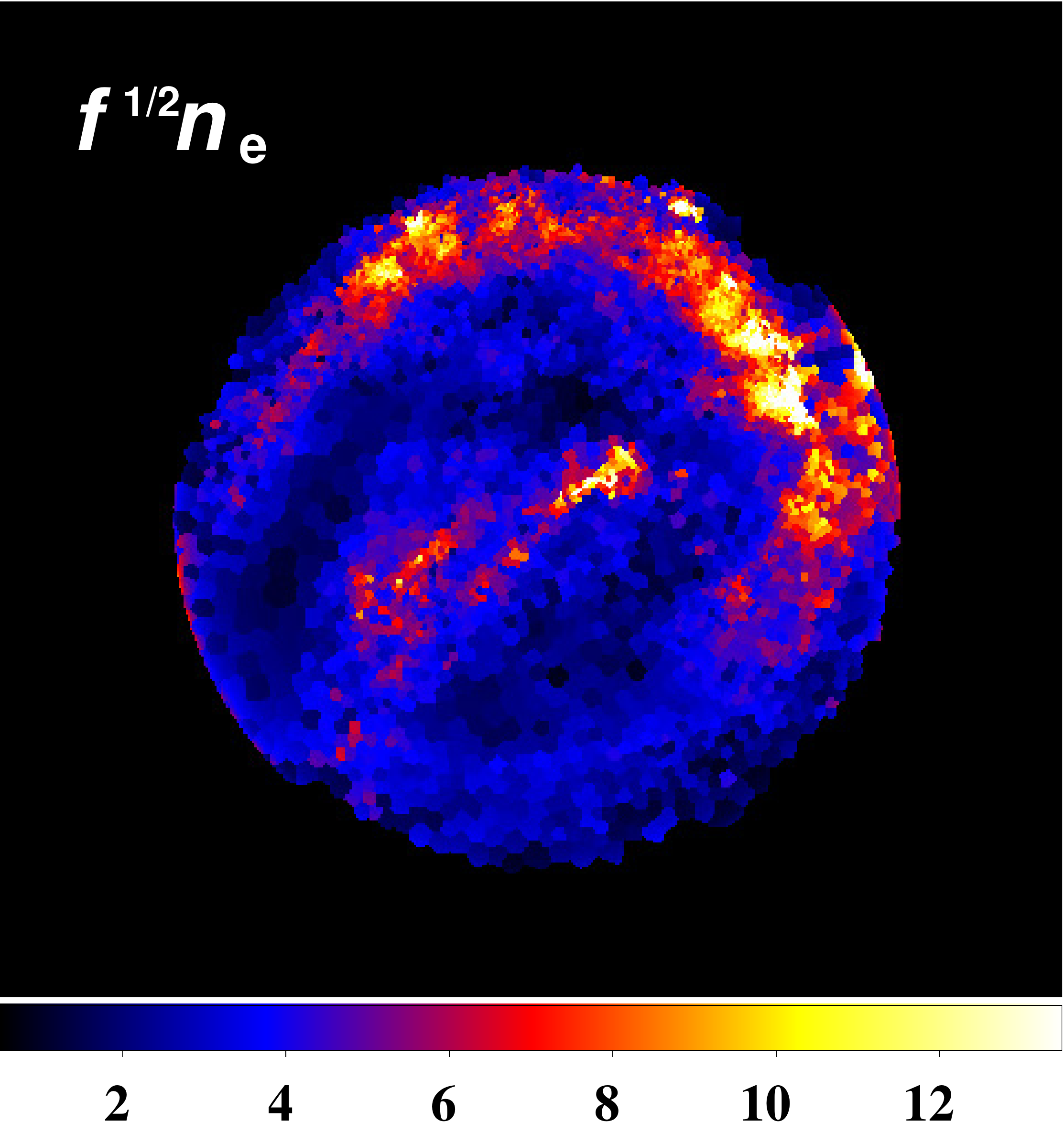}{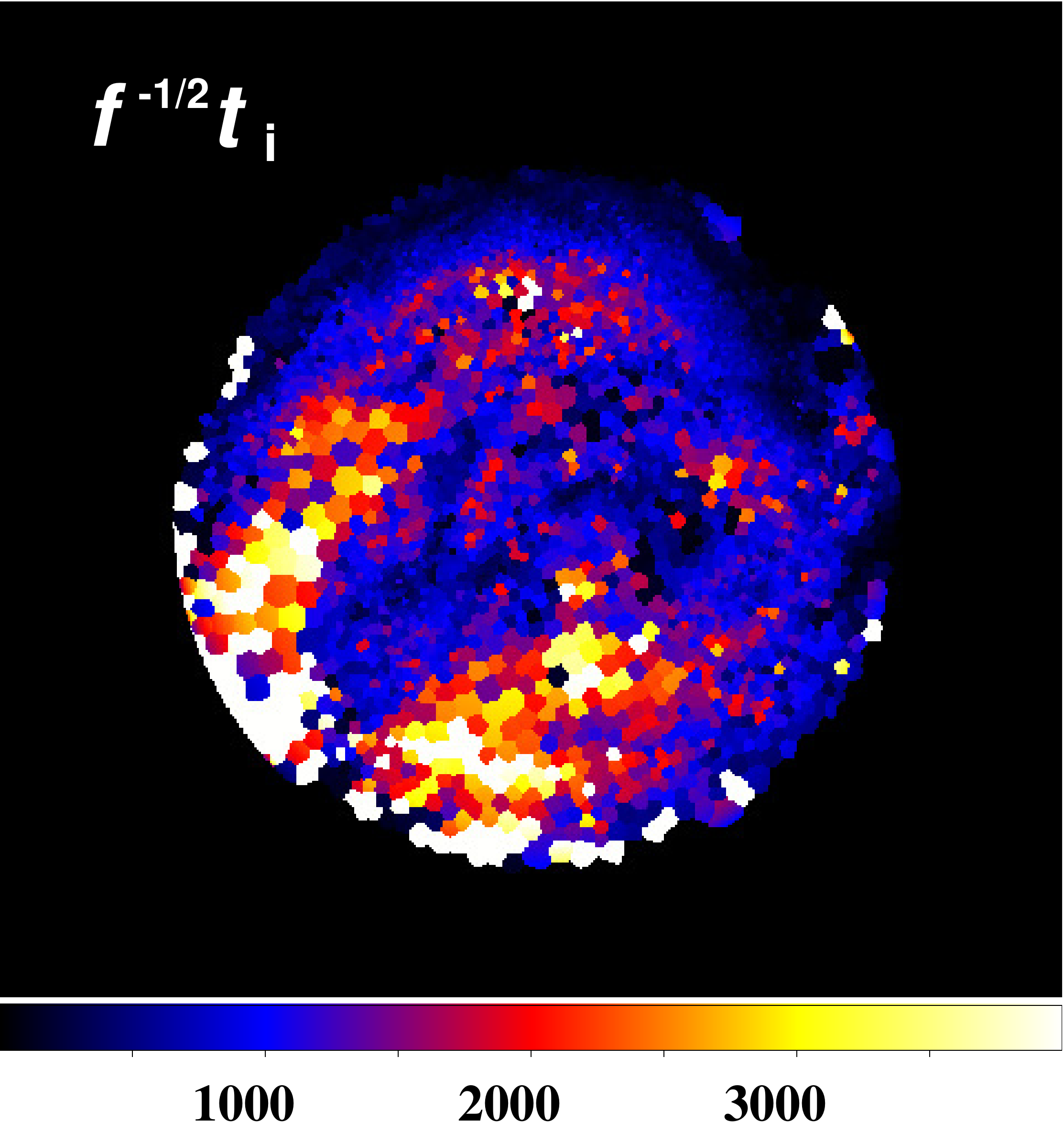}
		\caption{{Maps of electron density (cm$^{-3}$) and ionization age (year).} \label{fig:density}}
	\end{center}
\end{figure}

\begin{figure}
	\plottwo{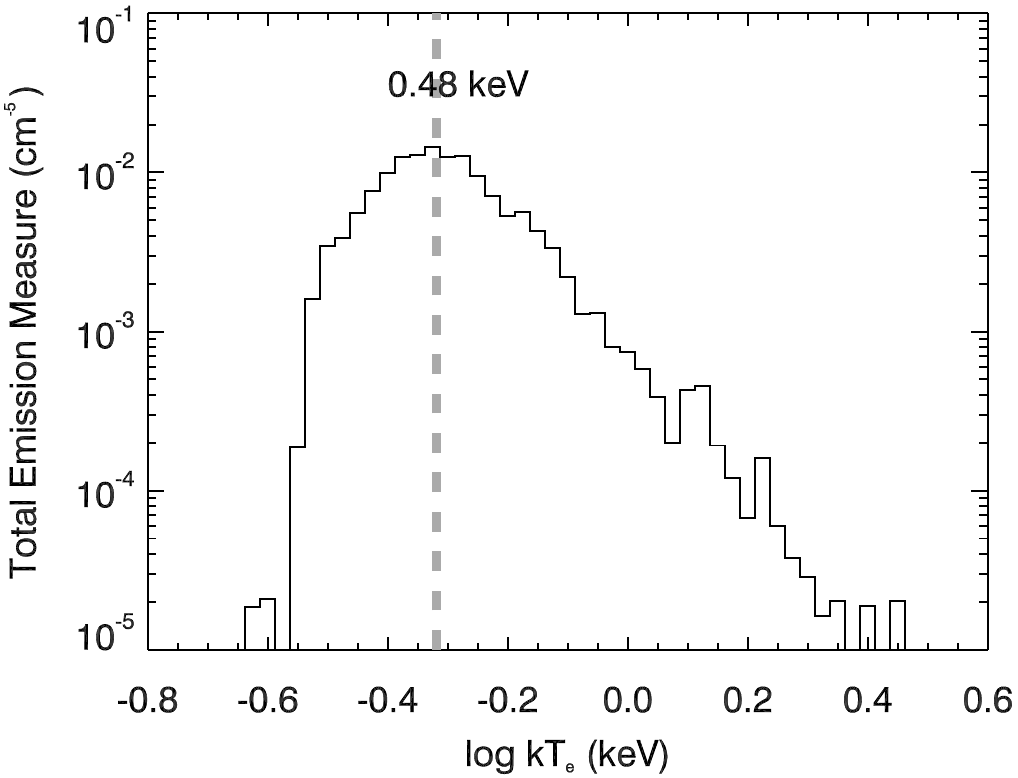}{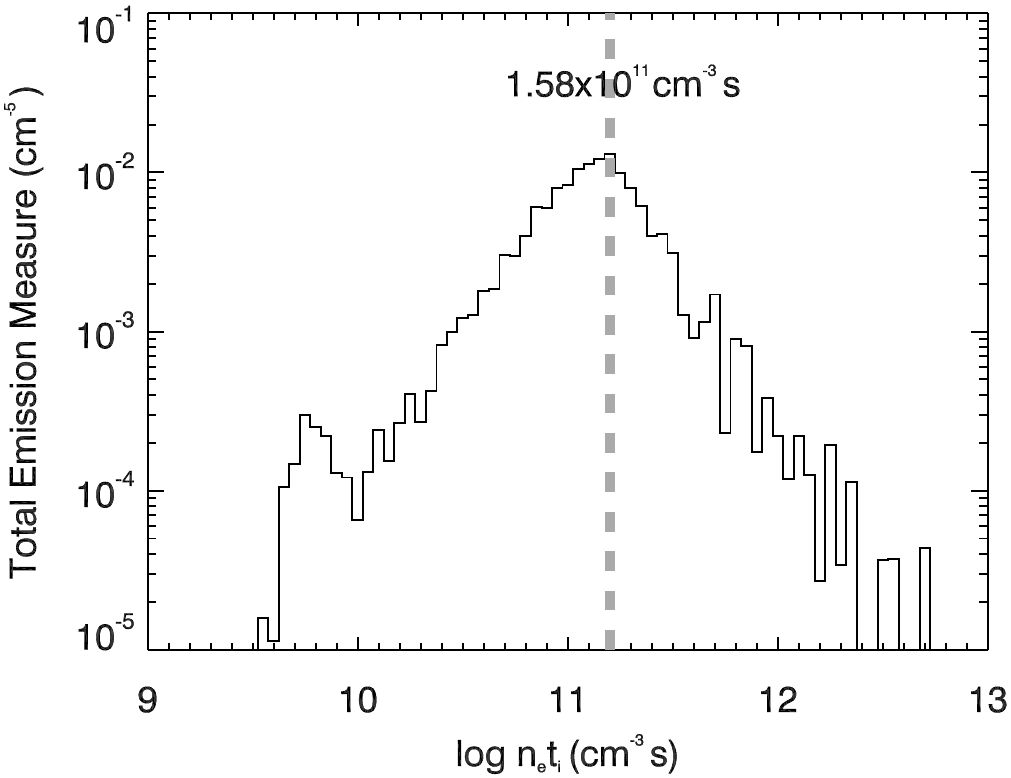}
	\caption{The PDFs of electron temperature $\kTe$ and ionization parameter $\net$.\label{fig:kT_Tau_hist}}
\end{figure}

\begin{figure}
	\plottwo{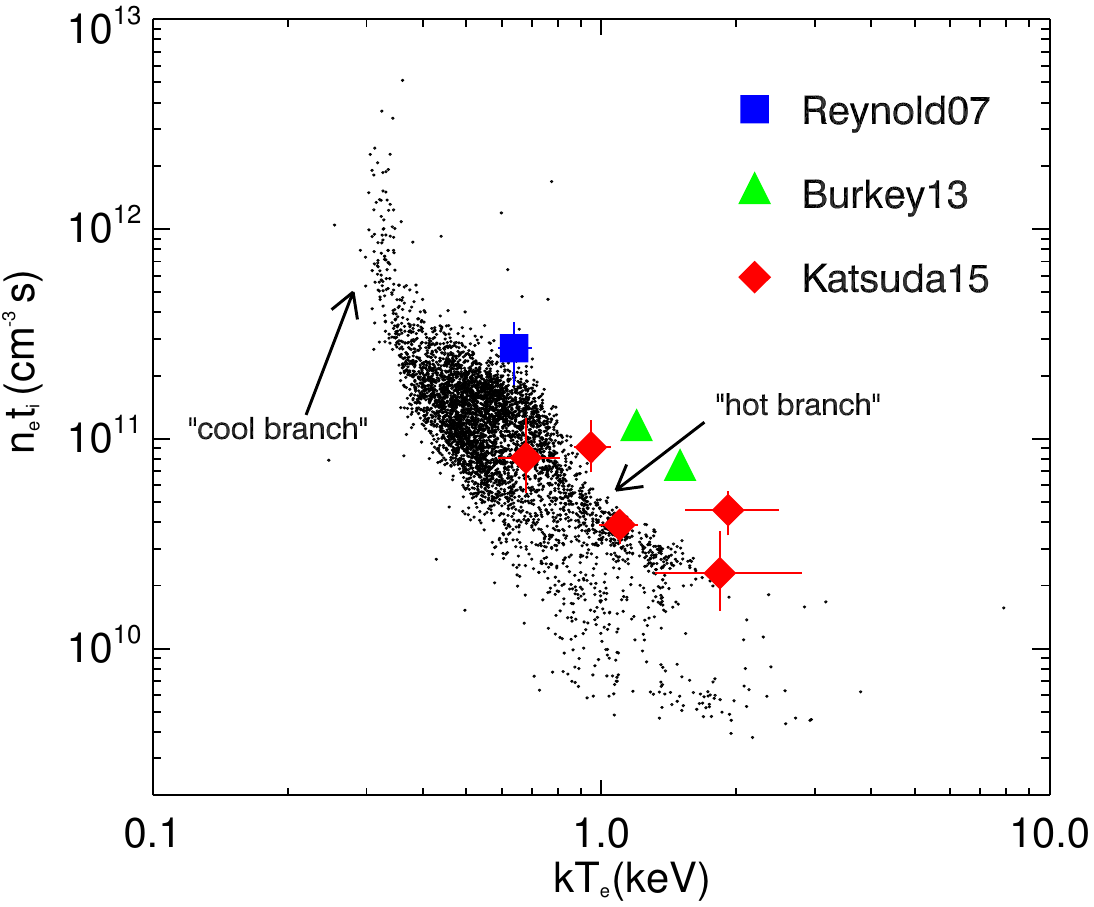}{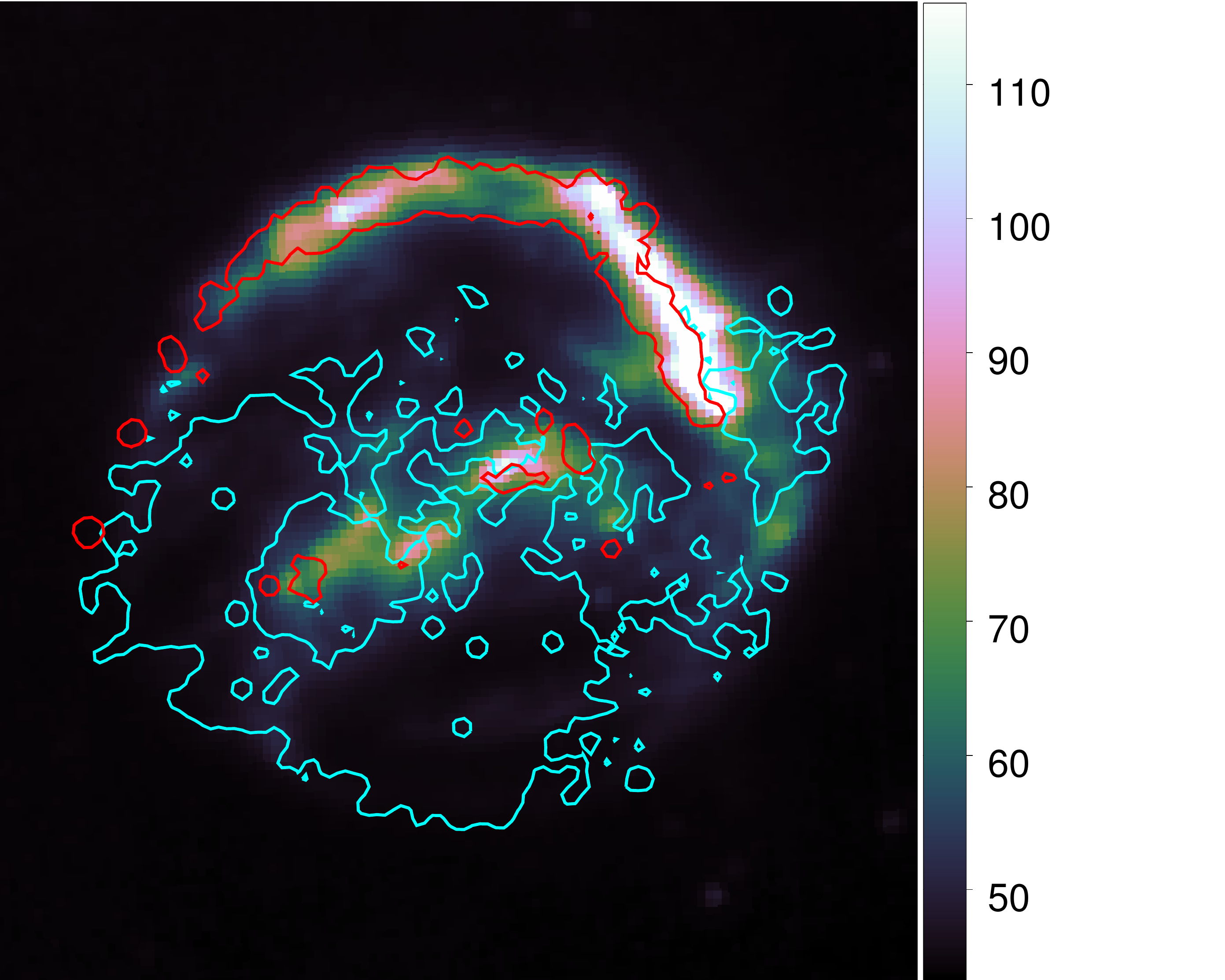}
	\caption{\textit{Left:} The $\kTe$-$\net$ diagram, with archival measurements \citep{2007ApJ...668L.135R,2013ApJ...764...63B,2015ApJ...808...49K}. Two arrows denote the two branches of the distribution of the $\kTe$ and $\net$.
		%Shown in this plot, temperature and ionization parameter of the plasma follow an anti-correlation, and further diverge into two branches: a hotter branch with higher $\kTe$ and lower $\net$, as well as a cooler branch with lower $\kTe$ and higher $\net$. 
		\textit{Right:} Spatial distributions of two branches, hot branch in red and cool branch in cyan. Color map is the \textit{Spitzer} 24 $\mu$m image \citep{2007ApJ...662..998B} which traces the distribution of warm CSM {dust}, {the color scale is in units of MJy sr$^{-1}$.} \label{fig:kT_Tau}}
\end{figure}

\begin{figure}
	\plottwo{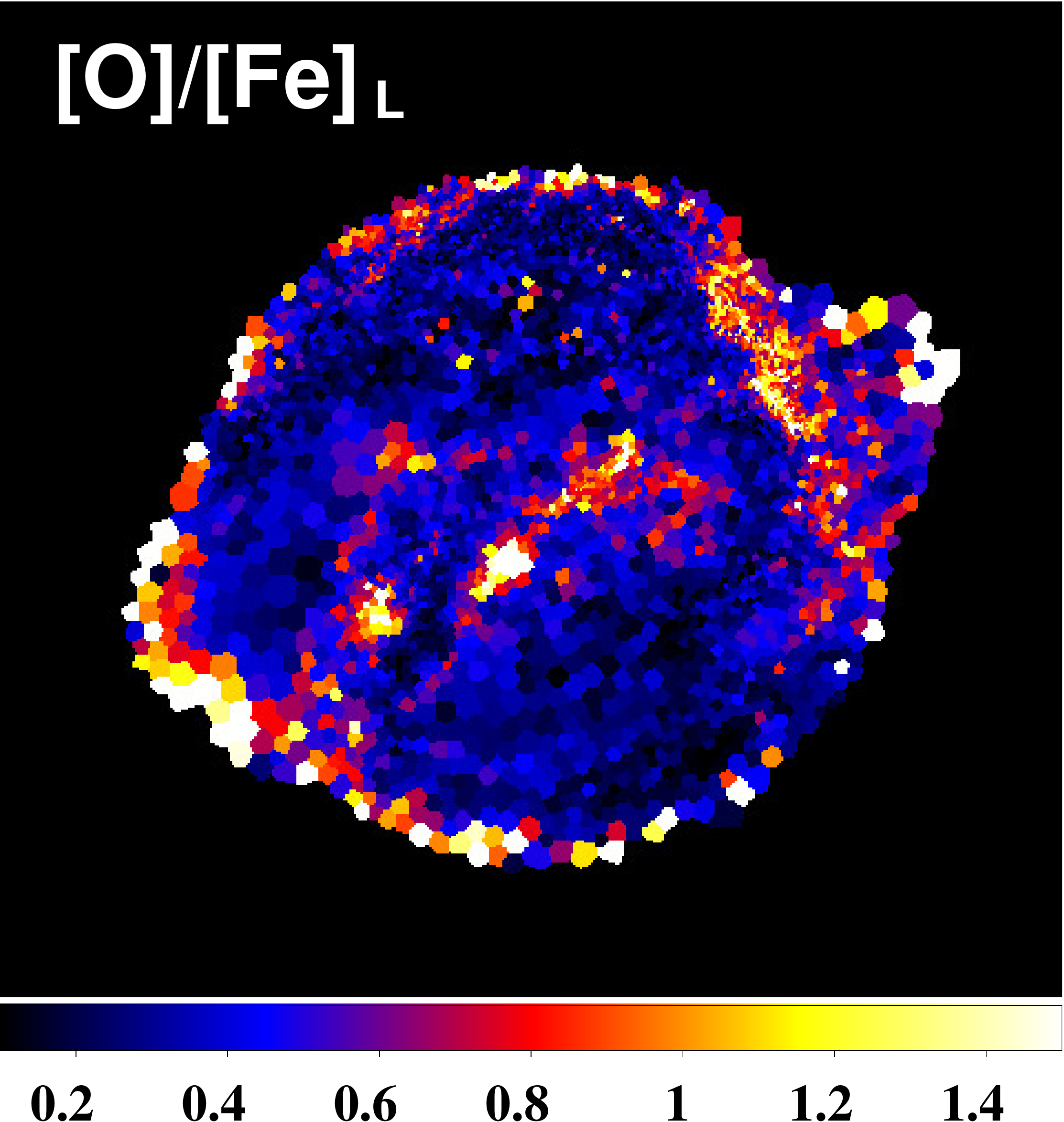}{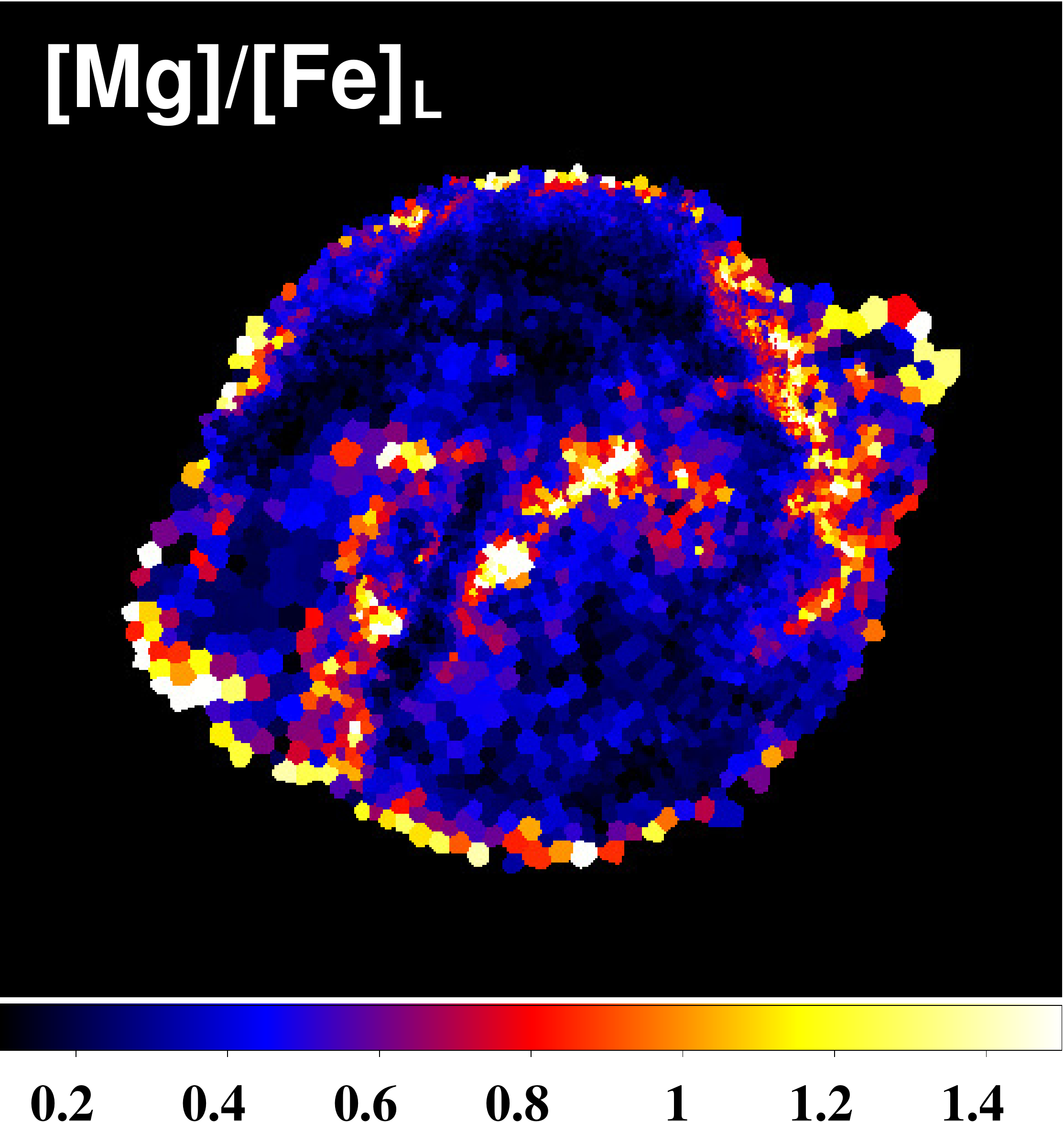}
	\caption{Maps of [O]/[Fe]$_{\rm  L}$ and [Mg]/[Fe]$_{\rm  L}$ ratios.  High [O]/[Fe]$_{\rm  L}$, [Mg]/[Fe]$_{\rm  L}$ ratios ($\sim 1$) mainly appear on NW, central bar-like region and outer regions of the remnant. \label{fig:O_Mg_Fe_map}}
\end{figure}

\begin{figure}
	\plottwo{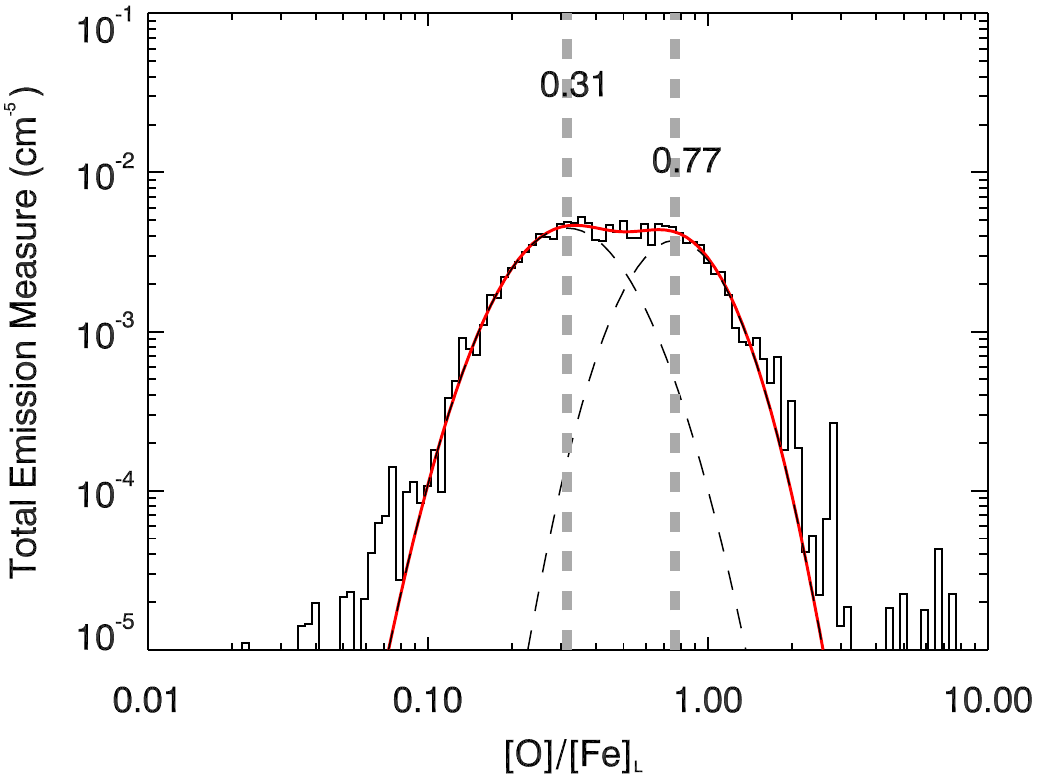}{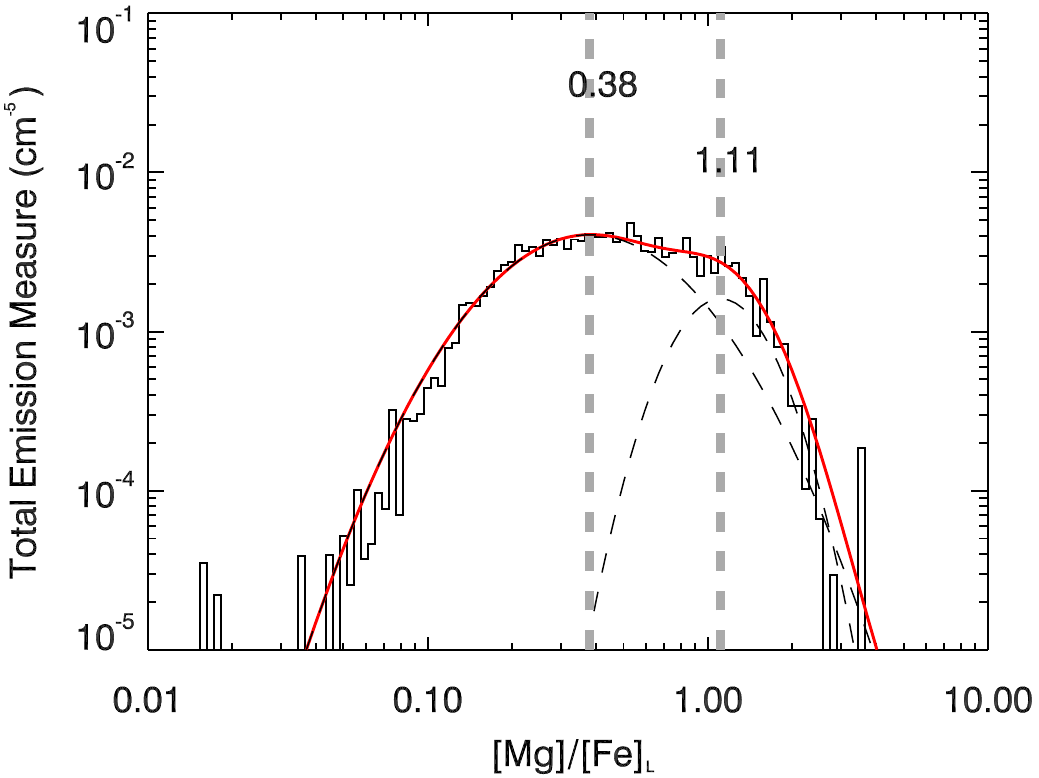}
	\caption{The PDFs and double-Gaussian fitting of [O]/[Fe]$_{\rm  L}$ and [Mg]/[Fe]$_{\rm  L}$ abundance ratios. Fitting results are summarized in Table \ref{tab:double_gaussian}. \label{fig:O_Mg_hist}}
\end{figure}

\begin{figure}
	\plotone{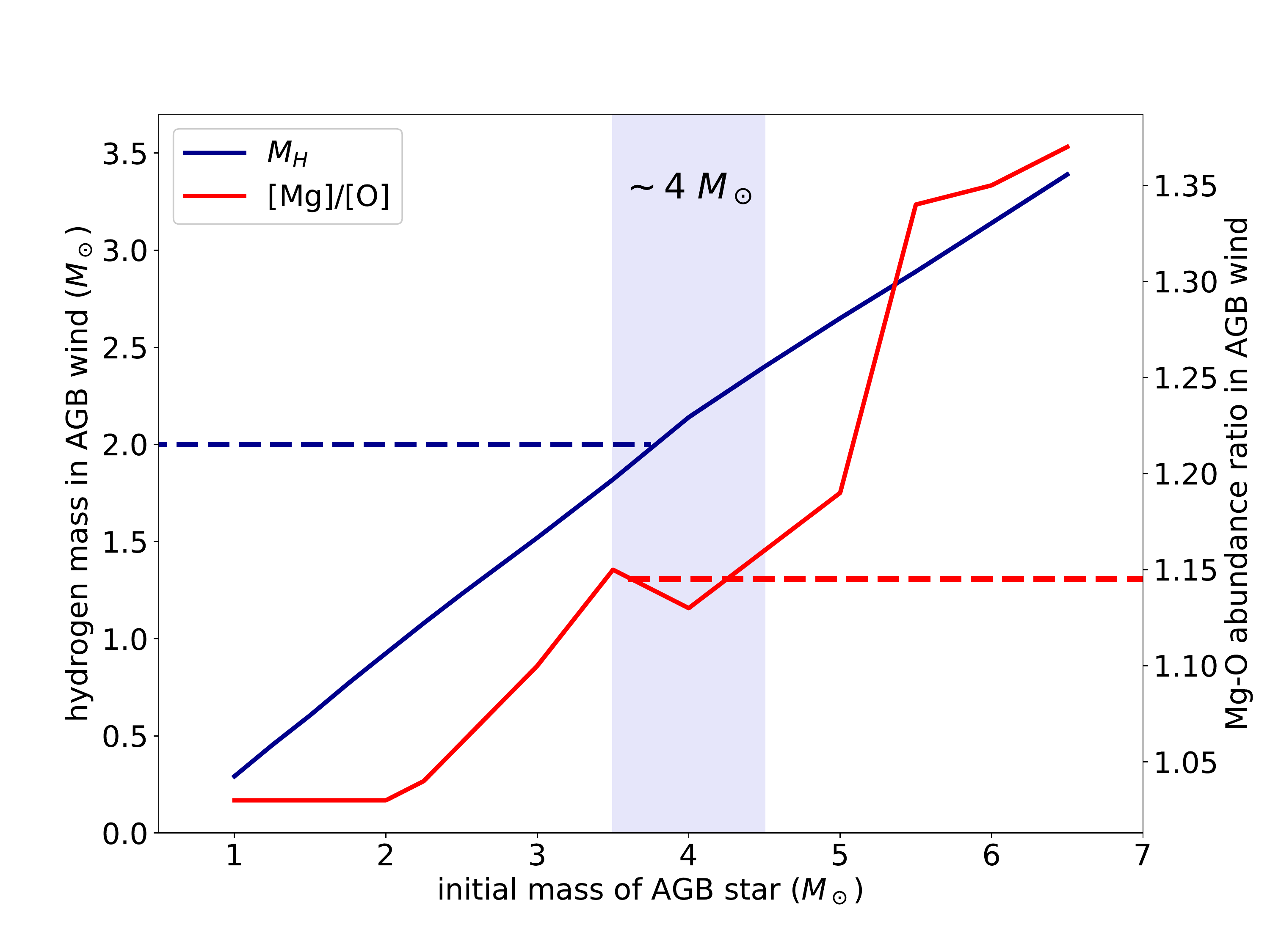}
	\caption{The hydrogen mass and Mg-to-O abundance ratio in AGB wind, based on the results of \citet{2010MNRAS.403.1413K}. A stellar wind with total mass $\gtrsim 2 \Msun$ and [Mg]/[O] $\sim$ 1.14 (denoted by dashed line) can be well reproduced by an AGB star with initial mass of $\sim 4 \Msun$.  \label{fig:AGB_mass}}
\end{figure}

\begin{figure}
	\plotone{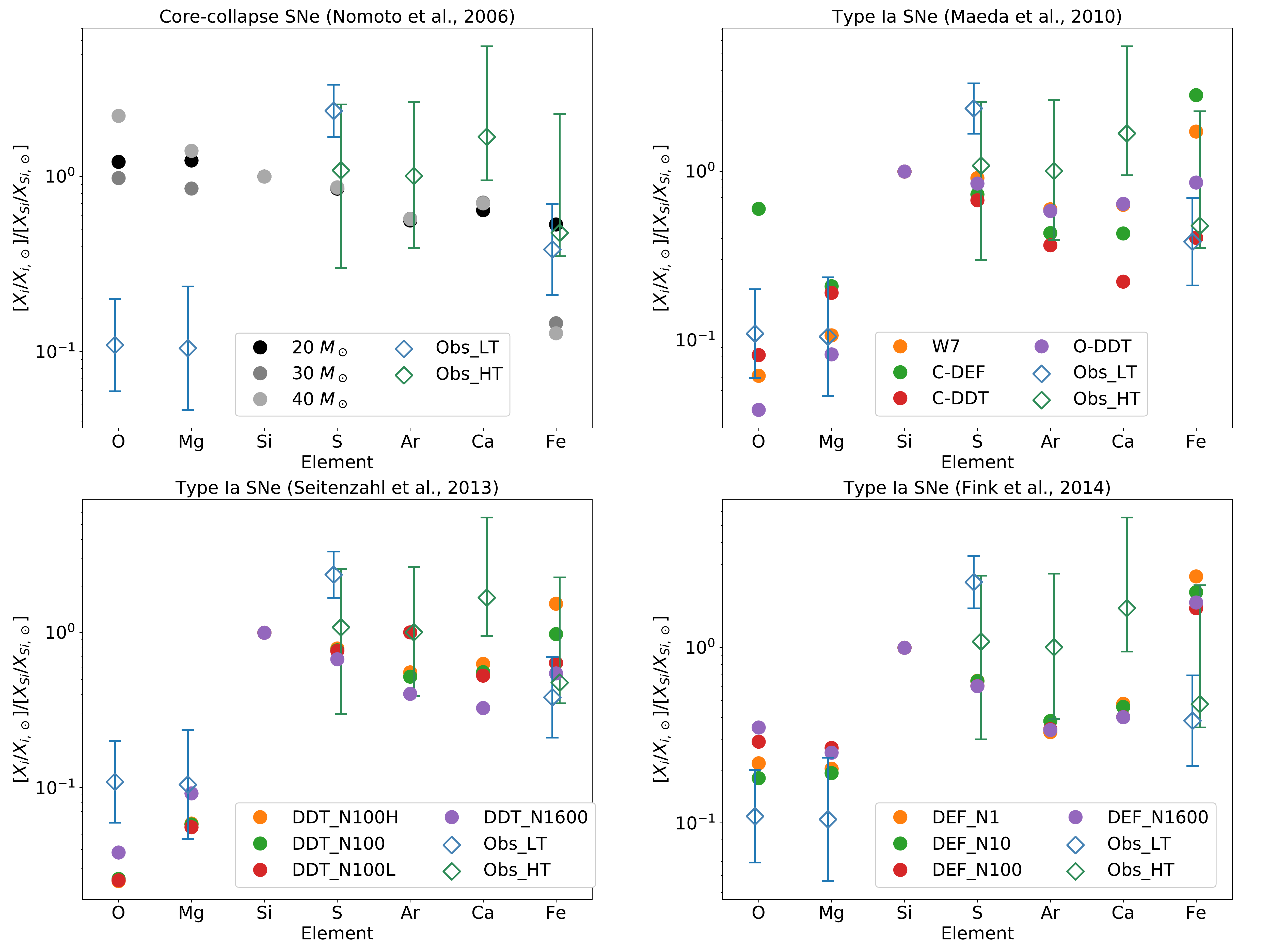}
	\caption{The metal abundance ratios of the ejecta in Kepler (empty diamonds: { blue for low temperature and green for high temperature ejecta, respectively}), compared with the predicted results of various SNe models (filled circles). \textit{Top left:} Core-collapse SNe models, in \citet{2006NuPhA.777..424N}. The initial mass of the progenitors are $20 \Msun$, $30 \Msun$ and $40 \Msun$, respectively, with the explosion energy of $10^{51}$ erg. \textit{Top right:} 2-D pure-deflagration (W7, C-DEF) and center/off-center delayed-detonation (C-DDT, O-DDT) models, in \citet{2010ApJ...712..624M}. \textit{Bottom left:} 3-D pure-deflagration models with multi-spot ignition, in \citet{2013MNRAS.429.1156S}. The numbers of central ignition spots are 1, 10, 100 and 1600, respectively. \textit{Bottom right:} 3-D pure-deflagration models with multi-spot ignition, in \citet{2014MNRAS.438.1762F}. 
		%The numbers of central ignition spots are 1, 10, 100 and 1600, respectively.
		 \label{fig:CC_Ia_ratio}}
\end{figure}

\begin{figure}
	\plottwo{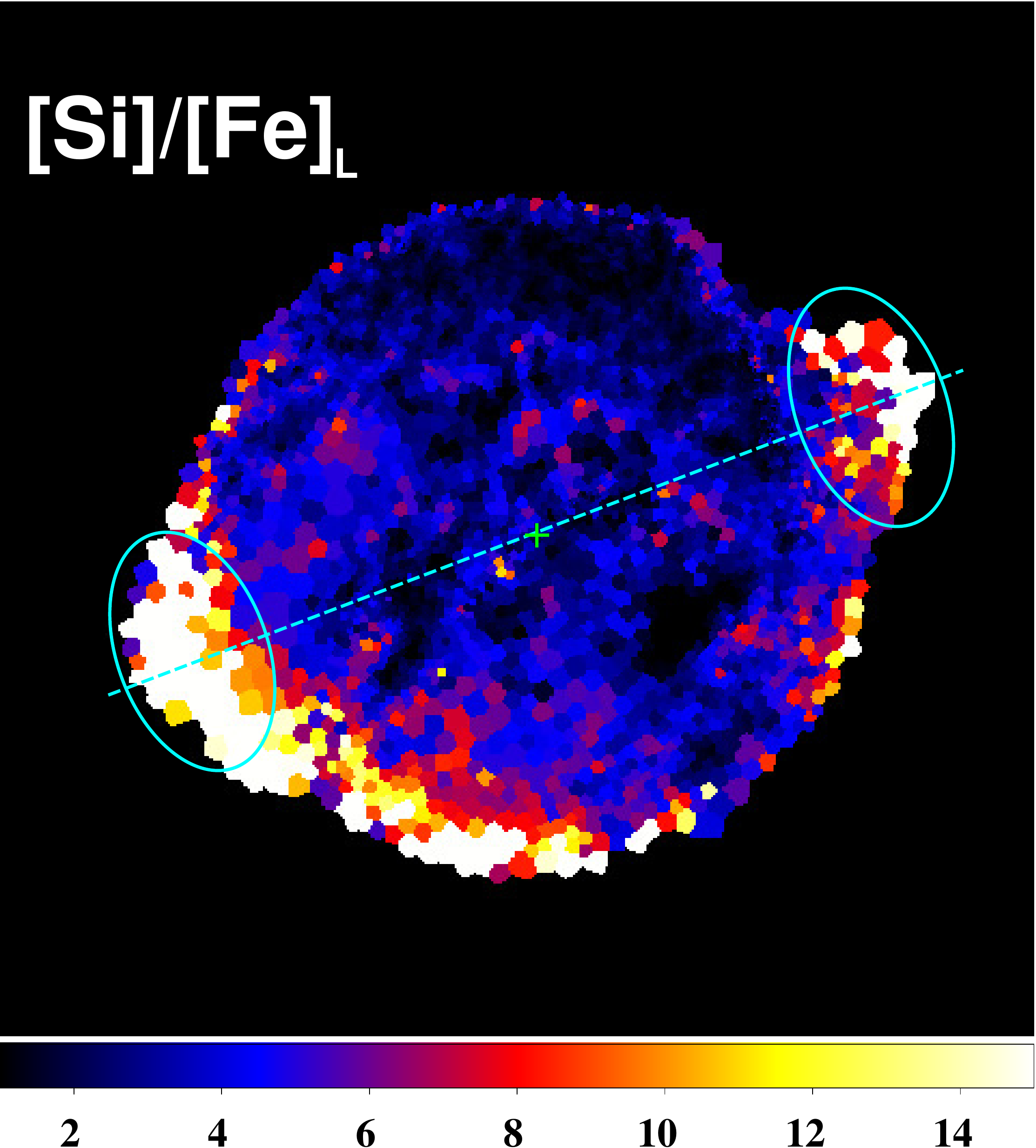}{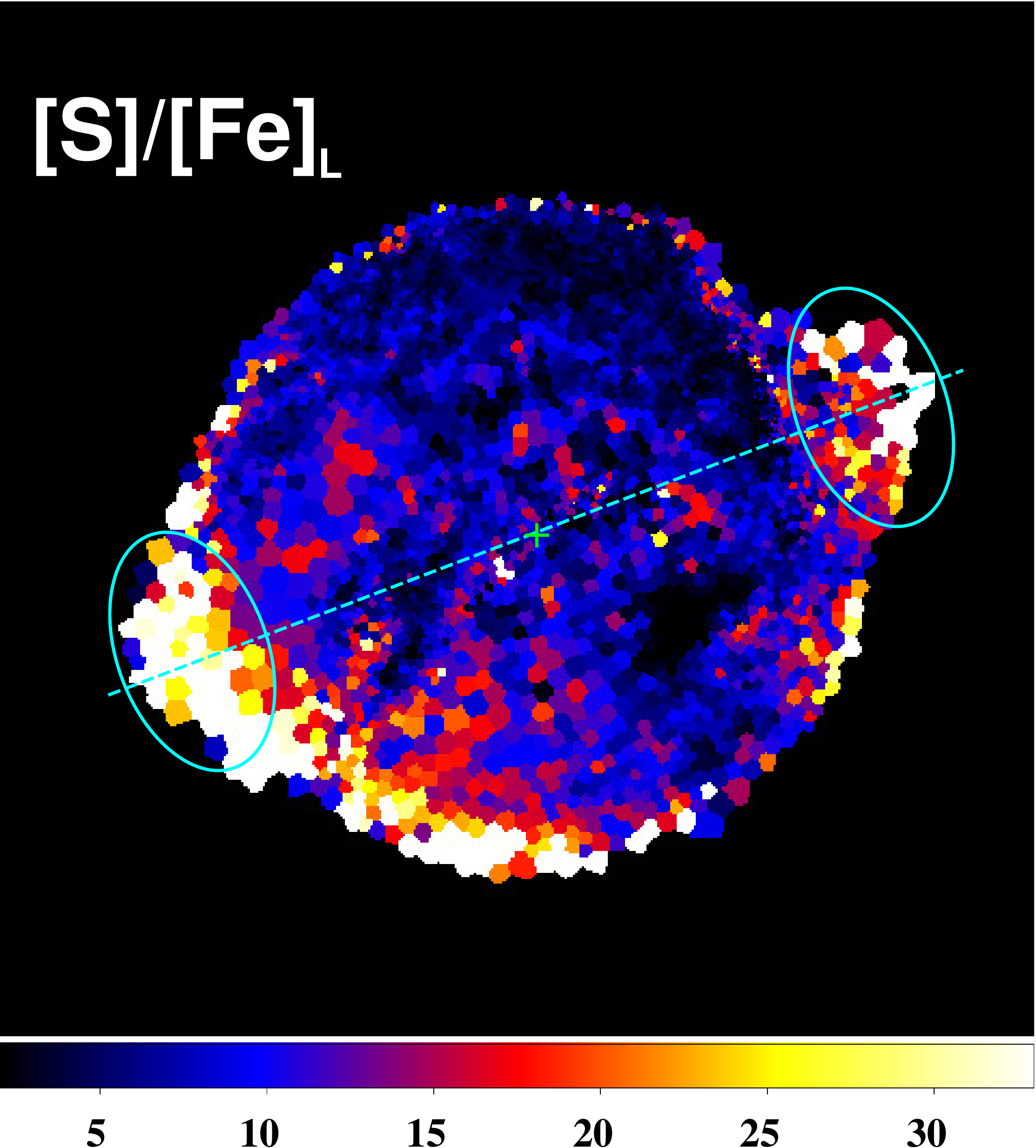}
	\caption{Maps of [Si]/[Fe]$_{\rm  L}$ and [S]/[Fe]$_{\rm  L}$ abundance ratios. Cyan ellipses denote the ``ears'' of the remnant. \label{fig:Si_S_Fe_map}}
\end{figure}

	The whole SNR is divided into 139 tessellated regions with the WVT binning algorithm. Each region contains at least 400 counts in the 6.1--6.8\,keV to guarantee enough counting statistics for the Fe K$\alpha$ complex. We then extract the spectra from the tessellated regions in all observations. The background spectra are extracted from the same area {shown} in Figure \ref{fig:SNR}. For each tessellated region, we stack the spectra extracted from different observations as well as the corresponding backgrounds and response files with the CIAO task \textit{combine\_spectra}. The 2.2--8.0\,keV background-subtracted spectrum of each region is fitted with a ``2 \texttt{vnei} $+$ \texttt{powerlaw}'' model. The hydrogen absorption has little effect on the hard band spectra, thus is ignored. The parameter settings of the spectral model are mainly based on \citet{2015ApJ...808...49K}. We assume one of the \texttt{vnei} component to be high-temperature ejecta consisting of almost pure metal elements. The abundance of S, Ar, Ca and Fe are set to be free, the abundance of Ni is tied to that of Fe, and the abundance of Si is fixed to be $10^5$ times the solar value. We assume another \texttt{vnei} component to be pure Fe ejecta which produces most of the K$\alpha$ emission and fix the abundances of other metals to zero. We find that the electron temperature of this pure Fe component is hard to be constrained in the spectral fitting. Thus we link it to the temperature of \texttt{vnei}1, with a factor of 1.3 based on \citet{2015ApJ...808...49K}. The Fe abundance are tied to each other between two components. The redshifts are also set to be free if necessary.
	% In addition, we multiply the two \texttt{vnei} components by the \texttt{gsmooth} model to account for the line-broadening effects.
	
\begin{figure}
	\plottwo{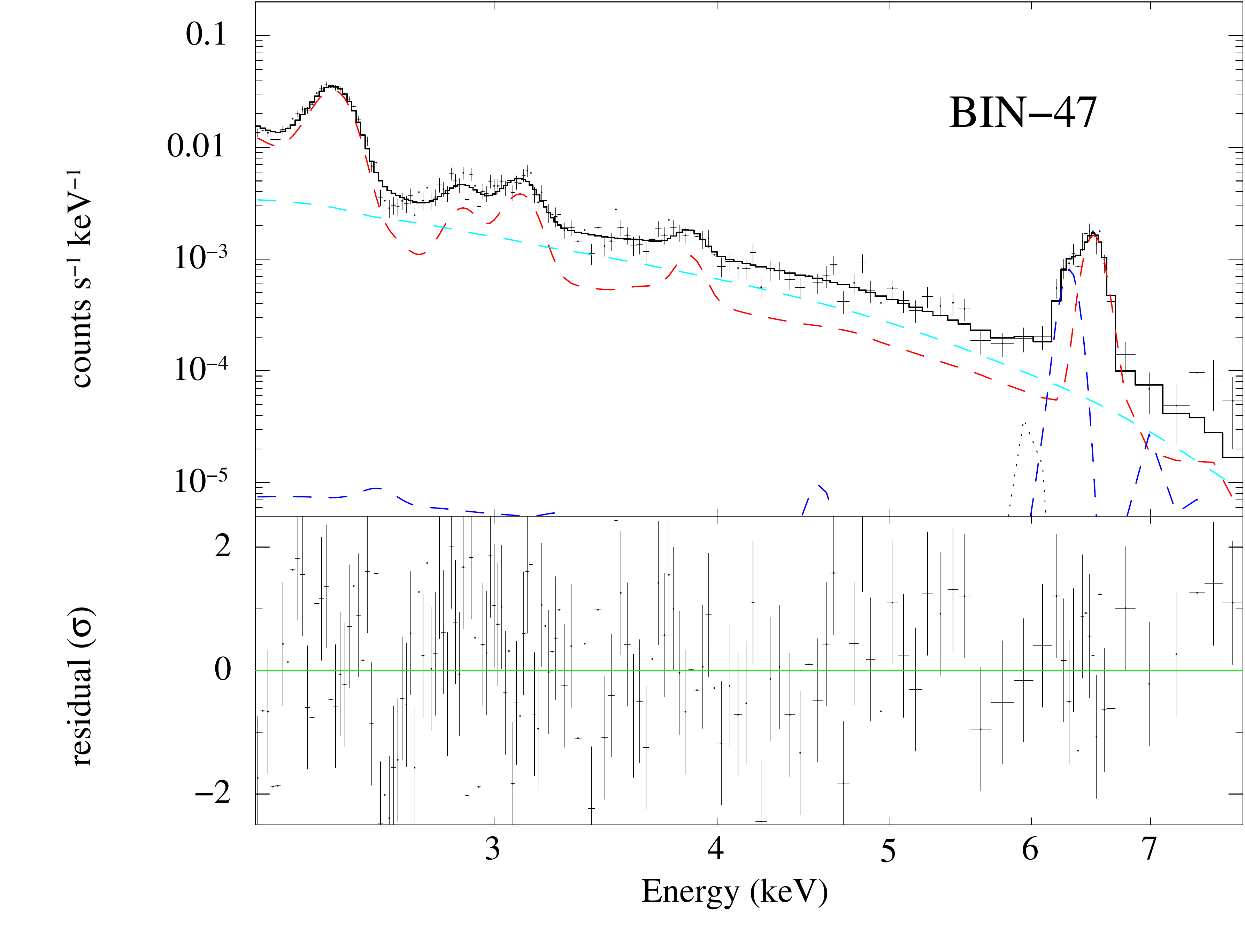}{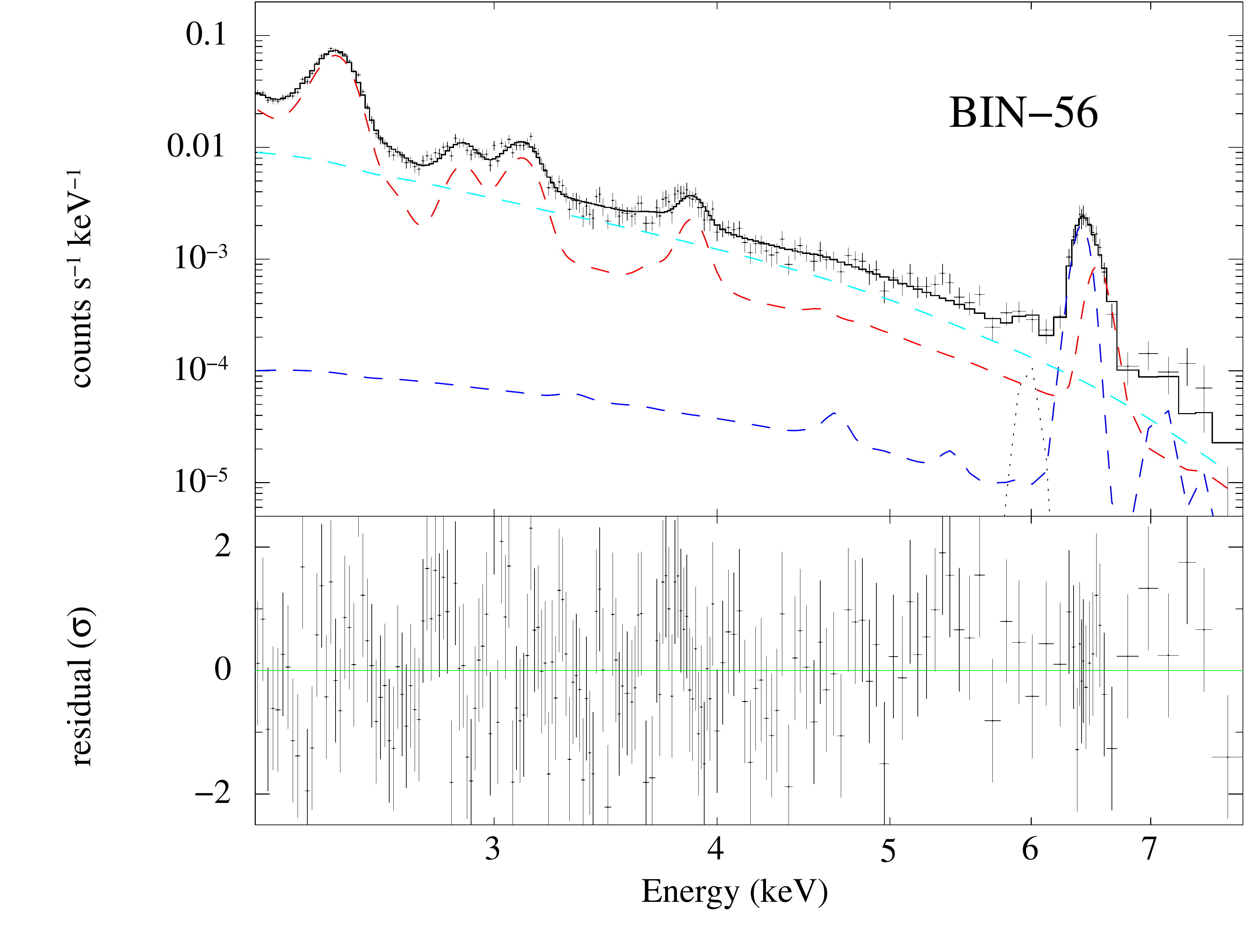}
	\caption{{Examples of fitted 2.2--8.0\,keV spectra from two tessellated regions (BIN-47 and BIN-56, as marked in the first panel of Figure \ref{fig:hard_par_map}). The black data points are the background-subtracted spectra, and the solid lines represent their best-fit models. Dashed lines with different colors represent different model components: hot ejecta \texttt{vnei}1 in red, pure Fe ejecta \texttt{vnei}2 in blue, and \texttt{powerlaw} in cyan.} \label{fig:hard_spec_exp}}
\end{figure}

	The ``2 \texttt{vnei} $+$ \texttt{powerlaw}'' model gives acceptable fitting results for most of the regions. Examples of spectra extracted from two tessellated regions are shown in Figure \ref{fig:hard_spec_exp} and the best-fit	parameters of them are summarized in Table \ref{tab:hard_spec_example}. There are 6 regions have \RCS$> 1.8$. In most of the cases, the residuals concentrate in 2.2--3.0\,keV, which are mainly caused by emission from low temperature S. We exclude these regions in subsequent analysis.
	
\begin{figure*}
	\gridline{
		\fig{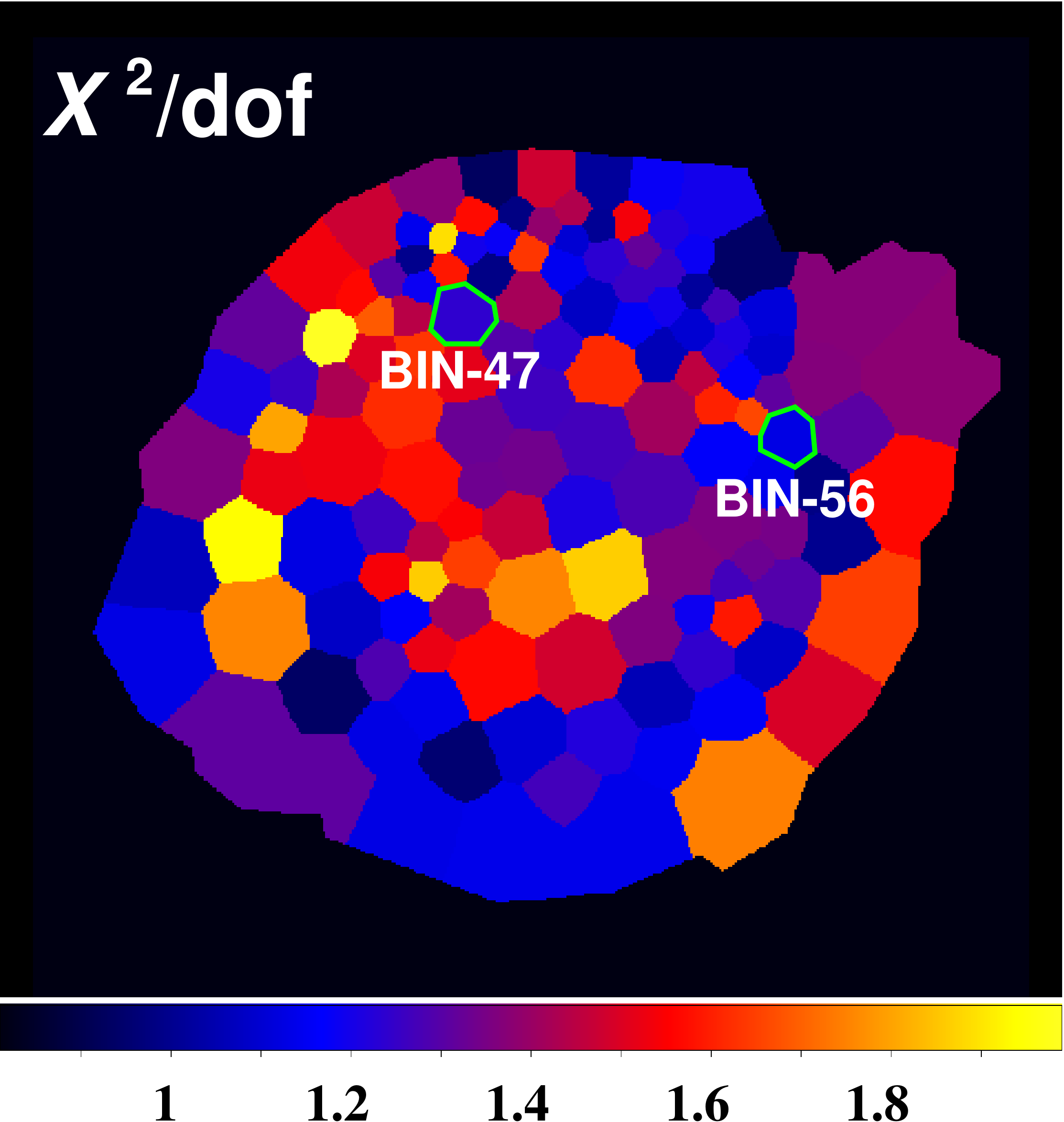}{0.2\textwidth}{(a)}
		\fig{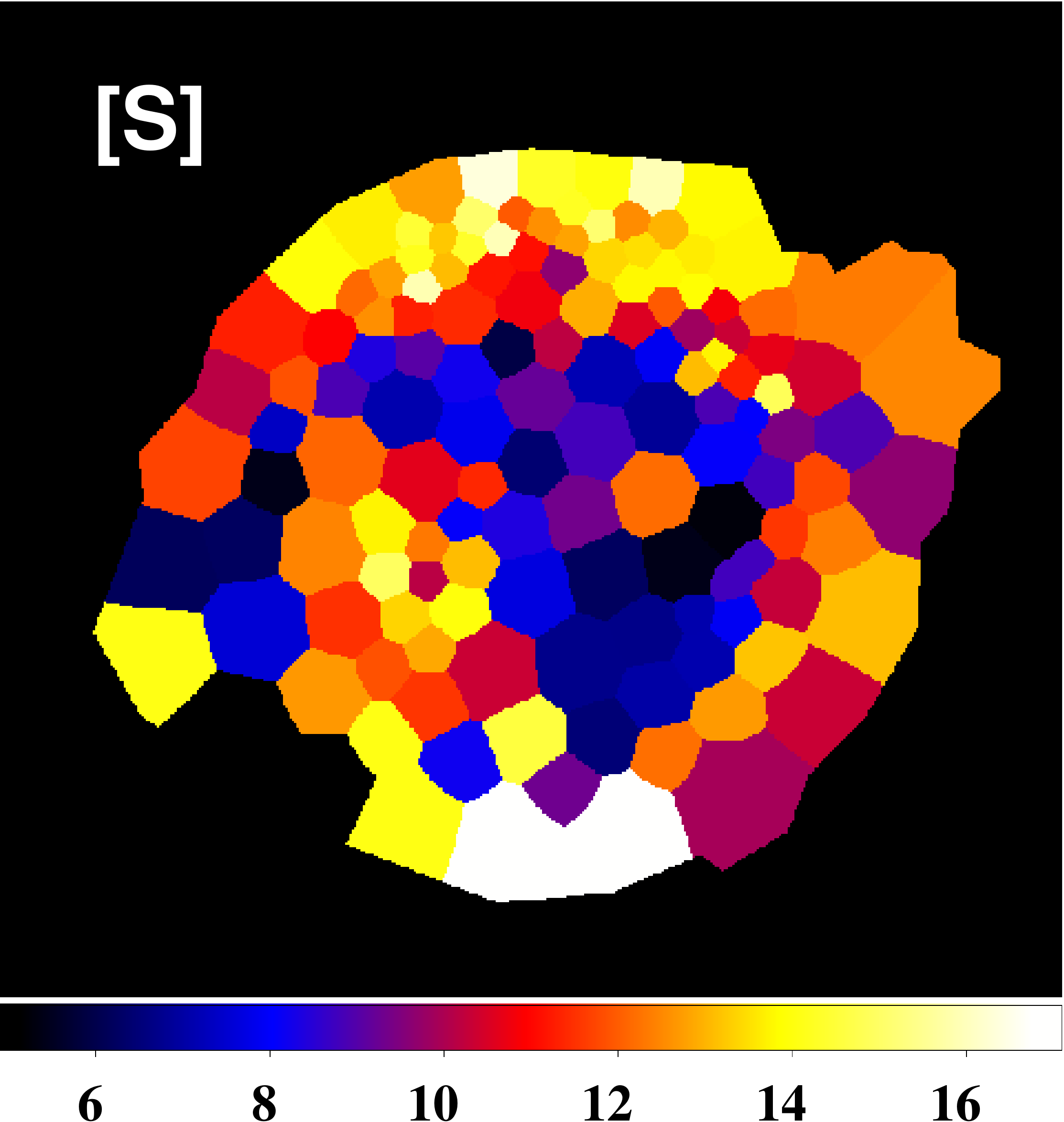}{0.2\textwidth}{(b)}
		\fig{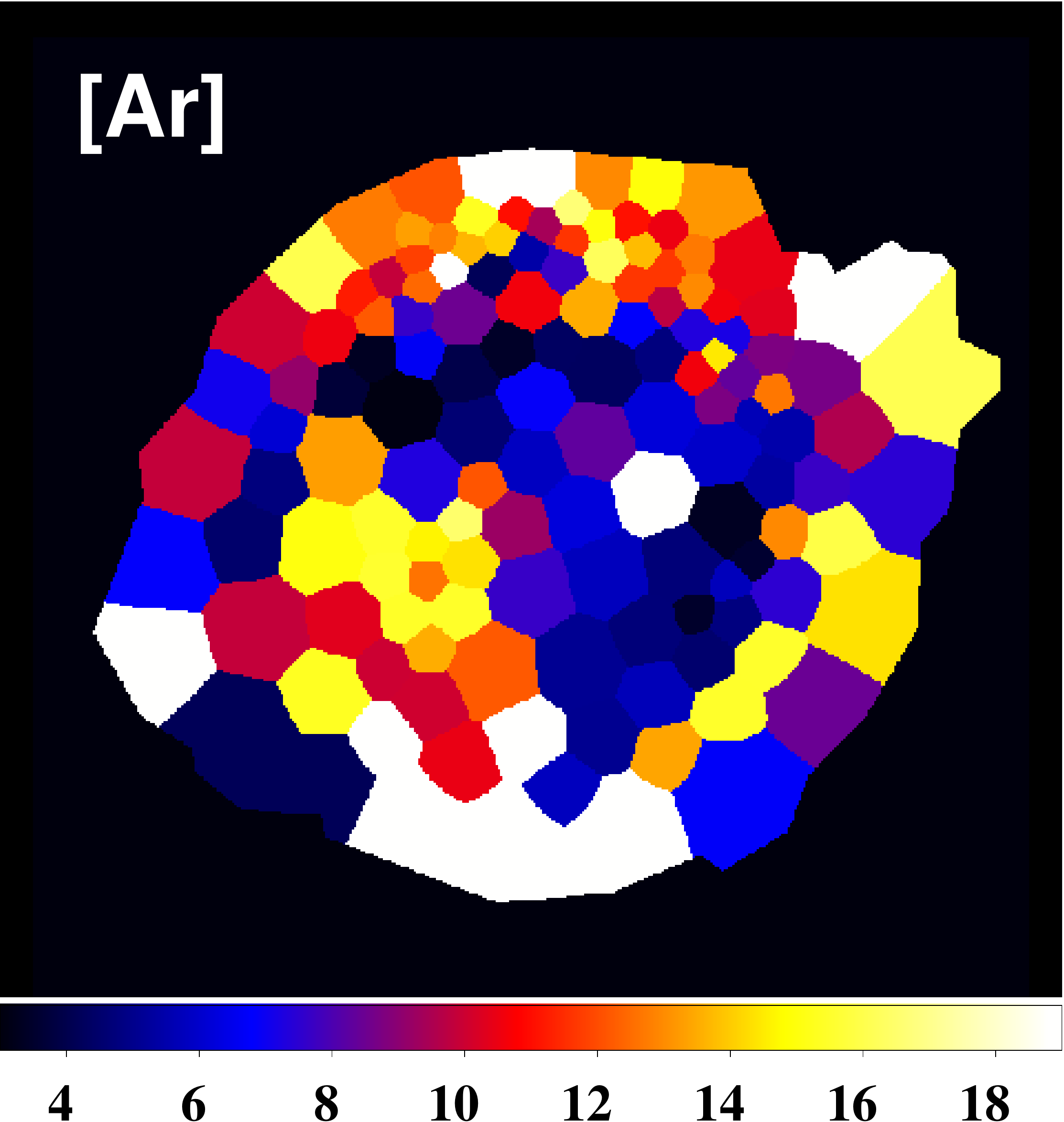}{0.2\textwidth}{(c)}
		\fig{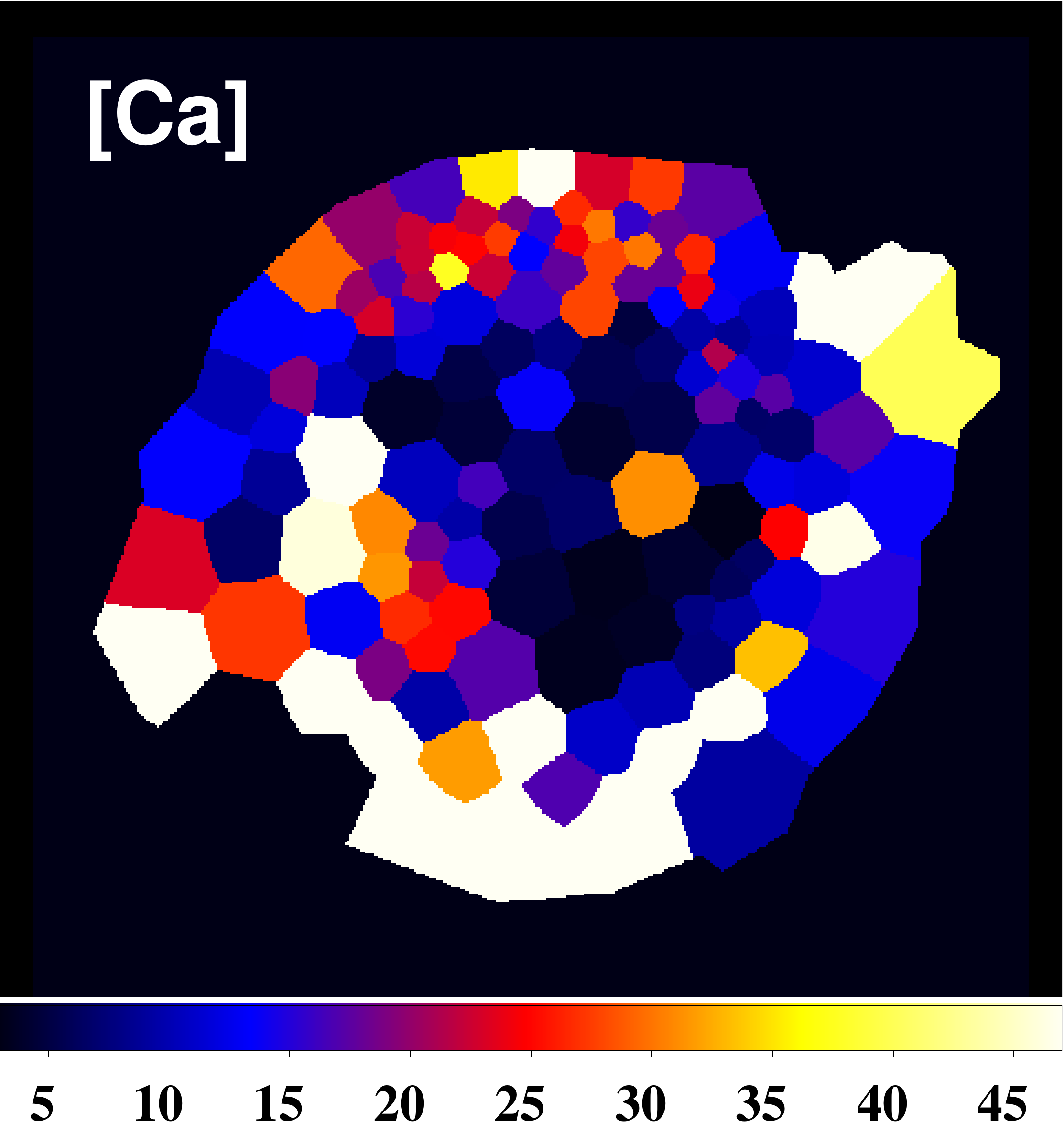}{0.2\textwidth}{(d)}
		\fig{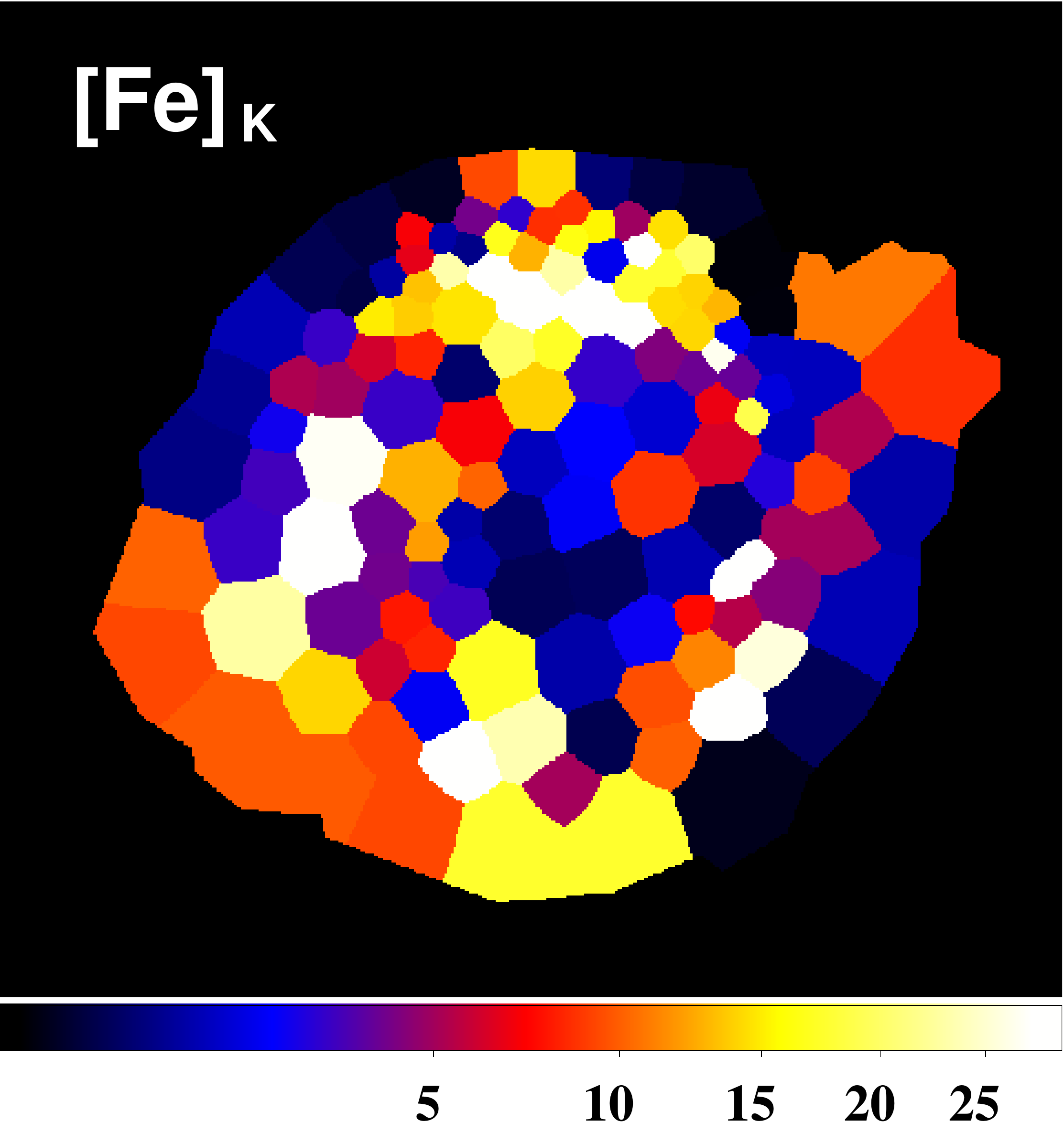}{0.2\textwidth}{(e)}
		}
	\gridline{
		\fig{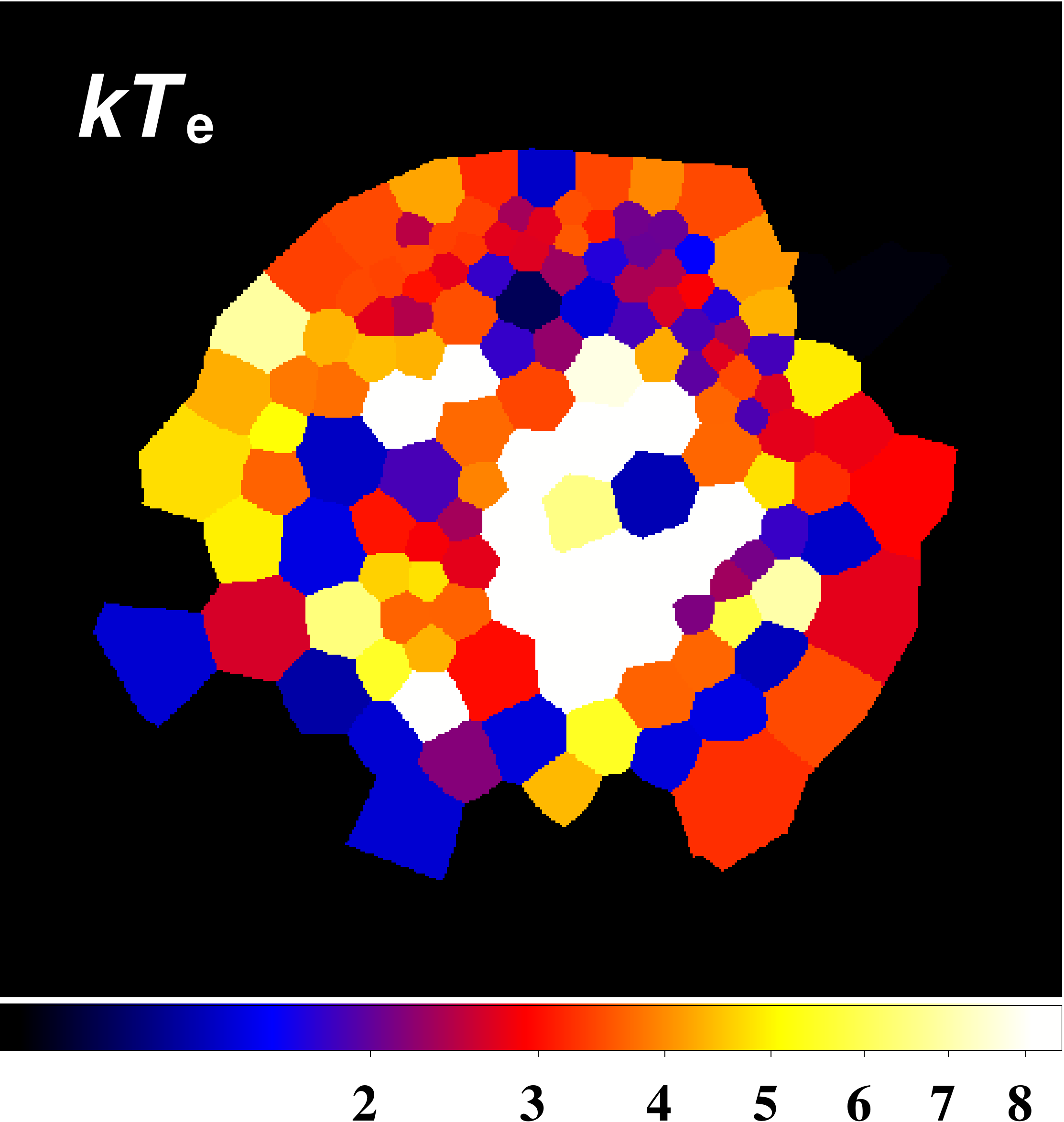}{0.2\textwidth}{(f)}
		\fig{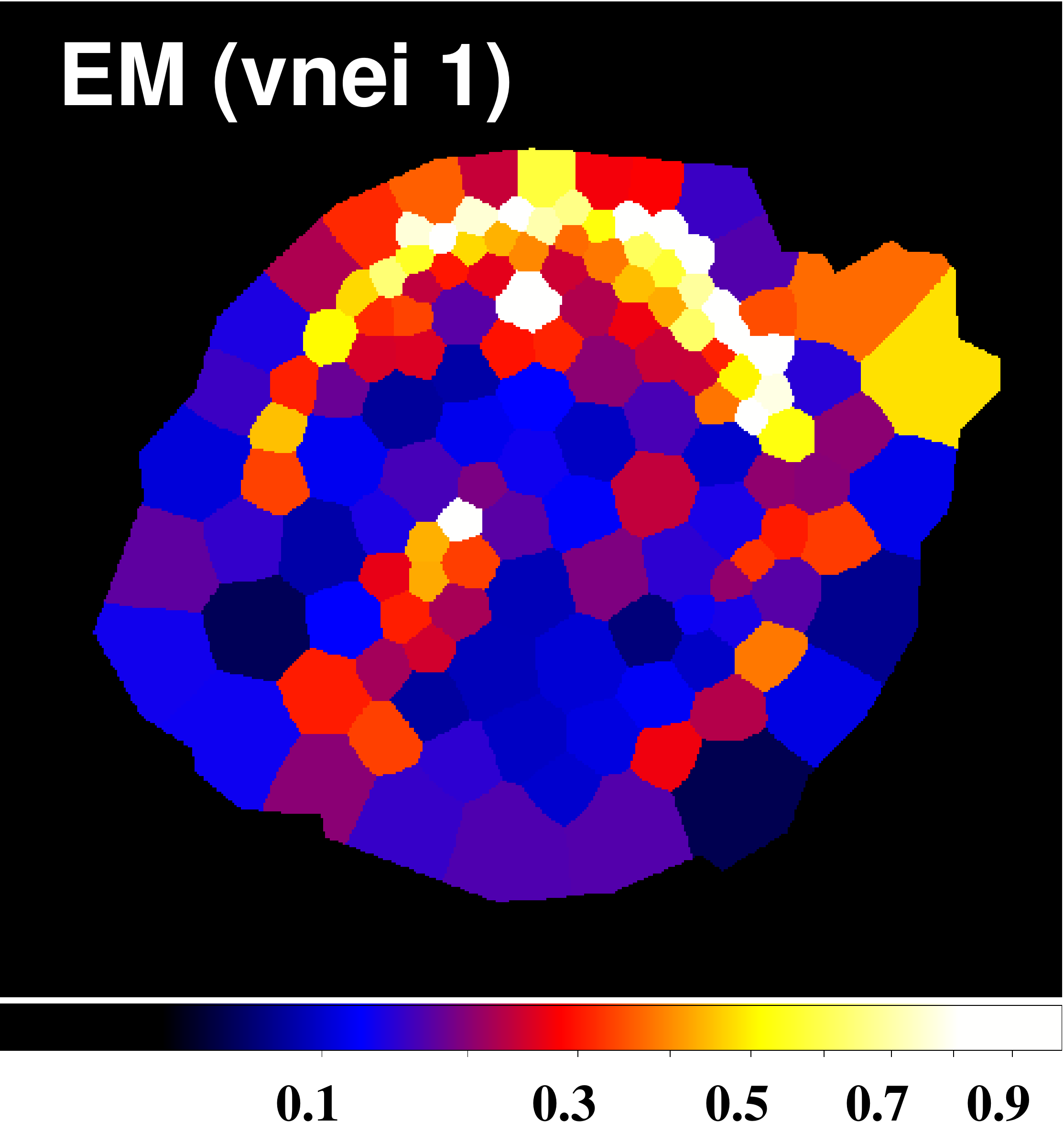}{0.2\textwidth}{(g)}
		\fig{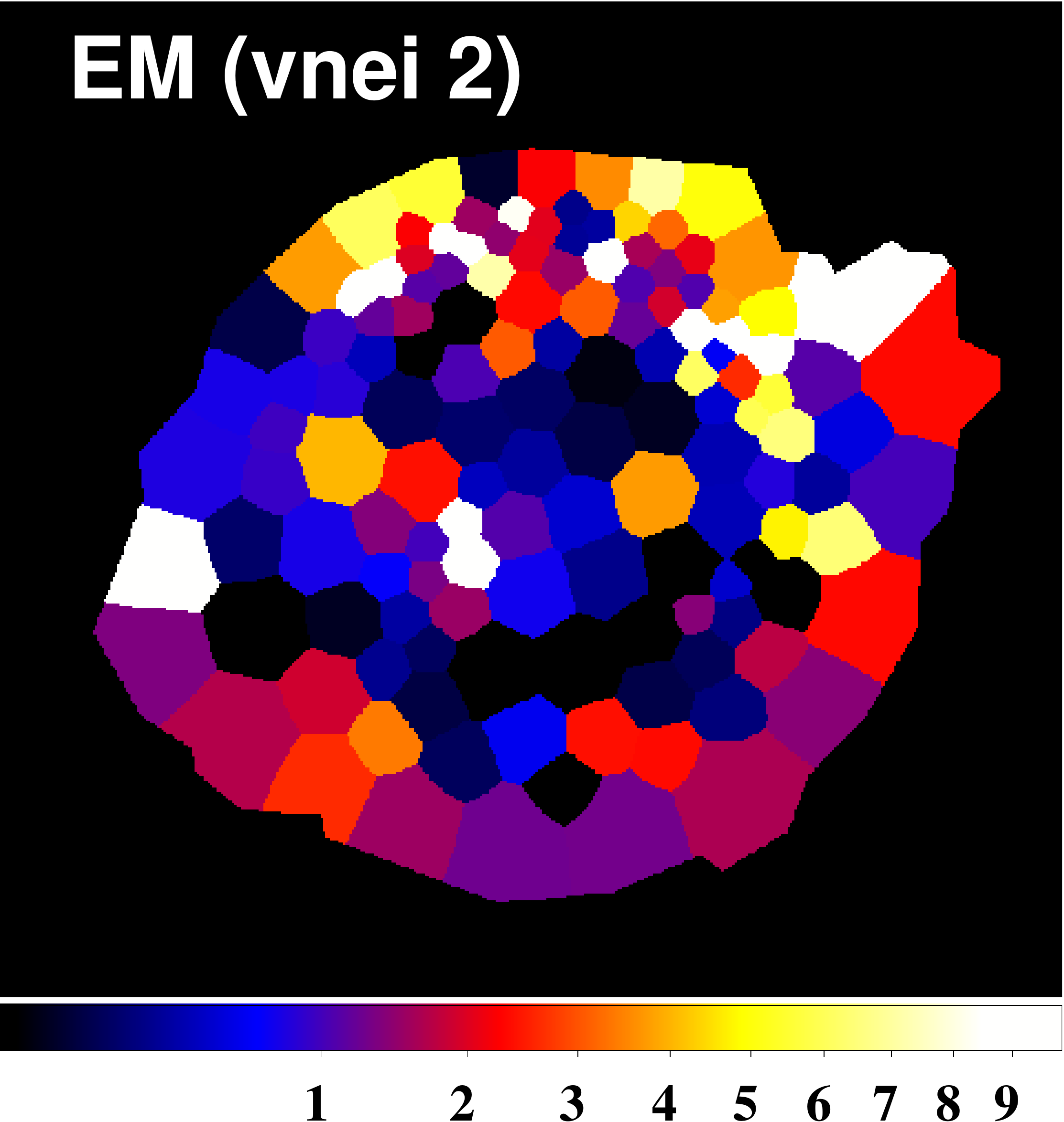}{0.2\textwidth}{(h)}
		\fig{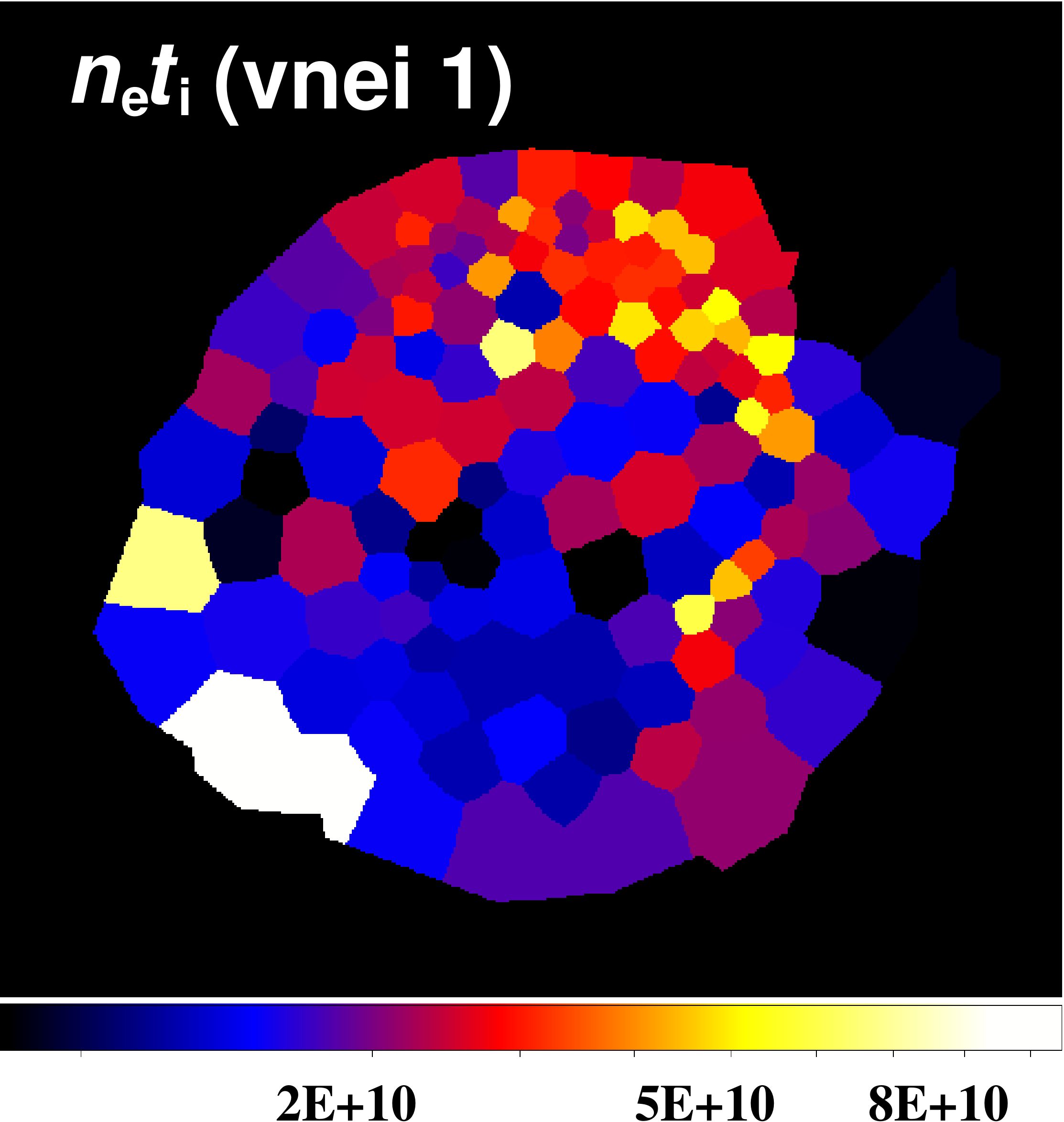}{0.2\textwidth}{(i)}
		\fig{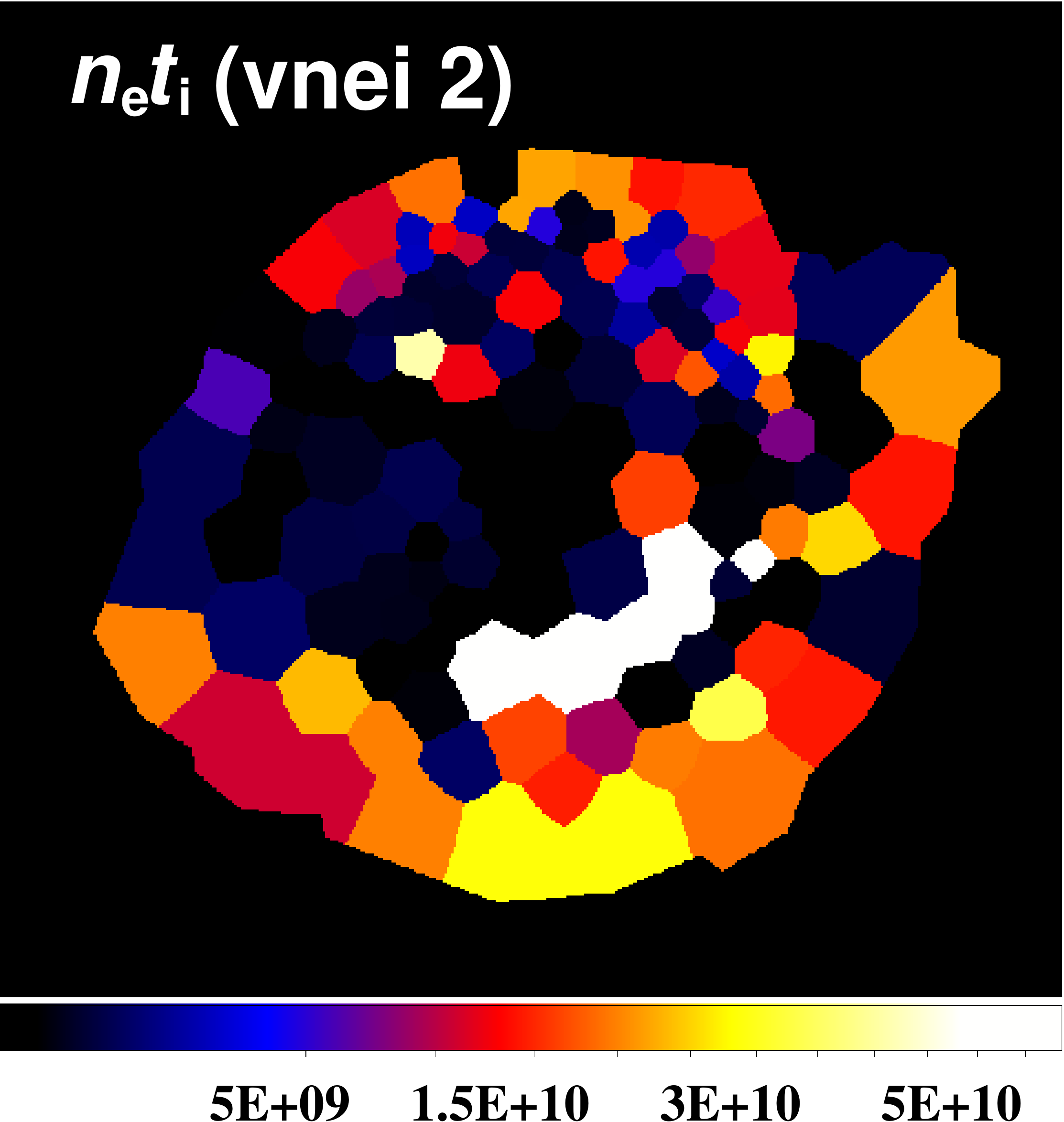}{0.2\textwidth}{(j)}
	}
	\caption{{Maps of 2.2--8.0\,keV spectral fitting parameters: (a) \RCS, the two green polygon denote the regions used to extract the example spectra in Figure \ref{fig:hard_spec_exp}, (b)-(e) metal abundances ($10^4$ solar value), (f) electron temperature (keV), (g)-(h) specific EM of two \texttt{vnei} components ($10^{-10}$\,cm$^{-5}$\,arcsec$^{-2}$) (i)-(j) ionization parameters of two \texttt{vnei} components (\netunit).} \label{fig:hard_par_map}}
\end{figure*}

	The maps of spectral fitting parameters are shown in Figure \ref{fig:hard_par_map}. We use similar method described in Section \ref{subsec:density} and the same geometry to estimate the electron and hydrogen densities in each region (for \texttt{vnei}1 and \texttt{vnei}2, respectively). Here, $n_{\rm e}/n_{\rm H}$ are determined by the metal abundances and the ionization state at the given $\kTe$ and $\net$ by using the SPEX\footnote{https://www.sron.nl/astrophysics-spex} code. The masses of metal species are calculated using hydrogen density and their abundances, and are summarized in Table \ref{tab:mass}.

\clearpage

\clearpage

\begin{deluxetable*}{cccc}
	\tablecaption{Observations Information \label{tab:obs}}
	\tablenum{1}
	\tablehead{
		\colhead{Observation ID}&\colhead{Instrument}&\colhead{Start Date}&\colhead{Exposure Time}
	}
	\startdata
	6714&ACIS-S&2006-04-27 23:12:44&157.82 ks\\
	6715&ACIS-S&2006-08-03 16:51:04&159.13 ks\\
	6716&ACIS-S&2006-05-05 19:17:13&158.02 ks\\
	6717&ACIS-S&2006-07-13 19:03:21&106.81 ks\\
	6718&ACIS-S&2006-07-21 14:54:10&107.80 ks\\
	7366&ACIS-S&2006-07-16 05:23:02&51.46 ks\\
	\enddata
\end{deluxetable*}

\clearpage

\clearpage

\clearpage

\begin{deluxetable*}{llcccc}
	\tablecaption{{Examples of Spectral Fitting Results} \label{tab:spec_example}}
	\tablenum{2}
	\tablehead{
		\colhead{}&\colhead{}&\colhead{BIN-715}&\colhead{BIN-1471}&\colhead{BIN-2402}&\colhead{BIN-3992}
	}
	\startdata
	{VNEI}&{$\kTe$ (keV)}&$1.00^{+0.55}_{-0.23}$&$0.73^{+0.05}_{-0.07}$&$0.50^{+0.09}_{-0.08}$&$0.53^{+0.01}_{-0.03}$\\
	{    }&{[O]}&$< 0.31$&$0.65^{+1.09}_{-0.47}$&$0.40^{+0.09}_{0.07}$&$< 0.44$\\
	{    }&{[Mg]}&$< 0.27$&$0.68^{+0.30}_{-0.17}$&$0.67 \pm 0.09$&$0.16^{+0.14}_{-0.12}$\\
	{    }&{[Si]}&$5.20^{+1.45}_{-1.06}$&$2.26^{+0.63}_{-0.41}$&$1.18^{+0.29}_{-0.22}$&$3.06^{+0.74}_{-0.58}$\\
	{    }&{[S]}&$23.5^{+10.3}_{-7.9}$&$3.90^{+0.77}_{-1.40}$&$5.18^{+2.97}_{-3.20}$&$9.56^{+3.49}_{-2.93}$\\
	{    }&{[Fe]$_{\rm  L}$}&$3.53^{+1.59}_{-0.88}$&$0.85^{+0.37}_{-0.25}$&$0.61^{+0.12}_{-0.10}$&$1.70^{+0.48}_{-0.33}$\\
	{    }&{$\net$ ($10^{10}$ \netunit)}&$1.17^{+0.54}_{-0.44}$&$13.6^{+5.8}_{-3.9}$&$6.54^{+4.75}_{-2.49}$&$24.7^{+15.6}_{-8.1}$\\
	{    }&{EM ($10^{-5}$ cm$^{-5}$)}&$0.96^{+0.43}_{-0.34}$&$2.55^{+0.81}_{-0.65}$&$7.21^{+2.82}_{-1.95}$&$3.35^{+0.83}_{-0.77}$\\
	{SRCUT}&{$S_{\rm 1GHz}$ ($10^{-3}$ mJy)}&$5.21^{+1.12}_{-1.32}$&$4.15^{+1.86}_{-2.02}$&$< 2.2$&$< 2.0$\\
	{Gaussian}&{Centroid (keV)}&$1.27^{+0.01}_{0.02}$&$1.23^{+0.03}_{-0.02}$&$1.21^{+0.04}_{-0.02}$&$1.23^{+0.01}_{-0.02}$\\
	{    }&{Flux ($10^{-6}$ photons s$^{-1}$ cm$^{-2}$ keV$^{-1}$)}&$1.10^{+0.17}_{-0.19}$&$0.39 \pm 0.18$&$0.48^{+0.23}_{-0.22}$&$1.01^{+0.20}_{-0.18}$\\
	$\chi^2/$dof&{    }&$245.27/195$&$190.86/193$&$265.89/256$&$214.98/205$\\	\enddata
\end{deluxetable*}

\clearpage

\clearpage

\clearpage

\begin{deluxetable*}{cccc}
	\tablecaption{{Masses of Metal Species ($\Msun$)} \label{tab:mass}}
	\tablenum{3}
	\tablehead{
		Elements&\multicolumn{2}{c}{Low-Temperature Component}&High-Temperature Components\tablenotemark{a}\\
		&Case 1&Case 2&
	}
	\startdata
		O&0.07&0.04&-\\
		Mg&$<0.01$&$<0.01$&-\\
		Si&0.05&0.02&0.05\tablenotemark{b}\\
		S&0.06&0.03&0.03\\
		Ar&- &- &$<0.01$\\
		Ca& -&- &0.01\\
		Fe&0.04&0.02&0.50\\
	\enddata
	\tablenotetext{a}{Assuming a filling factor of 1.}
	\tablenotetext{b}{Abundance is fixed at $10^5$ solar in the spectral fitting.}
	\tablecomments{Two cases: (1)  $f = 1$ assumed for all the regions; (2)  $f = 1$ adopted only for the regions with $f^{-1/2} t_{\rm i} < 400$ yr, but $f = [400~{\rm yr} /(f^{-1/2} t_{\rm i})]^2$ adopted for the others.}
\end{deluxetable*}

\clearpage

\clearpage

\clearpage

\clearpage

\clearpage

\begin{deluxetable*}{ccccc}
	\tablecaption{Double-Gaussian Fitting Results \label{tab:double_gaussian}}
	\tablenum{4}
	\tablehead{
		\colhead{}&\multicolumn{2}{c}{CSM}&\multicolumn{2}{c}{Ejecta}\\
		\cline{2-5}
		\colhead{}&\colhead{[O]/[Fe]$_{\rm  L}$}&\colhead{[Mg]/[Fe]$_{\rm  L}$}&\colhead{[O]/[Fe]$_{\rm  L}$}&\colhead{[Mg]/[Fe]$_{\rm  L}$}
	}
	\startdata
	Peak Ratio&$0.77^{+0.30}_{-0.23}$&$1.11^{+0.46}_{-0.32}$&$0.31^{+0.17}_{-0.10}$&$0.38^{+0.36}_{-0.19}$\\
	Total $n_{\rm e}M_{\rm gas}$ ($\Msun$ cm$^{-3}$)\tablenotemark{a}&$\sim$15&$\sim$6&$\sim$22&$\sim$31\\
	\enddata
	\tablenotetext{a}{Assuminig a distance to Kepler of 5.1 kpc.}
\end{deluxetable*}

\clearpage

\clearpage

\clearpage

\clearpage

\clearpage

\begin{deluxetable*}{llcc}
	\tablecaption{{Examples of 2.2--8.0\,keV Spectral Fitting Results} \label{tab:hard_spec_example}}
	\tablenum{5}
	\tablehead{
		\colhead{}&\colhead{}&\colhead{BIN-47}&\colhead{BIN-56}
	}
	\startdata
	\texttt{vnei}1& $\kTe$ (keV)& $2.92^{+2.17}_{-0.57}$&$2.66^{+0.80}_{-0.75}$\\
	 & [S]\tablenotemark{a}& $9.82^{+0.77}_{-1.00}$& $10.95^{+1.21}_{-1.32}$\\
	 & [Ar]\tablenotemark{a}& $6.98^{+1.57}_{-1.58}$& $6.50^{+1.44}_{-1.23}$\\
	 & [Ca]\tablenotemark{a}& $9.56^{+3.67}_{-3.41}$& $9.23^{+3.25}_{-2.43}$\\
	 & [Fe]$_{\rm K}$\tablenotemark{a} & $18.1^{+4.9}_{-11.5}$& $5.38^{+4.39}_{-2.73}$\\
	 & $\net$ ($10^{10}$\,\netunit) & $2.44^{+0.45}_{-0.36}$& $3.65^{+1.34}_{-0.62}$\\
	 & EM ($10^{-9}$\,cm$^{-5}$)& $6.60^{+1.22}_{-0.76}$& $11.6^{+4.2}_{-2.3}$\\
	\texttt{vnei}2&  $\kTe$ (keV)& $=$ \texttt{vnei}1$\times1.3$& $=$ \texttt{vnei}1$\times1.3$\\
	 &[Fe]$_{\rm K}$ & $=$ \texttt{vnei}1& $=$ \texttt{vnei}1 \\
	 & $\net$ ($10^{10}$\,\netunit)& $< 0.90$& $< 0.15$\\
	 & EM ($10^{-9}$\,cm$^{-5}$)& $8.32^{+4.84}_{-5.01}$& $46.5^{+49.4}_{-20.4}$\\
	 & redshift& $1.81^{+0.24}_{-0.31} \times 10^{-2}$& -\\
%	\texttt{gsmooth}& Gaussian Sigma at 6\,keV (eV)& $23.9^{+4.5}_{-5.2}$& $18.5^{+3.9}_{-4.7}$\\
	\texttt{powerlaw}& Photon index $\alpha$& $3.14\pm0.22$& $3.76^{+0.22}_{-0.25}$\\
	 & Norm ($10^{-4}$ photons s$^{-1}$ cm$^{-2}$ keV$^{-1}$ at 1\,keV)& $1.33^{+0.77}_{-0.31}$& $5.97^{+2.09}_{-1.84}$\\
	$\chi^2/$dof& & 176.76/147 & 186.09/181\\
	\enddata
	\tablenotetext{a}{In units of $10^4$ solar value.}
\end{deluxetable*}

\clearpage

\clearpage
\end{CJK}
\end{document}